\documentclass[a4paper,usenatbib]{mnras}

\usepackage{newtxtext,newtxmath}

\usepackage[T1]{fontenc}
\usepackage{ae,aecompl}

\usepackage{url}
\usepackage{lscape}
\usepackage{longtable}

\usepackage{graphicx}	
\usepackage{amsmath, amsfonts}	
\usepackage{amssymb}	

\title[Persistence of Efficient Environmental Quenching to $z\sim1.4$]{Persistence of the Color-Density Relation and Efficient Environmental Quenching to $z\sim1.4$}
\author[B.~C.\ Lemaux et al.]{B.C. Lemaux$^{1}$\thanks{E-mail:
bclemaux@ucdavis.edu},  A.~R.\ Tomczak$^{1}$, L.~M.\ Lubin$^{1}$, R.~R.\ Gal$^{2}$, L.\ Shen$^{1}$, D. Pelliccia$^{1}$, \newauthor P-F.\ Wu$^{3,4}$, D.\ Hung$^{2}$,
S. Mei$^{5,6,7}$, O. Le F\`{e}vre$^{8}$, N.\ Rumbaugh$^{1,9}$, D.~D.\ Kocevski$^{10}$, \newauthor \& G.~K.\ Squires$^{11}$\\
$^{1}$ Department of Physics, University of California, Davis, One Shields Ave., Davis, CA 95616, USA \\
$^{2}$ University of Hawai'i, Institute for Astronomy, 2680 Woodlawn Drive, Honolulu, HI 96822, USA \\
$^{3}$ Max-Planck Institut f\"{u}r Astronomie, K\"{o}nigstuhl 17, D-69117, Heidelberg, Germany\\
$^{4}$ EACOA Fellow, National Astronomical Observatory of Japan, Osawa 2-21-1, Mitaka, Tokyo 181-8588, Japan \\
$^{5}$ University of Paris Denis Diderot, University of Paris Sorbonne Cit\'{e} (PSC), 75205 Paris Cedex 13, France \\
$^{6}$ Sorbonne Universit\'{e}, Observatoire de Paris, Universi\'{e} PSL, CNRS, LERMA, F-75014, Paris, France \\
$^{7}$ Jet Propulsion Laboratory, Cahill Center for Astronomy \& Astrophysics, California Institute of Technology, 4800 Oak Grove Drive, Pasadena, California, USA\\ 
$^{8}$ Aix Marseille Universit\'{e}, CNRS, LAM (Laboratoire d'Astrophysique de Marseille) UMR 7326, 13388, Marseille, France \\
$^{9}$ National Center for Supercomputing Applications, University of Illinois, 1205 West Clark St., Urbana, IL 61801, USA\\ 
$^{10}$ Department of Physics and Astronomy, Colby College, Waterville, ME 04961, USA \\
$^{11}$ Spitzer Science Center, California Institute of Technology, M/S 220-6, 1200 E. California Blvd., Pasadena, CA 91125, USA} 

\date{Received: 2018 November 28; Revised: 2019 August 30; Accepted 2019: 2019 September 9}

\pubyear{2019}

\begin{document}
\label{firstpage}
\pagerange{\pageref{firstpage}--\pageref{lastpage}}
\maketitle

\begin{abstract}
Using $\sim$5000 spectroscopically-confirmed galaxies drawn from the Observations of Redshift Evolution in Large Scale Environments (ORELSE) survey we investigate 
the relationship between color and galaxy density for galaxy populations of various stellar masses in the redshift range $0.55 \le z \le 1.4$. The fraction of 
galaxies with colors consistent with no ongoing star formation ($f_q$) is broadly observed to increase with increasing stellar mass, increasing galaxy density, and 
decreasing redshift, with clear differences observed in $f_q$ between field and group/cluster galaxies at the highest redshifts studied. We use
a semi-empirical model to generate a suite of mock group/cluster galaxies unaffected by environmentally-specific processes and 
compare these galaxies at fixed stellar mass and redshift to observed populations to constrain the efficiency of environmentally-driven
quenching ($\Psi_{convert}$). High-density environments from $0.55  \le z \le 1.4$ appear capable of efficiently quenching galaxies with 
$\log(\mathcal{M}_{\ast}/\mathcal{M}_{\odot})>10.45$. Lower stellar mass galaxies also appear efficiently quenched at the lowest redshifts studied here, 
but this quenching efficiency is seen to drop precipitously with increasing redshift. Quenching efficiencies, combined  
with simulated group/cluster accretion histories and results on the star formation rate-density relation from a companion ORELSE study, are used to constrain 
the average time from group/cluster accretion to quiescence and the elapsed time between accretion and the inception of the quenching event. These timescales
were constrained to be $\langle t_{convert} \rangle=2.4\pm0.3$ and $\langle t_{delay} \rangle=1.3\pm0.4$ Gyr, respectively, for galaxies with 
$\log(\mathcal{M}_{\ast}/\mathcal{M}_{\odot})>10.45$ and $\langle t_{convert} \rangle=3.3\pm0.3$ and $\langle t_{delay} \rangle=2.2\pm0.4$ Gyr for lower
stellar mass galaxies. These quenching efficiencies and associated timescales are used to rule out certain environmental mechanisms 
as being the primary processes responsible for transforming the star-formation properties of galaxies over this 4 Gyr window in cosmic time.

\end{abstract}




\section{Introduction}

One of the most fundamental properties associated with the study of galaxy evolution is whether or not a galaxy is forming stars. This information,
when determined for a large and diverse enough population of galaxies, can be used to directly test and to calibrate simulations and models of galaxy 
evolution (e.g., \citealt{mcgee11, tal14, balogh16, kawinwanichakij16, kawinwanichakij17, mildmanneredmiguel18}). While the processes governing the likelihood that a particular 
galaxy is actively forming stars are extremely complex, and, further, while stochastic rather than smooth star-formation histories may be typical for 
galaxies (e.g., \citealt{pacifici13, teyssier13, kauffmann14, sparre17, emami19}), 
when averaging over an ensemble of galaxies there are physical parameters which can be used as strong predictors of how many galaxies in 
that population are likely to be forming stars. At least to $z\sim1$, it appears that three parameters in 
particular, cosmic epoch, the mass of the stellar content contained within the galaxy, and the environment in which a galaxy resides, correlate well with 
the star formation properties of galaxies. While there are many other properties beyond these three which may serve to modulate star formation 
including, but not limited to, the level of active galactic nuclei (AGN) activity (e.g., \citealt{dirtydale09, rum12, ofthewood13, juneau13, lem14}, 
shocks, merging, and other galaxy-wide processes (e.g., \citealt{birnboim03, keres05, carlos13, belfiore16, belfiore17}), kinematics (e.g., \citealt{genzel08, genzel14, hung18}), 
morphology (e.g., \citealt{martig09, bell12, whitaker17}), and the amount and metallicity of the gas, if any, contained within galaxies (e.g., 
\citealt{smokynicky16, smokynicky17, darvish18, sanders18, wang18}), all of these properties can be thought of in some way to be subordinate (i.e., correlated with) to 
redshift, stellar mass, or environment or some combination of the three. 

However, while studies exist in abundance probing each of these three parameters in tandem across a wide variety of redshifts, stellar masses, and 
environments to $z\sim1-1.5$, there are few studies which simultaneously and self-consistently probe a wide baseline in each of the three parameters. 
For example, \citet{wetzel13} presented an extensive investigation of the star formation rate (hereafter $\mathcal{SFR}$) distribution and the 
fraction of galaxies not forming stars, otherwise known as the quiescent faction ($f_q$), for a large sample of 
spectroscopically-confirmed galaxies from the Sloan Digital Sky Survey (SDSS). While this study, by virtue of the 
large number of galaxies, and the large dynamic range of stellar masses and environments probed, provided valuable insights on the star-formation
histories of galaxies in different environments, these insights were broadly limited to galaxy populations residing
in the local universe ($z\la0.3$). At higher redshift ($z\sim1$), samples from the VIMOS Very Deep Survey (VVDS, \citealt{dong05, dong13}), the Deep Extragalactic 
Evolutionary Probe 2 (DEEP2, \citealt{davis03,new13}), the zCOSMOS-Bright survey \citep{lilly07,lilly09}, and the VIMOS Public Extragalactic Redshift 
Survey (VIPERS; \citealt{bianca14, guzzo14}) have been extensively studied to measure variations of $f_q$. In such surveys, this quantity is typically 
proxied by the number of red galaxies relative to the total number of galaxies, as a function of redshift, stellar mass (or, somewhat equivalently, absolute 
magnitude), and environment (e.g., \citealt{olga06, mcoopz07, peng10, kovac14, olga17}). However, the range of environments probed by each survey is typically limited to, 
at most, massive group environments, and the spectroscopic populations are typically skewed toward bluer galaxies. As such, the conclusions drawn on these results 
necessarily share the same limitations. Conversely, observational campaigns exist 
which target the cores and intermediate
regions of massive clusters at a variety of redshifts, but such campaigns can suffer from a spectroscopic sampling issues, lack of a comparable field sample,
large sample variance from the small number of clusters probed, small redshift windows, and the lack of sampling in the sparser 
regions of clusters and the surrounding filamentary structure. 

While the number of conflicting results and the magnitude of these conflicts speak to the 
difficulty of overcoming the above difficulties completely, clever methods have been employed to overcome some of these limitations, and such campaigns have been
used to good effect to present compelling evidence that the relationship of the quiescent fraction of galaxies with environment, stellar mass, and redshift
persists to some extent up to $z\sim1.5$ (e.g., \citealt{strazzullo10, snyder11, quadri12, cooke16, nantais16}). Secular processes 
strongly tied to the stellar mass content of galaxies are generally thought to regulate star formation in lower-density environments, with star-forming galaxies 
observed to be less prevalent amongst galaxies at higher stellar mass. In these rarefied environments, small environmental effects may enter for galaxies that 
reside in or near voids and filaments, as the distance to the void or filament appears to modulate the colors and star-forming properties of such galaxies 
(e.g., \citealt{moorman16, kuutma17, katarina18, laigle18}), however, it generally appears that secular processes dominate. 
While the trend of increasing quiescent fraction with the increasing mass of stellar content in field environments appears to hold to $z\sim2-3$ 
(e.g., \citealt{ilbert13, lunchpailmcgee13, AA14}), the stellar mass at which quiescent galaxies become more abundant than star-forming galaxies is a strong function 
of redshift. Another metric used to probe the star-forming state of galaxies is the star formation rate normalized to the stellar mass content, otherwise 
known as the specific star formation rate ($\mathcal{SSFR}$). This metric appears, both for all types of galaxies and for star-forming galaxies only, to be 
strongly tied to both the redshift and stellar mass, in that the average $\mathcal{SSFR}$ of galaxies is seen to increase at increasing redshift and for 
galaxy populations with smaller average stellar masses (e.g., \citealt{AA16}). 

When a larger dynamic range of environments is probed, including the intermediate- to high-density regimes of groups and clusters, the situation becomes 
more complex. While, among star forming galaxies, at fixed $\mathcal{M}_{\ast}$ the average $\mathcal{SSFR}$ appears to be largely 
independent across all types of environments (e.g., \citealt{muz12, wijesinghe12, koyama13, zeimann13, darvish16, vulcani17, wagner17}, though see also 
\citealt{patel11, ziparo14, tran15, AA19}), the quiescent fraction is consistently found to be a strong function
of environment at least until $z\sim1$ (e.g., \citealt{balogh16, olga17, kawinwanichakij17, jian18}). This behavior indicates that the ability of galaxies to form stars 
is in some way 
related to the plethora of processes which are enhanced or exclusive to intermediate- to high-density environments. Indeed, the behavior of the quiescent fraction as a function
of stellar mass and environment at fixed redshift has been used in a number of studies to estimate timescales 
associated with the quenching of group and cluster galaxies as well as the processes involved in that quenching (e.g., \citealt{mok14, balogh16, fossati17, foltz18}). 
While promising, the limited number of cluster and group environments studied, the limited number of galaxies studied per system, and the lack of a comprehensive investigation
of quenching timescales as a function of both stellar mass and redshift, make it difficult to understand how applicable the conclusions drawn 
from these investigations are to the evolution of generic galaxy populations in higher density environments. 

In this paper we present a study of a large sample ($\sim$5000) of spectroscopically-confirmed galaxies from the Observations of Redshift Evolution
in Large Scale Environments (ORELSE; \citealt{lub09}) survey, an immense imaging and spectroscopic campaign over $\sim$5 $\Box^{\circ}$ probing the surrounding 
structure of 15 known large scale structures at $0.6\le z \le 1.3$ on $\sim$10-15 $h_{70}^{-1}$ proper Mpc scales. For this sample, which is 
largely representative of the overall galaxy population for galaxies with $\log(\mathcal{M}_{\ast}/\mathcal{M}_{\odot})>10$, we measure $f_q$ at 
different redshifts, stellar masses, and environments. These measurements, in conjunction with semi-empirical models and $N$-body simulations, 
are used to self-consistently infer the efficiency with which group and cluster environments are able to quench star formation in their constituent galaxies 
to $z\sim1.4$ and the quenching timescale for galaxies of different stellar masses. In tandem with a companion study \citep{AA19} in which a 
similar ORELSE sample is used to study the relationship between $\mathcal{SSFR}$ and environment for star-forming galaxies, we further estimate
the timescale that galaxies of differing stellar masses can remain in group and cluster environments without undergoing a galaxy-scale quenching 
event. These timescales are subsequently used to argue in favor or against certain processes as agents of this quenching. 

Throughout this paper all magnitudes, including those in the IR, are presented in the AB system \citep{okengunn83,fukugita96}.
All distances are quoted in proper units. We adopt a concordance $\Lambda$ cold dark matter cosmology with $H_{0}$ = 70 km s$^{-1}$ Mpc$^{-1}$, 
$\Omega_{\Lambda}$ = 0.73, and $\Omega_{M}$ = 0.27. 
While abbreviated throughout the paper for convenience, stellar masses are presented in units of $h_{70}^{-2} \mathcal{M}_{\odot}$, 
star formation rates in units of $h_{70}^{-2} \mathcal{M}_{\odot}$ yr$^{-1}$, total and halo masses in units of $h^{-1}_{70} \mathcal{M}_{\odot}$, 
ages in units of $h^{-1}_{70}$ Gyr or $h^{-1}_{70}$ Myr, absolute magnitudes in units of $M_{AB}+5\log(h_{70})$, proper distances in 
units of $h_{70}^{-1}$ kpc/Mpc, where $h_{70}\equiv H_{0}/70$ km$^{-1}$ s Mpc.

\begin{figure*}
\includegraphics[clip,angle=0,width=0.98\hsize]{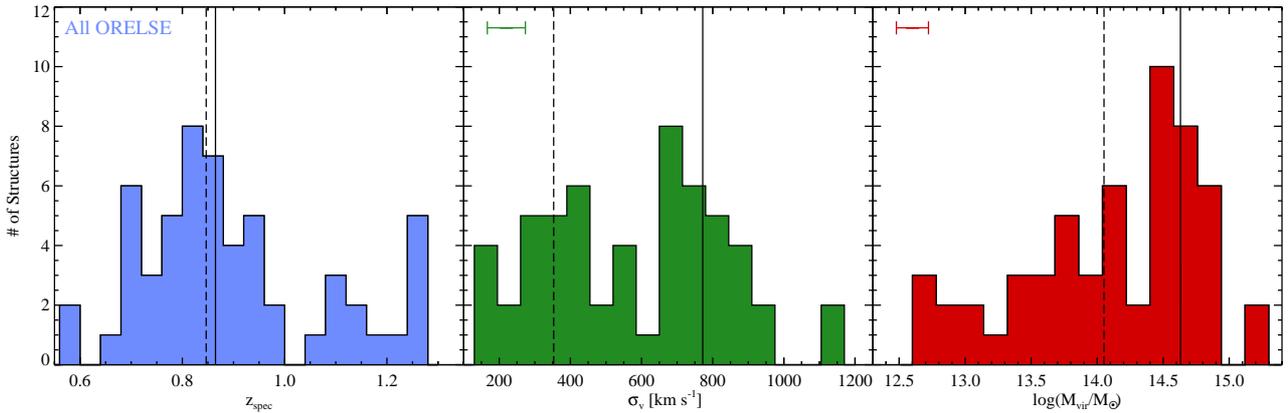}
\caption{\emph{Left:} Redshift distribution of the 56 groups and clusters detected in the 15 ORELSE fields presented in this paper. The systemic redshift of 
each structure is estimated by the average of all spectral members within 1 Mpc of the luminosity-weighted spectral member center (see \citealt{begona14}).
Solid and dashed vertical lines denote the median redshift of ORELSE clusters ($\sigma_{v, \, LOS}\geq550$ km s$^{-1}$) and groups ($\sigma_{v,\, LOS}<550$ km s$^{-1}$), 
respectively. \emph{Center:} Distribution of $\sigma_{v}$ for the same 56 structures. Velocity
dispersions are calculated from an average of 34 and 15 member galaxies for ORELSE clusters and groups, respectively. The median $\sigma_{v}$ of clusters and
groups are indicated by the solid and dashed vertical lines, respectively. The average $\sigma_{v}$ error, estimated through jackknife resampling for each structure, 
is shown in the top left. \emph{Right:} Virial mass of all 56 structures as estimated from $\sigma_{v}$. The median $\mathcal{M}_{vir}$ for clusters
and groups are indicated by solid and dashed vertical lines, respectively. The average $\mathcal{M}_{vir}$ error is shown in the top left.}
    \label{fig:structdist}
\end{figure*}

\section{The ORELSE Field and Large Scale Structure Sample}
\label{tarnobs}

In this paper we incorporate samples drawn from observations of all 15 of the final ORELSE fields. Adopting the naming convention of \citet{lub09}, these fields 
are Cl 0023+0423, RCS J0224-0002, XLSSC005a+b, Cl J0849+4452, Cl J0910+5422, RX J1053.7+5735, 1137+3000, RX J1221.4+4918, Cl 1324+3011/Cl1324+3059, Cl J1350+6007,
Cl J1429.0+4221, Cl 1604+4304/Cl 1604+4321, RX J1716.4+6708, RX J1757.3+6631, RX J1821+6827\footnote{While this list seemingly contains 17 entries, we consider Cl 1324+3011 and 
Cl1324+3059 as a single field, referred to collectively as SC1324, and Cl 1604+4304 and Cl 1604+4321, referred to collectively as SC1604, as a single field. The fields
Cl 0934+4804, Cl1325+3009, and RX J1745.2+6556 were dropped from the survey due to insufficient observations.}. The full galaxy sample taken from these fields is comprised 
of nearly 9000 spectroscopically confirmed galaxies
in the redshift range studied in this paper, $0.55\leq z \leq 1.4$, which span a range of environments from the sparsely populated field to the cores of 
massive, relaxed clusters. In total, 56 groups and clusters were discovered across all 15 of the ORELSE fields spanning the ranges of redshift, line-of-sight (LOS) galaxy 
velocity dispersion, and virial masses shown in the three panels of Figure \ref{fig:structdist}. The latter of these three measurements are made using the relation 
given in \citet{lem12}. While several 
thousand of the galaxies in the sample presented here are confirmed 
members of either these groups and clusters or the large scale structures (LSSs) that they are embedded in, the remainder of our sample, consisting of several thousand more 
galaxies, are completely unassociated with any massive structure. Such samples allow for an internal comparison of various subsets of galaxies across a 
a broad range of environments at different epochs. Imaging and spectroscopic observations made on nearly all of the final 15 ORELSE fields, as well 
as methods relating to the reduction of these data and the measuring of relevant quantities from these observations, have been discussed in other ORELSE studies 
(with the exception of one field, see \S\ref{Cl1137}). As such, we only briefly describe the characteristics of these observations and methods here. 

\subsection{Imaging and Photometry}
\label{phot}

Optical $riz$ imaging for most ORELSE fields was initially performed using either the Large Format Camera (LFC; \citealt{simcoe00}) situated on the prime focus of the Palomar 
5-m Hale telescope or Suprime-Cam \citep{miyazaki02} mounted on the prime focus of the Subaru 8-m telescope. In one case, XLSSC005a+b (hereafter XLSS005), initial optical imaging 
was drawn from Megacam \citep{boulade03} observations from the 3.6-m Canada-France-Hawaii Telescope (CFHT) taken as part of the ``Deep'' portion of the CFHT Legacy Survey (CFHTLS). 
Additional $B$ and $V$ imaging was taken for all ORELSE fields with Subaru/Suprime-Cam except for the XLSS005 field where $u^{\ast}$- and $g^{\prime}$-band imaging were
available. In some cases Subaru/Suprime-cam imaging in the Y band was also taken. The typical depths of the optical imaging ranged from $m_{AB}=26.4$ in the B band to 
$m_{AB}=24.6$ in the $z^{\prime}/Z^+$ bands as estimated from the method described in \citet{AA17}. All LFC data were reduced with Image Reduction and Analysis Facility
(\texttt{IRAF}, \citealt{tody93}) following the methods of \citet{gal08}. Reduction of the Suprime-Cam data were performed
with the \texttt{SDFRED2} pipeline \citep{ouchi04} supplemented by several Traitement \'{E}l\'{e}mentaire R\'{e}duction et Analyse des PIXels
(\texttt{TERAPIX\footnote{http://terapix.iap.fr}}) routines\footnote{C. D. Fassnacht, \emph{private communication}} (for more details, see \citealt{AA17}). In both cases, 
photometric calibration was performed from observations of
\citet{landolt92} standard star fields taken on the same night of each observation. CFHTLS observations were reduced and photometrically calibrated using TERAPIX routines 
following the methods described in \citet{ilbert06} and the T0006 CFHTLS handbook\footnote{http://terapix.iap.fr/cplt/T0006-doc.pdf}. Though ORELSE imaging observations in
the redder portion of the observed-frame optical employed a variety of different filter curves, for simplicity, we will usually use the generalized terms $r$, $i$, and $z$ 
throughout the paper to refer to all variants of these filters (i.e., $r^{\prime}/R_C/R^+$, $i^{\prime}/I_C/I^+$, and $z^{\prime}/Z^+$, respectively) except when specificity is
helpful.

All but one field in ORELSE (Cl1350) was observed in the near-infrared (NIR) $J$ and $K/K_{s}$ bands. These observations were taken from the Wide-Field Camera 
(WFCAM; \citealt{casali07}) mounted on the United Kingdom Infrared Telescope (UKIRT) and the Wide-field InfraRed Camera (WIRCam; \citealt{puget04}) 
mounted on CFHT and reached a typical depth of $m_{AB}=21.9$ \& 21.7 for the $J$ and $K/K_{s}$ bands, respectively. The UKIRT data were processed using 
the standard UKIRT processing pipeline provided courtesy of the Cambridge Astronomy Survey Unit\footnote{http://casu.ast.cam.ac.uk/surveys-projects/wfcam/technical}
and, the CFHT data, through the I'iwi pre-processing routines and \texttt{TERAPIX}. Both pipelines provided fully-reduced mosaics and associated weight maps. 
The photometric calibration of the mosaics output by both pipelines was done selecting bright ($m<15$), non-saturated objects with existing Two Micron All Sky Survey 
(2MASS; \citealt{skrutskie06}) photometry, with appropriate $k$-corrections made using stars drawn from the Infrared Telescope Facility (IRTF) spectral library \citep{rayner09}. 
Additional imaging in the NIR was taken from the \emph{Spitzer} \citep{wer04} space observatory using the InfraRed Array Camera (IRAC; \citealt{fazio04}) in the two 
non-cryogenic channels ($[3.6]/[4.5]$) for all 15 ORELSE fields and additionally in the two cryogenic channels ($[5.8]/[8.0]$) for four of the ORELSE fields (SC1604, RXJ1716, RXJ1053, 
and XLSS005 following the naming convention of \citealt{rum18, AA19}) to an average depth of 24.0, 23.8, 22.4 and 22.3 magnitudes, respectively. The basic calibrated data 
(cBCD) images provided by the \emph{Spitzer} Heritage Archive were reduced using the MOsaicker and Point source EXtractor (\texttt{MOPEX}; \citealt{makovoz06}) package in 
conjunction with several custom Interactive Data Language (\texttt{IDL}) scripts written by J. Surace. For more details on the reduction of these data see \citet{AA17}. 

For each field, all optical and non-\emph{Spitzer} images were registered to a common grid of plate scale 0.2$\arcsec$ pixel$^{-1}$ and convolved to the worst point 
spread function (PSF) for that field using the methods described in \citet{AA17}. Prior to this convolution, some images with exceptionally large PSFs for which we had the 
luxury of another image with broadly redundant spectral coverage (i.e., LFC $r^{\prime}$ vs. Suprime-cam $R^{+}$ imaging) with a smaller measured PSF were removed.
The worst PSF per field for all retained ground-based optical/NIR images ranged from $\sim$1.00$\arcsec$-1.96$\arcsec$ for the fields studied here, with only one 
field (Cl1350) having an image with a PSF that exceeded 1.4$\arcsec$. Source detection and photometry were obtained by running Source Extractor (\texttt{SExtractor}; 
\citealt{BertinArn96}) in dual-image mode using either a stacked $\chi^2$ optical image or an image in a single band as a detection image (for details on the 
specific image used for each field except for 1137+3000 (hereafter Cl1137), see
\citealt{AA17,rum18}. For details on Cl1137, see \S\ref{Cl1137}. Fixed-aperture photometry was performed on all PSF-matched images with \texttt{SExtractor} employing an aperture 
of 1.3$\times$ the FWHM of the homogenized PSF for each field and transformed to a total magnitude using the ratio of aperture and AUTO flux densities as measured in the 
detection image. Magnitude uncertainties were calculated from adding, in quadrature, \texttt{SExtractor} uncertainties to our own estimates of background noise drawn 
from the 1$\sigma$ root mean square (RMS) scatter of measurements in hundreds of blank sky regions for each band. \emph{Spitzer}/IRAC magnitudes were incorporated by 
running the software \texttt{T-PHOT} \citep{merlin15} on the fully reduced mosaics using the segmentation maps from the ground-based detection images as input, with flux
density uncertainties estimated from the scaled best fit model for each object. For more details on the reduction and measurements of ORELSE imaging data see \citet{AA17}.

\subsection{Optical Spectroscopy}
\label{spectra}

The optical imaging described above was used following the methods described in \citet{lub09} to select targets for spectroscopic followup with the DEep Imaging and 
Multi-Object Spectrometer (DEIMOS; \citealt{fab03}), located at the Nasmyth focus of the Keck \ion{}{II} telescope. In brief, the spectroscopic targeting scheme employed
a series of color and magnitude cuts that are applied to maximize the number of targets with a high likelihood of being on the cluster/group red sequence at the 
presumed redshift of the LSS in each field (i.e., priority 1 targets). However, as discussed extensively in \citet{AA17}, because of the relative rarity of such objects, 
the majority of targets for all ORELSE fields were objects with colors outside of these ranges (predominantly blueward). The fraction of priority 1 targets which entered 
into our final sample ranged from $\sim$10\% to $\sim$45\% across all ORELSE fields, a fraction which tended to  vary strongly with the density of spectroscopic 
sampling per field. We discuss the consequences of this targeting scheme for the results presented in this study in \S\ref{specrepresent}. Spectroscopic targets were generally 
limited to $i<24.5$, with a median of $\widetilde{i}\sim23$, though this magnitude limit was not imposed strictly and an appreciable number of objects 
were targeted below this limit. In addition, objects detected
at X-ray wavelengths in our \emph{Chandra} imaging \citep{rum17} or at radio wavelengths from our Very Large Array (VLA) 1.4 GHz imaging \citep{shen17} were also 
highly prioritized, though the number density of these objects is typically low and, thus, such objects only comprised a small fraction of targets on a given mask 
(typically $\la10$\%). 

\begin{table*}
        \centering
        \caption{ORELSE DEIMOS Spectral Observations$^{a}$}
        \label{tab:LSSs}
        \begin{tabular}{ccccccccccc}
                \hline
                Field$^{a}$ & $\lambda_{c}$ & Avg. Coverage & $t_{int}$ & Avg. Seeing & N$_{\rm{mask}}$ & $N_{\rm{target}}^b$ & $N_{\rm{spec}}^{c}$ & $N_{\rm{spec, gal}}^{d}$ & $\log(\mathcal{M}_{vir}/\mathcal{M}_{\odot})$$^{e}$ & $z_{spec}$ \\
                          & [\AA] & [\AA] & [s] & [$\arcsec$] & & & & &  \\
                \hline
                SG0023 & 7500-7850 & 6200-9150 & 5700-9407 & 0.45-0.81 & 9 & 1128 & 923 & 758 & 12.7-13.9 & 0.829-0.980 \\
                RCS0224 & 7300-7450 & 6000-8750 & 6840-7520 & 0.53-0.88 & 4 & 598 & 493 & 381 & 13.9-14.8 & 0.778-0.854 \\
                XLSS005 & 7900 & 6600-9200 & 1063-12600 & 0.39-1.12 & 9 & 999 & 733 & 586 & 14.5 & 1.056 \\
                SC0849 & 8700 & 7400-10000 & 6300-16200 & 0.51-1.50 & 8 & 977 & 556 & 390 & 12.7-14.7 & 0.568-1.270 \\
                RXJ0910 & 8000-8100 & 6700-9400 & 7200-11664 & 0.50-1.05 & 7 & 971 & 736 & 505 & 12.7-14.7 & 0.760-1.103 \\
                RXJ1053 & 8200 & 6900-9500 & 7200-9000 & 0.56-0.79 & 5 & 698 & 405 & 300 & 14.8 & 1.129-1.204 \\
                Cl1137 & 7850 & 6550-9150 & 2600-14400 & 0.53-1.20 & 6 & 827 & 539 & 438 & 14.1 & 0.955 \\
                RXJ1221 & 7200 & 5900-8500 & 4860-8400 & 0.55-1.20 & 5 & 663 & 519 & 411 & 13.9-14.7 & 0.700-0.702 \\
                SC1324 & 7200 & 5900-8500 & 2700-10800 & 0.44-1.00 & 12 & 1653 & 1332 & 985 & 12.8-14.8 & 0.696-1.098 \\
                Cl1350 & 7500 & 6200-8800 & 3600-10400 & 0.50-1.55 & 6 & 803 & 623 & 352 & 13.4-14.7 & 0.800-0.802 \\
                Cl1429 & 7400-7500 & 6100-8800 & 5000-7800 & 0.43-0.85 & 8 & 1017 & 835 & 563 & 14.8 & 0.987 \\
                SC1604 & 7700 & 6400-9000 & 3600-14400 & 0.50-1.30 & 18 & 2358 & 1801 & 1294 & 13.4-14.7 & 0.600-1.182 \\
                RXJ1716 & 7800 & 6500-9100 & 5400-9600 & 0.54-0.83 & 6 & 944 & 675 & 513 & 14.4-15.1 & 0.809-0.853 \\
                RXJ1757 & 7000-7100 & 5700-8400 & 6300-14730 & 0.47-0.82 & 6 & 945 & 742 & 397 & 13.4-14.8 & 0.693-0.946 \\
                RXJ1821 & 7500-7800 & 6200-9100 & 7200-9000 & 0.58-0.86 & 6 & 728 & 611 & 342 & 14.5-15.1 & 0.817-0.919 \\
                \hline
        \end{tabular}
        \begin{flushleft}
$a$: Field names are shorthand notation of the names given in section \S\ref{tarnobs} and can be matched by number
$b$: For numbers of targets, secure spectral redshifts, and secure spectral extragalactic redshifts, we also include numbers from all other spectral surveys incorporated into this study (see \S\ref{spectra}) with the exception of those from the VVDS survey.
$c$: Includes only those objects with a secure spectral redshift (see \S\ref{spectra}) and includes serendipitous detections which comprise $\sim3.6$\% of the entire sample.
$d$: These numbers include only those galaxies with a secure spectral redshift in the redshift range used in this study, $0.55 \le z \le 1.4$
$e$: Virial mass range of groups and clusters detected within each field. These groups and clusters only refer to those previously known groups and clusters reported in \cite{denise19} and not those additional overdensities found in that study. 
\end{flushleft}

\end{table*}

Spectral observations with DEIMOS taken as part of ORELSE exclusively used the 1200 l mm$^{-1}$ grating with 1$\arcsec$ slit widths for the $\sim$120 targets per slitmask 
and a central wavelength that ranged between 7000-8700\AA\ depending on the redshift of the LSS being targeted. These observations resulted in a pixel scale of 0.33 \AA\ 
pixel$^{-1}$, a wavelength coverage of $\pm$1300\AA\ roughly centered on the central wavelength of the observation\footnote{The true central wavelength of a given slit can vary 
$\pm$150\AA\ from the fiducial value depending on where the slit is placed on the DEIMOS mask in the direction parallel to the dispersion dimension.}, and a spectral resolution of 
$R\sim5000$ ($\lambda/\theta_{FWHM}$, where $\theta_{FWHM}$ is the FWHM spectral
resolution). This setup and resolution was required to allow us the possibility to detect several important spectral features at the redshift of the targeted LSS (e.g., 
[\ion{O}{II}] $\lambda$3726, 3729\AA, \ion{Ca}{II} K\&H $\lambda$3934, 3969\AA, $D_{n}(4000)$, and H$\delta$ $\lambda$4101\AA) and to ensure, in most cases, the 
spectral separation of the $\lambda$3726, 3729\AA\ [OII] doublet. Separating this doublet allows for a secure determination of the redshift of targets based on this feature alone. 
The number of slitmasks per field varied from 4 (RCS0224) to 18 (SC1604), with generally larger and more complex LSSs, as well as those at higher redshift, given more extensive 
coverage. Average integration times per mask ranged from $\sim$7000s to $\sim$10500s and were varied based on conditions and the faintness of the target population to roughly 
maintain the same median signal-to-noise ratio of all targets from mask to mask. For details on the observations of specific ORELSE fields, see Table \ref{tab:LSSs}. 

Observations from DEIMOS were reduced using a modified version of the Deep Evolutionary Extragalactic Probe 2 (DEEP2; \citealt{davis03, new13}) \texttt{spec2d} pipeline.
This package combines the individual exposures of the slit mosaic and performs wavelength calibration, cosmic ray removal and sky subtraction on a slit by slit basis, generating 
a processed two-dimensional spectrum for each slit. The \texttt{spec2d} pipeline also generates a processed one-dimensional spectrum for each slit. This extraction creates a 
one-dimensional spectrum of the target, containing the summed flux at each wavelength in both a boxcar and an optimized window \citep{horne86}. The accompanying \texttt{spec1d} 
package is then run on all resulting one-dimensional spectra. This package cross-correlates a suite of galactic and stellar templates to find 10 redshifts which correspond
local minima in $\chi^2$ space for different combinations of templates. These redshifts are used to inform the visual inspection process performed subsequent to the 
reduction and cross-correlation steps (see below)  

For the purposes of the reduction of the data presented in this paper, several modifications were made to the
official version of the \emph{spec2d} pipeline. These modifications included an improved methodology of interpolating over the $\sim10$ \AA\ chip gap between the red and blue
CCD arrays on DEIMOS, the implementation of an improved throughput correction, and an improvement in the functionality related to estimating the final wavelength solution.
More specifically, rather than interpolating over single pixel values on either side of the gap, the pipeline now smooths variations over relatively large windows
($\sim$50 pixels) on either side of the chip gap and linearly interpolates over these smoothed arrays. This implementation helps considerably in reducing artifacts resulting
from reduction or from improper identification of the gap. In addition, an improved method of applying the measured instrumental throughput was employed by combining throughput
data from multiple discrete observational setups\footnote{Taken from https://www2.keck.hawaii.edu/inst/deimos/ripisc.html} and estimating an appropriate throughput correction
at every wavelength for observational setups at all filter/grating/tilt combinations. The pipeline also now attempts fits to multiple optical models if the fiducial guess
fails, has an increased search window for tweaking the wavelength solution resulting from the initial optical model guess as well as an increased tolerance for offsets between
wavelength solutions derived from arc lamps and those derived from night sky lines. Additionally, a more complete night sky line list has been added to the code to allow for
more accurate matching to the airglow spectrum measured on the science frames. These latter changes were necessary as it was noticed early on in the reduction of some masks
that wavelength solutions were either failing catastrophically, or, worse, converging, without error, to incorrect values for some slits on our masks while converging to
correct values for other slits. Finally, a larger suite of empirical templates generated from a variety of observations \citep{lilly07, lem09, dong13, dong15} was added to 
the templates used in the \texttt{spec1d} cross-correlation process primarily to allow the extension of its functionality to higher ($z>2$) redshift. 

Following the reduction and template cross-correlation, each spectrum was then visually inspected using the publicly available DEEP2 redshift measurement program, 
\texttt{zspec} \citep{new13} to determine, 
if possible, the redshift of each target either from the list 10 redshifts determined by \texttt{spec1d} or from a custom redshift fit to visually-identified spectral features. 
Additionally, all two-dimensional spectra were searched for serendipitous detections both spatially coincident and separated from target
galaxies (see \citealt{lem09} for details on these types of 
detections and the method used for finding them). For those spectra which contain one or more serendipitous detections, one-dimensional spectra were extracted in the same manner
as was done with the \texttt{spec2d} pipeline and their redshifts were determined, when possible, in the \texttt{zspec} environment. Each target and serendipitous detection
was assigned a quality code, $Q$, which represents our confidence in the redshift measurement. The criteria required to assign each quality code is the same as that adopted in 
the DEEP2 survey (see \citealt{gal08, new13, AA17} for details on these criteria). For this study, we consider only those objects with $Q$=-1, 3, \& 4 to have secure
spectral redshifts. These three quality codes correspond to stellar ($Q=-1$) and extragalactic ($Q=3,4$) redshifts secure at the $>95$\% level. 

Additionally, a small number
of spectral redshifts ($\sim$250) were included from ORELSE precursor surveys or other, unrelated surveys designed specifically to observe the LSSs also targeted in ORELSE 
\citep{oke98, galnlub04, gioia04, tanaka08, mei12} that employed a variety of different instruments and setups. For spectra coming from these surveys we imposed, to the best
of our ability, the same quality code system as was applied to our DEIMOS data and accepted only those spectral redshifts with a high probability of being correct (i.e., 
the equivalent of $Q$=-1, 3, \& 4). Finally, an additional $\sim$1000 redshifts were drawn from the VIMOS Very Deep Survey (VVDS; \citealt{dong13}), of which $\sim$700 are 
within the redshift range used in this study ($0.55 \le z \le 1.4$). These objects are exclusively located in the XLSS005 field. For these data, we incorporated only those 
objects with spectroscopic reliability flags of X2, X3, X4, or X9, where X=0-2\footnote{X=0 is reserved for target galaxies, X=1 for broadline AGN, and X=2 for non-targeted objects 
that fell serendipitously on a slit at a spatial location separable from the target.}, which corresponds to a probability of being correct of $\ga75$\%. 
Our results do not change appreciably if we instead adopt only those objects in VVDS with highly secure ($>95$\%) spectral redshifts. Galaxies suspected of containing 
an AGN due to either detection in X-rays, radio, broadline spectral features, or power-law infrared spectral energy distributions (SEDs) were kept in our final sample
as the redshift of such sources still remains useful for a variety of purposes. Further, the presence of the AGN does not necessarily preclude the possibility that some
or all of the physical parameters derived from our SED fitting process (see \S\ref{SEDfitting}) remain valid as, in many cases, the restframe ultraviolet/optical/NIR portion 
of the SED remains dominated by the stellar component of such galaxies. Regardless, our results do not change appreciably if these objects are excluded from all analysis. The 
number of spectral targets, secure spectral redshifts, and with the total number of secure spectral extragalactic redshifts in the redshift range studied in this paper are 
given in Table \ref{tab:LSSs}. These numbers include both serendipitous detections and all redshifts from non-ORELSE surveys with the exception of VVDS. Also in
Table 1 we list the properties of the known ORELSE groups and clusters listed in \cite{denise19}. For a full account of the groups and clusters detected in the ORELSE fields 
see \cite{denise19}. 

\subsection{Cl 1137+3000}
\label{Cl1137}
The only ORELSE field included in the sample presented in this paper that has not appeared in detail in previous ORELSE studies is Cl1137. 
The remainder of the fields as well as the associated observations on each field are discussed extensively in other ORELSE papers (e.g., 
\citealt{lem12, lem17a, rum17, rum18, AA17, AA19}). The Cl1137 field is unique among the ORELSE sample as it is the only cluster selected at radio wavelengths.
This selection was based on the presence of a large ($\sim$240 kpc) wide-angle tailed radio source (WAT; \citealt{owen76}) detected in the Very Large Array (VLA) Faint Images of the 
Radio Sky at Twenty Centimeters (FIRST; \citealt{thebob95}) survey, which is indicative of a Fanaroff-Riley type I \citep{FR74} active galactic nucleus (AGN) or a similar 
phenomenon interacting at high relative	
velocities with ambient material, typically that formed by the presence of a group or cluster \citep{odon93,blanton03}. An optical/NIR imaging and Keck \ion{}{II}/Low 
Resolution Imaging Spectrometer (LRIS; \citealt{oke95}) spectroscopic campaign 
in the region surrounding the location of the WAT confirmed the presence of a structure at a systemic\footnote{Calculated by the biweight mean of the
galaxies presented as members in \citet{blanton03}} redshift of $\langle z \rangle = 0.9559$ with a line of sight (LOS) galaxy velocity dispersion of 
$\sigma_{v}=530^{+190}_{-90}$ km s$^{-1}$ based on ten members \citep{blanton03}. The projected distance of the WAT with respect to the structure center as 
determined by the ORELSE data (see below), $\sim$0.25 h$_{70}^{-1}$, is well within $R_{500}$ and consistent with the typical observed location of WATs in 
lower redshift clusters making it extremely likely that its origins result from interaction with a dense medium. 

Both proprietary and archival optical/NIR imaging were collected on the Cl1137 field from Subaru/Suprime-Cam, UKIRT/WFCAM, and \emph{Spitzer}/IRAC
to the depths listed in Table \ref{tab:limits}. These data were reduced, and PSF-matched photometry was measured for all ten bands in a manner identical to 
that presented in \S\ref{phot} and \citet{AA17} using the $Z^{+}$-band image as the detection image. In total, six masks were observed with the Keck \ion{}{II}/DEIMOS 
from January 2011 to April 2015 as part of the ORELSE survey. Integration times varied from 2600-14400s per mask\footnote{The shallowest mask was used primarily to select targets 
for subsequent masks. The next shortest integration time was 7200s.} under mostly photometric conditions and seeing which varied from 0.5-1.2$\arcsec$.
All observations were taken with the 1200 l mm$^{-1}$ grating tilted to a central wavelength of 7850\AA\ and used the OG550 order blocking filter. Spectroscopic
targets were assigned following the scheme presented in \citet{lub09}, with objects that had $R_CI^+Z^+$ colors consistent with those expected for a generic 
cluster red sequence at $z\sim0.95$ were highly prioritized (i.e., priority 1). In the Cl1137 field, out of 827 targets, only
116 (14.0\%) had colors consistent with our priority 1 color window (i.e., $1.0 \leq R_C-I^+ \leq 1.4$, $0.6 \leq I^+-Z^+ \leq 1.0$). In total, these observations yielded
539 (438) high-quality (extragalactic) redshifts, of which 66 fell in the redshift range of the main LSS of 
$0.937 < z < 0.963$. Only one massive structure was found in this field, a structure centered at [$\alpha_{J2000}$, $\delta_{J2000}$] = [174.39786, 30.00893]\footnote{This 
center represents the $Z^{+}$ luminosity-weighted center of all spectroscopic member galaxies. For more details on this calculation see \citet{begona14}.}, with a systemic
redshift of $\langle z \rangle = 0.9553 $ and a LOS galaxy velocity dispersion of $\sigma_{v}=535\pm81$ km s$^{-1}$ based on 28 members within 1 h$_{70}^{-1}$ Mpc
from the structure center. Following the formalism of \citet{lem12}, these measurements indicate a structure with a virial mass of 
$\log(\mathcal{M}_{vir}/\mathcal{M}_{\odot}) = 14.1\pm0.2$, consistent with the presence of either a low-mass cluster or a high-mass group. 

\begin{table}
    \begin{center}
    \caption{Photometry}
    \label{tab:limits}
    {\vskip 1mm}
    \begin{tabular}{c @{\hskip 15mm} c}

        \begin{tabular}{llll}

        \hline \\[-3.3mm]

        Filter & Telescope & Instrument & Depth$^a$ \\[0mm]
        \hline \\[-3mm]
        Cl1137 \\[0.5mm]
        \hline \\[-3mm]

        $B$    &   Subaru   &   Suprime-Cam   &   25.9   \\
        $V$    &   Subaru   &   Suprime-Cam   &   26.0   \\
        $R_C$  &   Subaru   &   Suprime-Cam   &   25.3   \\
        $I_C$  &   Subaru   &   Suprime-Cam   &   24.4   \\
        $I^+$  &   Subaru   &   Suprime-Cam   &   25.1   \\
        $Z^+$  &   Subaru   &   Suprime-Cam   &   24.6   \\
        $J$   &   UKIRT   &   WFCAM   &   22.5   \\
        $K$   &   UKIRT   &   WFCAM   &   21.6   \\
        $[3.6]$  &   {\it Spitzer}  &   IRAC   &   23.2   \\
        $[4.5]$  &   {\it Spitzer}  &   IRAC   &   23.3   \\[1mm]

        \hline \\[-3mm]
	\end{tabular}

    \end{tabular}

    \end{center}
$^a$ 80\% completeness limits derived from the recovery rate of artificial sources inserted at empty sky regions.

\end{table}

\begin{figure*}
\includegraphics[clip,angle=0,width=0.48\hsize]{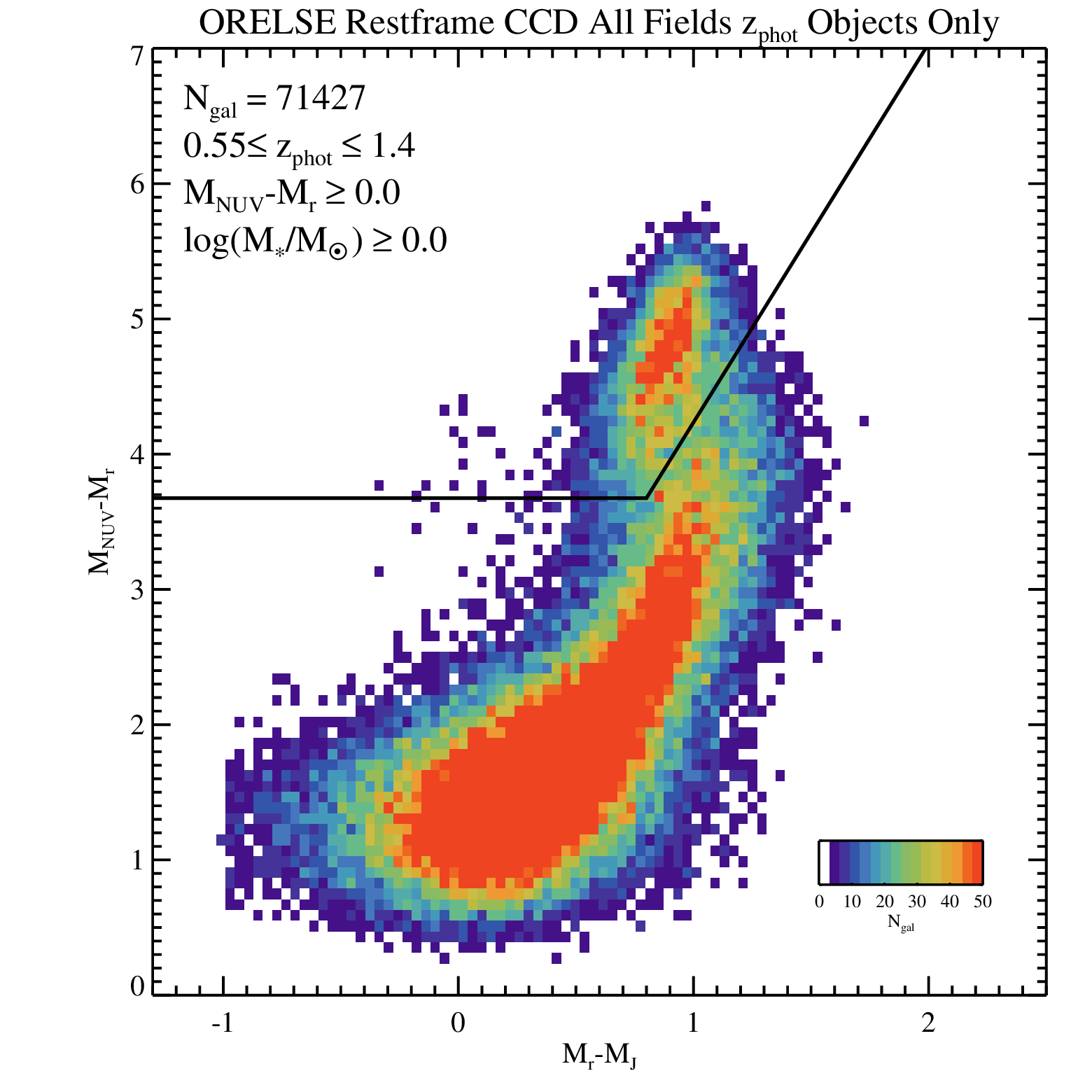}
\includegraphics[clip,angle=0,width=0.48\hsize]{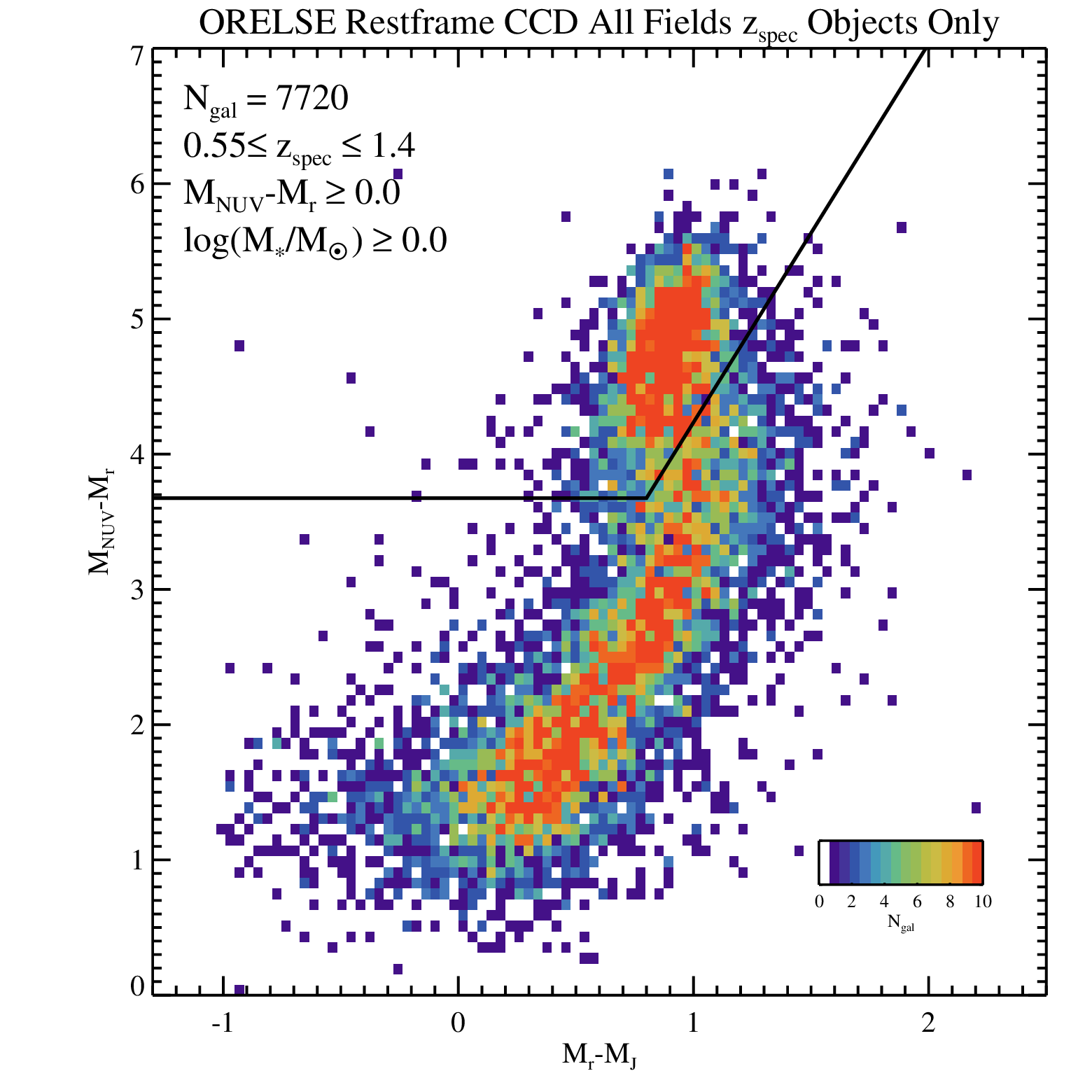}
\caption{\emph{Left}: Rest-frame $M_{NUV}-M_{r}$/$M_{r}-M_{J}$ color-color diagram (CMD) for all objects in ORELSE in the photometric 
redshift range indicated that fall within $\leq60\arcsec$ of a spectral target and have an apparent magnitude in the range
$18.5\leq i \leq 24.5$ and $18.5\leq z\leq 24.5$ for objects with $z_{phot}< 1$ and $z_{phot}\ge 1$, respectively. The
number of $z_{phot}$ objects satisfying these criteria is given in the top left. The color bar in the bottom right of 
the panel indicates the associated color for the number of objects in each two-dimensional bin. The solid line shows
an average of the color-color criteria used to delineate quiescent galaxies (top left region, above the lines) and star-forming 
galaxies (bottom right region, below the lines) at $z<1$ and $z\ge1$ (see \S\ref{SEDfitting}). \emph{Right:} The same as the 
left panel, but for galaxies spectroscopically confirmed in the redshift range indicated subject to the same apparent magnitude 
cuts as the $z_{phot}$ objects.} 
    \label{fig:NUVrJ}
\end{figure*}

\subsection{Spectral Energy Distribution Fitting and Stellar Mass Limits} 
\label{SEDfitting}

\begin{figure*}
\includegraphics[clip,angle=0,width=0.43\hsize]{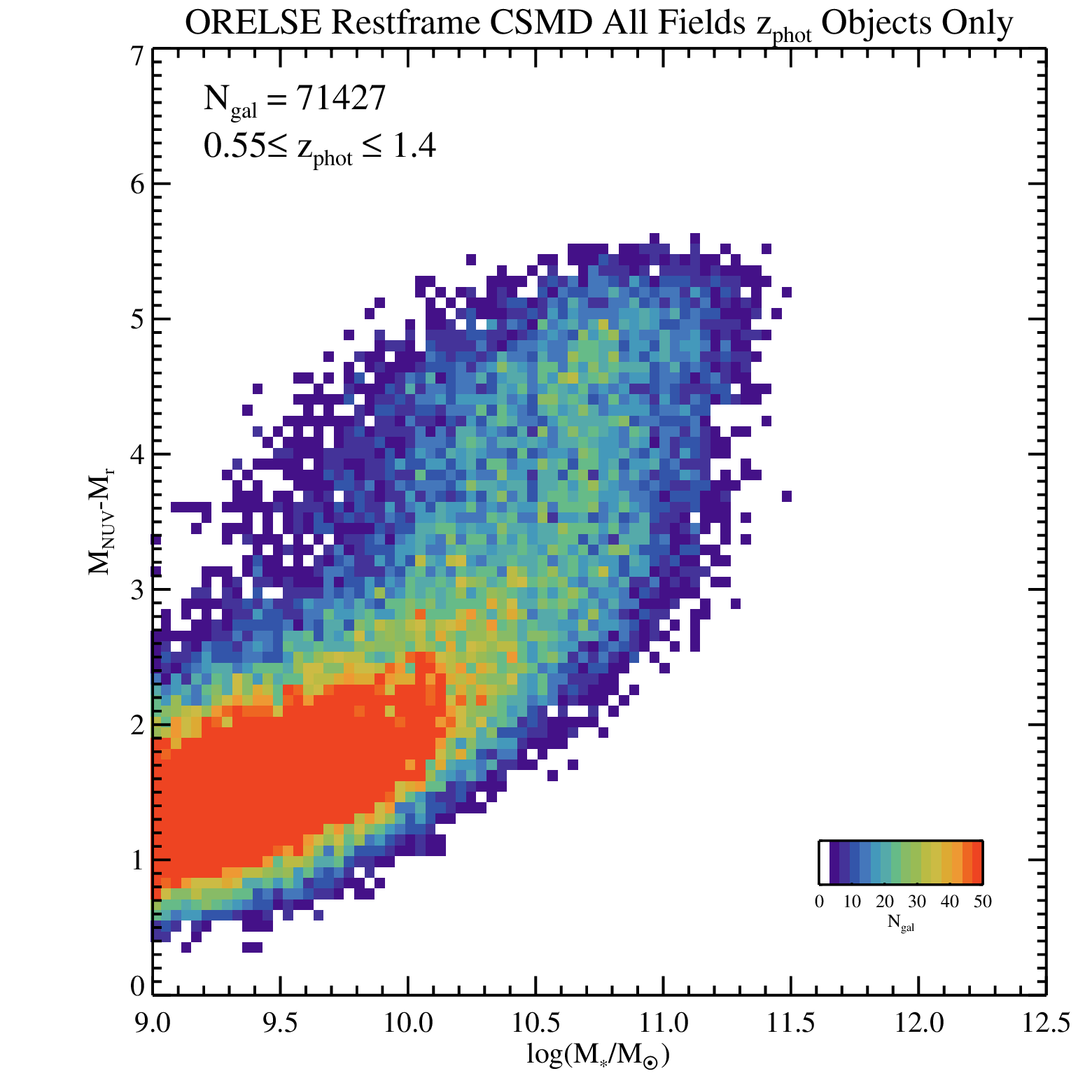}
\includegraphics[clip,angle=0,width=0.43\hsize]{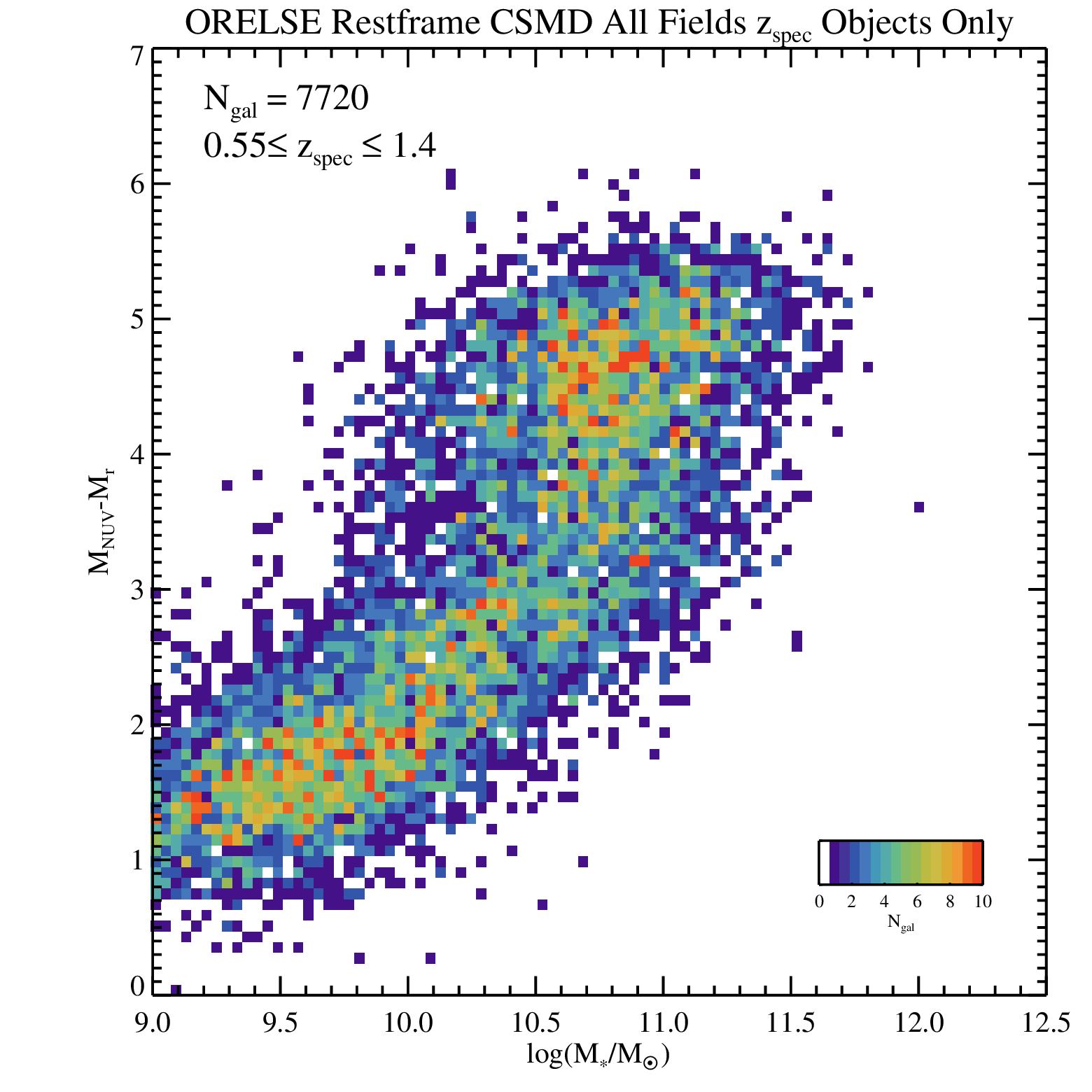}
\includegraphics[clip,angle=0,width=0.43\hsize]{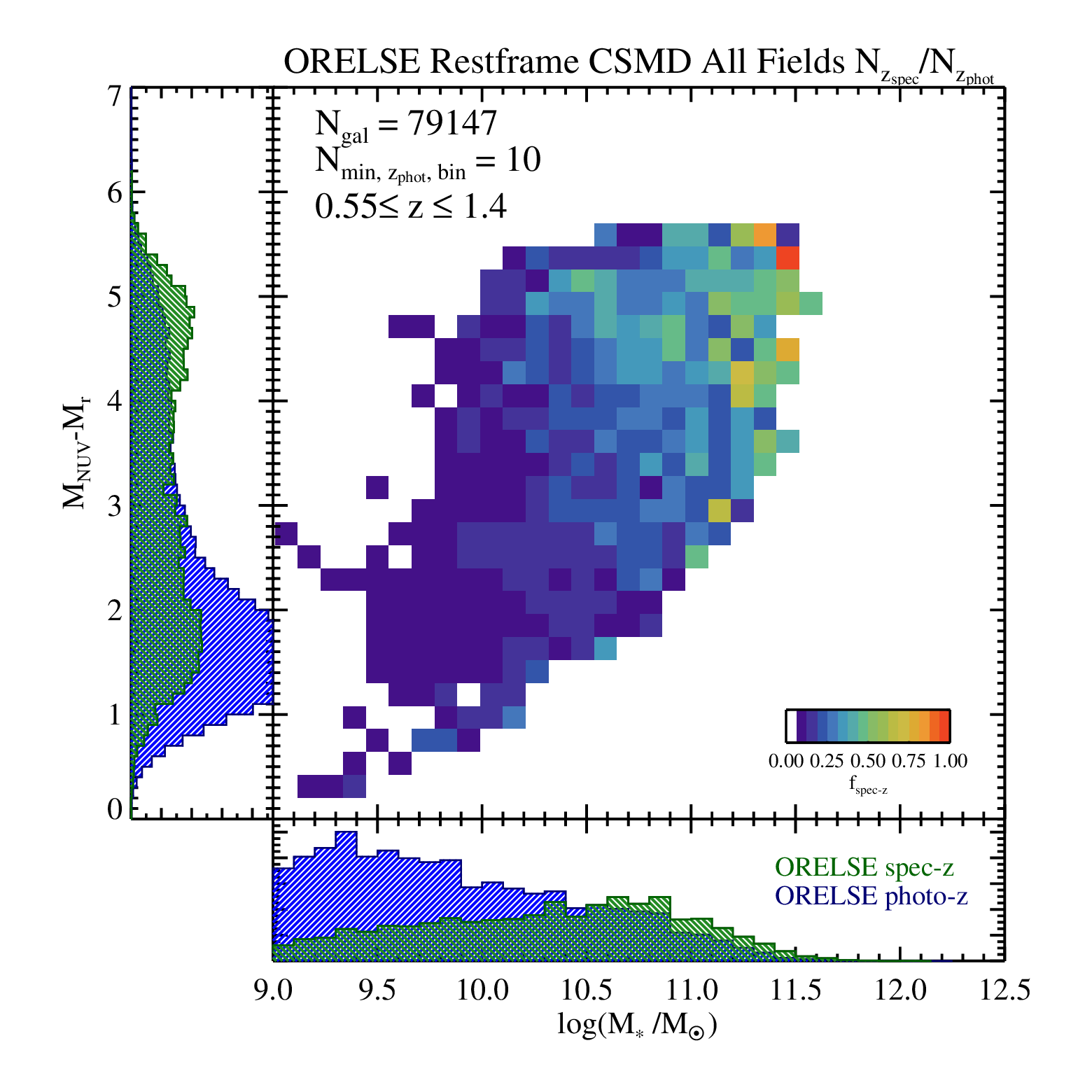}
\caption{\emph{Top Left}: Rest-frame $M_{NUV}-M_{r}$ color-stellar mass diagram (CSMD) for all objects in ORELSE in the photometric redshift range indicated 
in \S\ref{specrepresent} that fall within $\leq60\arcsec$ of a spectral target and have an apparent magnitude in the range 
$18.5\leq i \leq 24.5$ and $18.5\leq z\leq 24.5$ for objects with $z_{phot}< 1$ and $z_{phot}\ge 1$, respectively. The 
number of $z_{phot}$ objects satisfying these criteria is given in the top left. The color bar in the bottom right of the panel indicates 
the number of objects in each two-dimensional bin. 
\emph{Top Right:} The same as the left panel, but for galaxies spectroscopically confirmed in the redshift range indicated subject to the same 
apparent magnitude cuts as the $z_{phot}$ objects in the left panel. \emph{Bottom:} Fraction of $z_{spec}$ to $z_{phot}$ objects from the top 
two panels as a function of rest-frame color and stellar mass. At least 10 $z_{phot}$ objects are required to fall within a given
color/stellar mass bin in order to be plotted. Accompanying histograms of the ORELSE $z_{spec}$ and $z_{phot}$ samples in stellar mass and 
rest-frame color are shown and are normalized such that the area of two histograms for each parameter are equal for
stellar masses in excess of $10^{10} M_{\odot}$ and colors bluer than $M_{NUV}-M_{r}=2$. As can be seen in both the one- and two-dimensional 
histograms, the fraction of spec-$z$s in ORELSE drops dramatically below these two limits. Above these limits, the spectral sample is found
to be broadly representative of the underlying $z_{phot}$ population in terms of these two parameters (see \S\ref{specrepresent}).} 
    \label{fig:representativeness}
\end{figure*}

Fitting of the SED of each object detected in the detection image of each field was first performed on aperture magnitudes measured on the PSF-homogenized images
using the Easy and Accurate $z_{phot}$ from Yale (\texttt{EAZY}, \citealt{brammer08}) code. This fitting was done for the purposes of estimating photometric 
redshifts (hereafter $z_{phot}$) of each object and an associated probability distribution function (PDF). The PDFs were generated by minimizing the $\chi^2$ of the observed flux
densities and a set of basis templates at each redshift (including emission lines, see \citealt{brammer11}). See \citet{AA17} for more details on 
this fitting process. The parameter ``$z_{\rm{peak}}$'' was adopted as the measure of $z_{phot}$, with the uncertainties on this parameter estimated from the PDF of 
each source. Additional fitting for each source was done to the
\citet{pickles98} stellar library and was used, in conjunction with a variety of other criteria, to create a ``use flag" for each source (see \citealt{AA17} for 
more details on this flag). Objects with a use flag set to zero, except for those objects which were spectroscopically determined to be a star at high confidence, were removed 
from all subsequent analysis. These objects typically totaled $\sim$5-10\% of the total number of objects in a given ORELSE field. 
The precision and accuracy of the photometric redshifts were estimated from fitting a Gaussian to the distribution of $(z_{spec}-z_{phot})/(1+z_{spec})\def\Delta z/(1+z)$ 
for those galaxies with high-$Q$ $z_{spec}$ measurements in the range 0.5$\le z \le$1.2 and a use flag=1. This exercise resulted in a precision of 
$\sigma_{\Delta z/(1+z)}=0.017-0.038$, with an average of 0.029, and a catastrophic outlier rate ($\Delta z/(1+z)>0.15$) that ranges from $\eta\sim1.7-9.5$\%, with an 
average of 5.8\%, across all ORELSE fields to limit of 
$i\le24.5$. At this stage, if applicable, systematic offsets from zero in the $(z_{spec}-z_{phot})/(1+z_{spec})$ distribution in each field 
were cataloged and corrected by applying the opposite of this offset to all $z_{phot}$ values. These offsets ranged from -0.006 to 0.005 with a median offset of 0.001. 
Subsequently, an additional phase of SED fitting was run with \texttt{EAZY} in 
which either the high-$Q$ $z_{spec}$, when available, or the $z_{phot}$ was set as a redshift prior in order to estimate extinction-uncorrected rest-frame magnitudes 
estimated from the best-fit template following the methodology of \citet{brammer11}. These rest-frame magnitudes are primarily used in this paper to determine, in a binary 
fashion, the state of star formation in objects that enter our final sample. For all such objects, we adopt the rest-frame $M_{NUV}-M_{r}$ vs. $M_{r}-M_{J}$ cuts used in
\citet{lem14} to delineate between galaxies whose rest-frame colors are dominated by older and younger stellar populations used in \citet{lem14}. These two colors, or similar proxies, have been 
found to be intimately tied to the level of star formation activity \citep{animatedarnouts07,cmart07, wyder07} and dust content of galaxies (e.g., 
\citealt{animatedarnouts13}) and are extremely effective in separating the two populations independent of dust concerns (e.g., \citealt{williams09, ilbert10, ilbert13, 
lunchpailmcgee13, thibaud16}). We use the following redshift-dependent criteria to define objects dominated by older stellar populations:

\begin{equation}
\begin{split}
M_{NUV}-M_{r}\ge 2.8(M_{r}-M_{J}) + 1.51 \, \, \wedge \\
\quad M_{NUV}-M_{r} \ge 3.75 \, \, \, \, \, \, \quad \mathbf{(z<1)} \\
\\
M_{NUV}-M_{r}\ge 2.8(M_{r}-M_{J}) + 1.36 \, \, \wedge \\ 
\quad M_{NUV}-M_{r} \ge 3.60 \, \, \, \, \quad \mathbf{(z\ge1)}
\end{split}
\label{eqn:SFquiescent}
\end{equation}

\noindent Note that at all redshifts investigated in this study in all ORELSE fields there existed imaging in bands whose filter curves in the rest-frame 
appreciably overlapped with each of the three rest-frame bands used here such that template-based $k$-corrections are minimized.
Galaxies meeting their redshift-appropriate criteria above were designated as ``quiescent'', while all remaining galaxies were classified as ``star-forming''. 

As a rough estimate of the level of $\mathcal{SSFR}$ a classification of quiescent/star-forming corresponds to, in \cite{ilbert13} it is shown that the 
quiescent region of this diagram, defined in a similar way to our own, contains nearly all galaxies with $\log(\mathcal{SSFR_{SED}}$ yr)$<10^{-11}$ for the redshift
range considered in this study. While we were 
not able to estimate $\mathcal{SFR}$ values from our DEIMOS/LRIS spectroscopy for galaxies classified as quiescent due to fear that the main emission line present 
in our spectroscopy across the full redshift range studied here, [\ion{O}{II}], does not faithfully trace star formation activity (e.g., \citealt{yan06, lem10, lem17a}),
we did measure the extinction-corrected [\ion{O}{II}]-derived $\mathcal{SSFR}$ for the galaxies in the spectral sample classified as $NUVrJ$ star forming. The procedure
to estimate $\mathcal{SSFR{\rm{[\ion{O}{II}]}}}$ values generally followed the 
formalism of \cite{lem14}, which combines rest-frame magnitudes as well as other SED-fit parameters along with the equivalent width of the [\ion{O}{II}] as measured from 
our DEIMOS/LRIS spectroscopy. Rest-frame $U-$band magnitudes were taken from our \textsc{EAZY} fitting, stellar masses and extinctions from our \textsc{FAST} 
fitting (see below), and the \cite{wuyts13} relation was adopted to translate between stellar and nebular extinction. The equivalent widths of [\ion{O}{II}] were measured 
using the bandpass method described in \cite{lem10}. Of the $\sim4300$ ORELSE spectral galaxies classified as $NUVrJ$ star-forming across all redshifts, the vast majority of 
galaxies, $\sim$97\%, were found to have extinction-corrected [\ion{O}{II}]-derived $\mathcal{SSFR}$ values in excess of $\log(\mathcal{SSFR_{\rm{[\ion{O}{II}]}}}$ yr)$>10^{-11}$. 

The final stage of the fitting process employed the code Fitting and Assessment of Synthetic Templates (\texttt{FAST}, \citealt{kriek09}) for the purposes of estimating
galaxy physical parameters. This time aperture-corrected magnitudes are used (see \S\ref{phot}) with the same redshift priors as were used in the final stage of the 
\texttt{EAZY} fitting. Exponentially declining stellar population synthesis (SPS) \citealt{bc03} models (hereafter BC03) were adopted with a \citealt{chab03} initial mass
function, a \citet{calz00} extinction law, and a stellar-phase metallicity fixed at $Z=Z_{\odot}$. The ranges of allowed parameters are identical to those of \citet{AA17}, 
with the imposition that the age of a given objects could not exceed the age of the universe at the estimated redshift of that object. The value of each parameter is taken 
from the best-fit value and uncertainties are derived through 100 realizations of re-fitting to an SED with photometry that has been tweaked by a Gaussian random multiple 
of its photometric errors for each band (as in, e.g., \citealt{Rusl13}). 

\subsection{Spectral Representativeness}
\label{specrepresent}
As discussed in \citet{AA17}, the ORELSE imaging of the fields presented in that paper
is broadly to a sufficient depth to detect all galaxy types to a stellar mass limit of $\log(\mathcal{M}_{\ast}/\mathcal{M}_{\odot})\ga 10$ over the redshift
range studied here. Here we test that statement for all ORELSE fields using a method similar to that described in \citet{AA17}. Stellar mass limits are determined by 
scaling the brightness of an exponentially declining BC03 model galaxy at each redshift, in steps of $\Delta z=0.1$, to the 80\% magnitude completeness limits of 
the detection image in each field. The BC03 template was generated with a formation redshift of $z_{f}=5$, a star-formation history (SFH) with an e-folding time 
($\tau$) of 1 Gyr, a \citet{chab03} 
IMF, and no internal dust extinction. The scaling factor used to match the brightness of the model galaxy at each redshift to the magnitude completeness limits is 
similarly used to scale the stellar mass of the model galaxy that then sets the mass completeness limit. Note that this template will contain a higher stellar mass-to-light
ratio than the vast majority of galaxies at all redshifts studied here and, thus, represents a conservative upper limit to the stellar mass completeness of the 
ORELSE images. We find that the statement made in \citet{AA17} generally holds for the full ORELSE sample presented here, with most fields having stellar mass
limits of $\log(\mathcal{M}_{\ast}/\mathcal{M}_{\odot})\sim10$ or below from $0.55 \le z \le 1.4$. While a few fields have stellar mass limits in excess of 
$10^{10} \mathcal{M}_{\odot}$ at or near the highest redshifts presented in this study, our results do not change meaningfully if we tailor the redshift range included
in each field to set comparable stellar mass limits across the entire sample. As such, we ignore any effects of the differential loss to our sample of lower mass, older 
quiescent galaxies in a few of the ORELSE fields at the highest redshifts studied in this paper. Note that extremely dust-reddened galaxies at these stellar 
masses broadly comparable rest-frame optical/UV fluxes to galaxies dominated by a near-maximal age stellar population. As the detection bands\footnote{see, \citealt{AA17,rum18} 
and \S\ref{Cl1137} for details on detection bands for each ORELSE field.} employed in ORELSE probe the rest-frame optical or redder-portion of the UV for galaxies in 
the redshift range considered in this study, such dusty populations should generally be detected by our imaging. Such a statement is corroborated by the routine 
detection of dusty star-forming galaxies in both our imaging and spectroscopy (see, e.g., \citealt{dirtydale11b, shen17}). 

\begin{figure*}
\includegraphics[clip,angle=0,width=0.85\hsize]{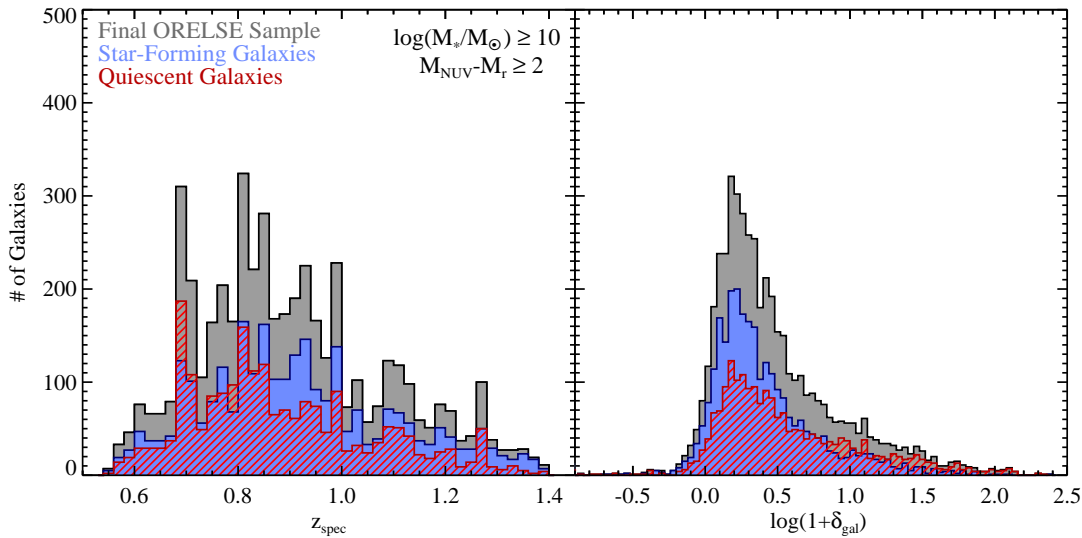}
\caption{\emph{Left:} Redshift histogram of the $\sim$2500 star-forming (filled light blue) and $\sim$2000 quiescent (hatched red)
galaxies in our final spectroscopic sample as well as the combined sample (gray histogram). All galaxies included in these histogram are subject to the stellar
mass and rest-frame $M_{NUV}-M_{r}$ cuts indicated as well as the redshift and apparent magnitude cuts given in \S\ref{specrepresent}, cuts that are imposed to
ensure the representativeness of the spectral sample. Peaks corresponding to individual LSSs are clearly visible in the distribution. \emph{Right:} Distribution in
local overdensity, $\log(1+\delta_{gal})$, as estimated by our Voronoi Monte Carlo technique (see \S\ref{voronoi}) of the same galaxy samples shown in the left panel.
The galaxies in the final spectral sample span a wide spread in local density that ranges from the sparse field to the core of massive clusters. The fraction of
quiescent galaxies appears to increase systematically with increasing $\log(1+\delta_{gal})$, with such galaxies appearing almost exclusively in the highest density
environments $\log(1+\delta_{gal})\ga1.5$.}

    \label{fig:finalsample}
\end{figure*}

In \citet{shen17}, a spectral sample drawn from five ORELSE LSSs was compared to an underlying photometric population at various stellar mass and color limits. This
comparison was made in order to understand to what limits the spectral sample in that study could be considered representative of the galaxy population at these 
redshifts as a whole. It was found for an apparent magnitude cut of $i\le24.5$ that the ORELSE spectral sample presented in that paper could be broadly 
characterized as being representative of the underlying photometric population in the range $\log(\mathcal{M}_{\ast}/\mathcal{M}_{\odot})\ge 10.2$ and $M_{NUV}-M_{r}\ge2.5$. Here 
we expand this comparison to all of ORELSE following a method similar to that presented in \citet{shen17}. Spectral and photometric samples from all ORELSE fields were 
selected within the redshift range $0.55\le z_{spec}/z_{phot} \le 1.4$ and to an apparent magnitude limit of $18.5\le i\le24.5$ or $18.5\le z\le24.5$ 
depending on whether the object is at $z\le1$ or $z\ge1$, respectively. The $z_{phot}$ sample in each ORELSE field consisted only of those objects without a high-$Q$ $z_{spec}$ 
that were within 60$\arcsec$ of an object targeted for spectroscopy, a scheme which roughly selects all those objects which were potential spectroscopic targets which 
were not targeted or which were targeted but for which we had no secure $z_{spec}$. The value 60$\arcsec$ was chosen to ensure that the $z_{phot}$ objects selected broadly 
shared the same depth and breadth of imaging as the actual spectral targets, but large enough that it does not excise objects that fell within the spectral footprint that 
were separated from the spectral targets only by virtue of the chosen slit geometry (i.e., 60$\arcsec$ is roughly half of the maximum interslit distance for our masks). 
In addition, only those objects with estimated stellar masses and rest-frame magnitudes and with a use flag=1 (see \S\ref{SEDfitting})
were included in this comparison. Rest-frame $NUVrJ$ color-color diagrams for these spectroscopic and $z_{phot}$ samples are shown in Figure \ref{fig:NUVrJ} along with the
delineation lines for quiescent and star-forming galaxies described earlier in this section. Note that the upper right portion of the star-forming locus, i.e., 
objects with $M_{r}-M_{J}>1$ and $M_{NUV}-M_{r}>3.5$, a region where the dustiest galaxies are thought to lie (e.g., \citealt{animatedarnouts13}) are well represented 
in both samples. Further, the relative abundance of objects in this region of color-color space is comparable to legacy fields like, e.g., COSMOS, where the detection band(s) 
employed are well in the rest-frame NIR at these redshifts where dust effects are minimal (e.g., \citealt{ilbert13, laigle16}). 

We compared the stellar mass and rest-frame $M_{NUV}-M_{r}$ color distributions for the spectroscopic and $z_{phot}$ samples in bins of 0.2 dex and 0.5 mags, respectively, by 
means of a Kolmogorov-Smirnov (KS) test for all values of $\log(\mathcal{M}_{\ast})$ and $M_{NUV}-M_{r}$. For objects with stellar masses in excess of 
$\log(\mathcal{M}_{\ast}/\mathcal{M}_{\odot})= 10.0$ and colors redder than $M_{NUV}-M_{r}=2.0$ the KS tests were unable to discern the two distributions in all bins of stellar 
mass and color at a confidence of $\ga3\sigma$. In addition, for all bins of $\mathcal{M}_{\ast}$ and $M_{NUV}-M_{r}$ color within these ranges, the fraction of $z_{phot}$ objects
with secure spectroscopic redshifts exceeded 10\% and generally fell within the range $15-35$\%. Figure \ref{fig:representativeness} shows the $M_{NUV}-M_{r}$ rest-frame color-stellar mass diagram (CSMD) of each of these two samples as well as the fraction of secure spectral redshifts as a function of stellar mass and $M_{NUV}-M_{r}$ color. 

Regardless of the results above, there still appears to be an excess of the fraction of $NUVrJ$ quiescent galaxies in the ORELSE spectral sample relative
to that of the full underlying photometric population. Such an excess can be explained by our preference to spectroscopically target objects with a higher likelihood of 
being in LSS environments, and, thus, conventionally, more likely to be quiescent. However, this relative excess is only an issue for our analysis if it persists in
different environments, as all analysis presented in this study relies on the assumption that the fraction of quiescent galaxies is being measured representatively 
by our spectral sample in all environments probed by our data. To test this, we compared the fraction of quiescent galaxies in the photometric and spectroscopic 
sample in the three different environmental bins defined in \S\ref{quiescentfrac}. In each of the three cases, the quiescent fractions of the 
two samples are similar, never exceeding a difference of 0.05. A similar level of concordance is observed if the three environmental bins are further broken down into 
two redshift bins. This high degree of concordance implies that no bias is induced through our spectral sampling scheme in terms of the fraction of quiescent galaxies 
in a given environment at a given redshift. 

Through these comparisons, in conjunction with the knowledge that the imaging 
in all fields is of sufficient depth to detect galaxies of all types to $\log(\mathcal{M}_{\ast}/\mathcal{M}_{\odot})\ge 10$, we conclude that the ORELSE spectral sample shown
here is broadly representative of the full galaxy population in the range $\log(\mathcal{M}_{\ast}/\mathcal{M}_{\odot})\ge10.0$, $M_{NUV}-M_{r}>2.0$, and 
$0.55\le z \le 1.4$.
These limits, along with the associated apparent magnitude and quality flag cuts, set the final spectral sample that is presented in the remainder of the paper. This final 
sample consists of 4552 galaxies with secure spectral redshifts and reliable photometry. Since we have shown that this sample is representative of the full
galaxy population in the range of colors, stellar masses, and redshifts we are probing, and, further, since $z_{phot}$ objects have relatively large uncertainties 
in all three of the parameters were are interested in ($z$, $\mathcal{M}_{\ast}$, and $\log(1+\delta_{gal})$), we chose for the remainder of the paper to exclusively 
use this final spectral sample. We note that none of the results presented in this paper change meaningfully if we slightly change the stellar mass and color limits that 
define our final sample (i.e., by $\pm0.2$ dex and $\pm$ 0.5 mags, respectively).

\subsection{Local Overdensity}
\label{voronoi}

In order to estimate the local environment of the galaxies in our sample, we employ the Voronoi Monte-Carlo (VMC) technique described in \citep{lem17a}. This estimator 
has been employed in a variety of different works at both low and high redshift (e.g., \citealt{shen17, shen19, lem17a, lem18, AA17, AA19, debz18, olga18}), and
its precision and accuracy are discussed briefly in \citet{AA17}. Further quantitative measures of the precision and accuracy of this reconstruction for the detection
and characterization of groups and clusters will be presented in \citet{denise19}. Here, the VMC method is described briefly. 

The VMC attempts to combine the information provided by the high-precision spectroscopic redshifts with the full $P(z)$ information contained within each object
without a spectroscopic redshift to determine the local overdensity in thin redshift slices that span the redshift range where both our spectroscopic and photometric 
data are sensitive ($0.55 \le z \le 1.4$). In this scheme, all objects with secure spectroscopic redshifts are considered to be at the measured redshift. For 
those objects without a secure spectroscopic redshift, multiple realizations are run for each redshift slice to sample from the statistics of the photometric redshift
assigned to each object to determine the range of possible density fields. In practice, for each Monte-Carlo realization, Gaussian sampling is performed to determine 
a new set of $z_{phot}$ values for each object without a high quality $z_{spec}$ (but with a good use 
flag, see \S\ref{SEDfitting}) for that realization. The sampled value, in units of $\sigma$, is then multiplied by either the effective $1\sigma$ lower or upper uncertainty 
on $z_{phot}$ for that object depending on which side of the peak of the Gaussian sample fell. This value is then either subtracted from or added to the original $z_{phot}$ 
for each object to create a new set of $z_{phot,\ MC_{i}}$ for that realization. A thin redshift slice is cut from the combined catalog which includes both objects 
with new $z_{phot}$ values as well as all galaxies with high quality extragalactic $z_{spec}$, and Voronoi tessellation is performed for the realization of that slice. 
This process is performed, in total, for 
100 realizations for each of the 85 redshift slices running from $0.55 \le z \le 1.4$, which is the number of slices required to span this redshift range for a 
slice width of $\pm$1500 km s$^{-1}$ and steps between the central redshift of adjacent slices that are half of the slice depth (i.e., 1500 km s$^{-1}$). 

For each realization of each slice, a grid of 75$\times$75 kpc is created to sample the underlying local density distribution. The local density
at each grid value for each realization and slice is set equal to the inverse of the Voronoi cell area (multiplied by $D_{A}^2$) of the cell that encloses the
central point of that grid. Final local densities, $\Sigma_{VMC}$, for each grid point in each redshift slice are then computed by median combining the values
of the 100 realizations of the Voronoi maps for that slice. The local overdensity value for each grid point is then computed as
$\log(1+\delta_{gal}) \equiv \log(1+ (\Sigma_{VMC}-\tilde{\Sigma}_{VMC})/\tilde{\Sigma}_{VMC})$, where $\tilde{\Sigma}_{VMC}$ is the median $\Sigma_{VMC}$
for all grid points over which the map is defined (i.e., where there is coverage in a sufficient number of imaging bands). By adopting local overdensity rather
than local density as a proxy of environment, we largely mitigate issues of sample selection and differential bias as a function of redshift. For VMC overdensity maps in all 
ORELSE fields, photometric and spectroscopic catalogs were cut at $18\le i \le 24.5$ or $18 \le z\le 24.5$ depending on the field. 
Uncertainties associated with the overdensity estimate for each object are taken from the 16th and 84th percentile of the $\log(1+\delta_{gal})$ distribution generated from
all 100 Monte Carlo iterations. Figure \ref{fig:finalsample} shows the redshift and $\log(1+\delta_{gal})$ distribution of all galaxies contained in our final spectral 
sample defined in \S\ref{specrepresent}. 

\section{The Persistence of the Color-Density Relation and Efficient Environmental Quenching to $z\sim1.4$}

\subsection{The Quiescent Fraction in ORELSE}
\label{quiescentfrac}

\begin{figure*}
\includegraphics[clip,angle=0,width=0.49\hsize]{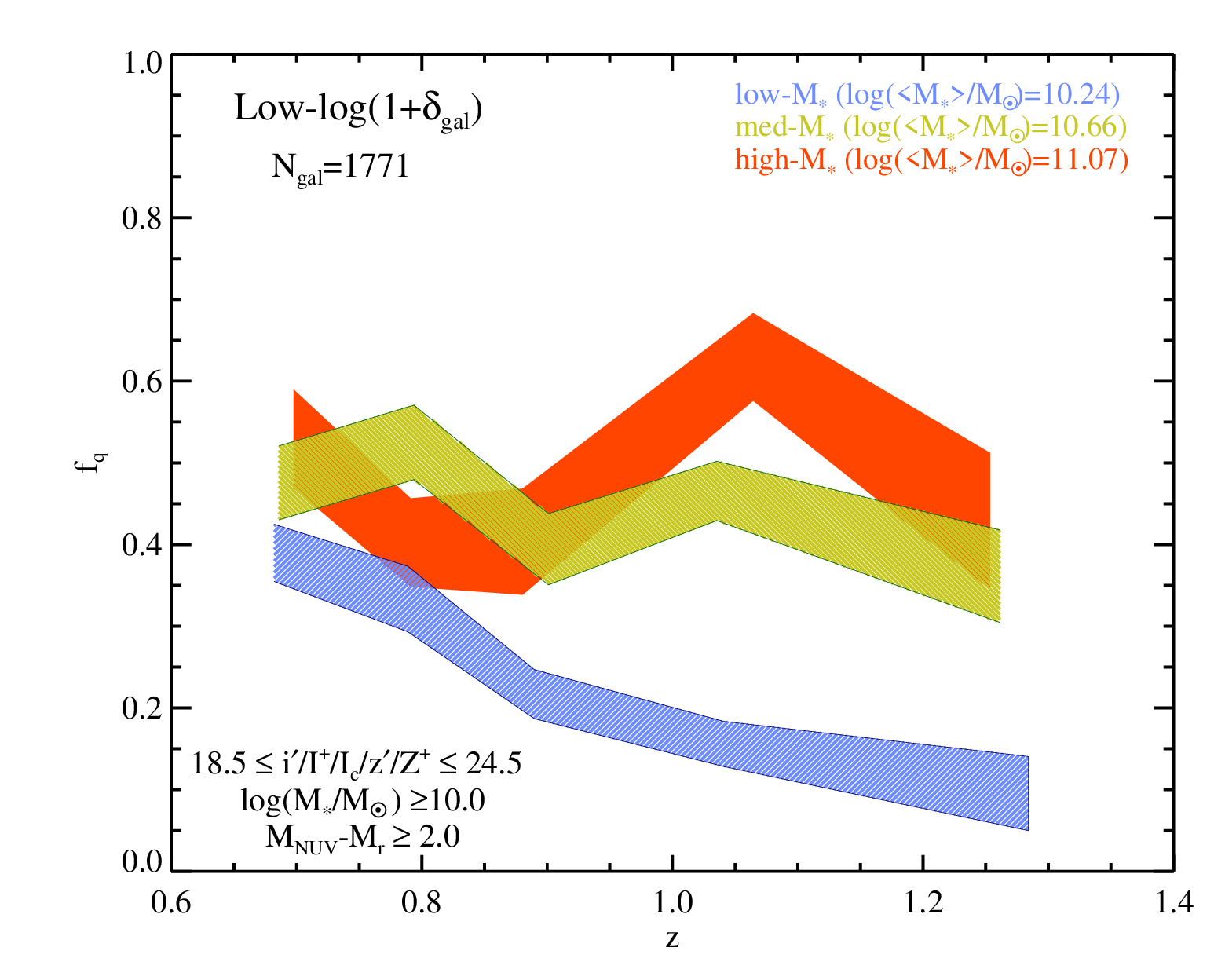}
\includegraphics[clip,angle=0,width=0.49\hsize]{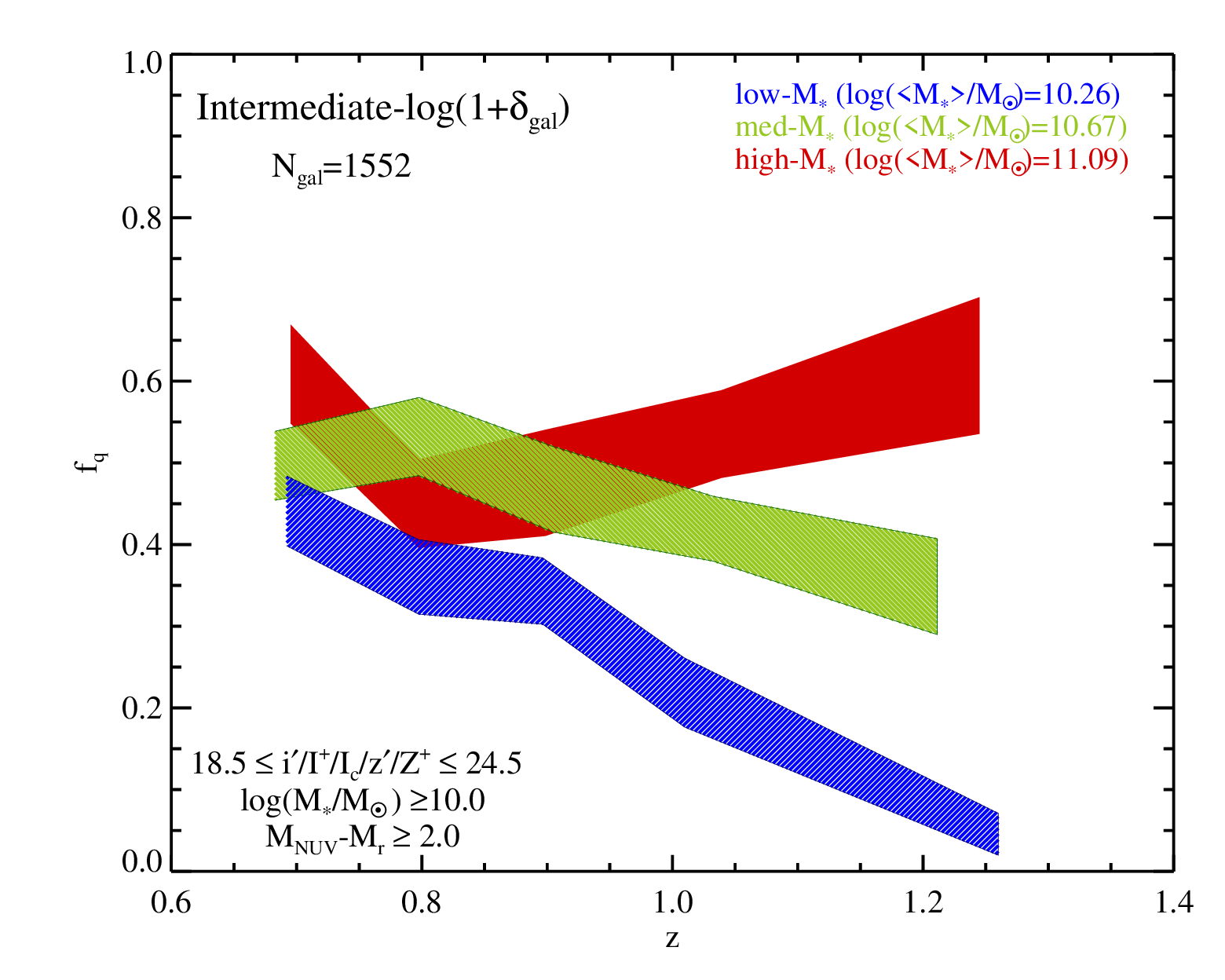}
\includegraphics[clip,angle=0,width=0.49\hsize]{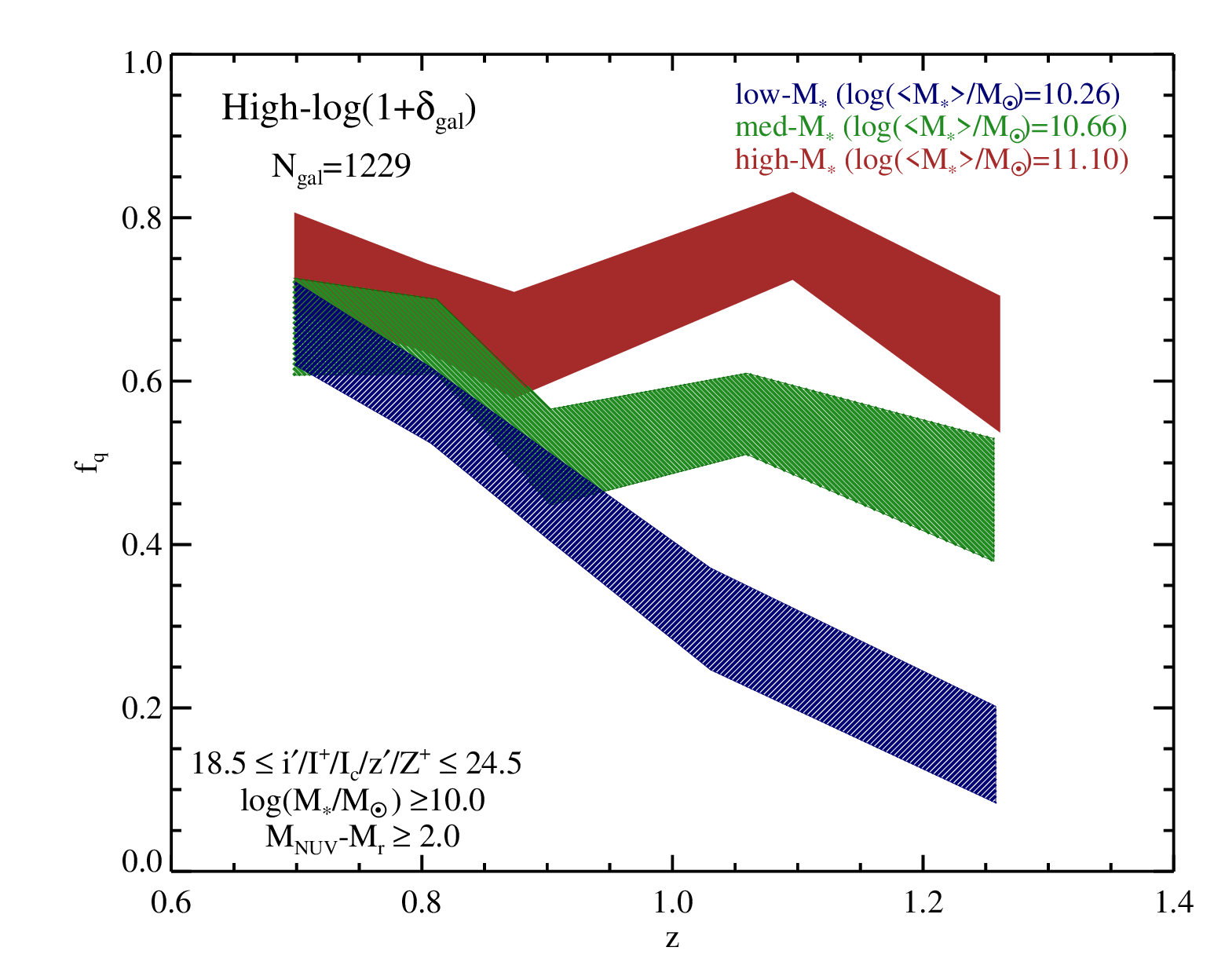}
\caption{\emph{Top Left:} Quiescent fraction, $f_{q}$, of galaxies in the lowest overdensity regime of the final spectral sample 
($\log(1+\delta_{gal}) \la 0.3$, i.e., field environments) as a function of redshift in three different stellar mass ranges (roughly 10-10.45, 10.45-10.9, 
and $>$10.9 in $\log(\mathcal{M_{\ast}}/\mathcal{M}_{\odot})$ for the low-, med-, and high-$\mathcal{M}_{\ast}$ samples, respectively). The median stellar mass
of the galaxies in each stellar mass bin across all redshifts is shown in the top right. The apparent magnitude,
stellar mass, and rest-frame color criteria imposed on the final spectral sample is shown in the bottom left, and the number of galaxies in this
environmental bin is given in the top left. Uncertainties are derived from Poissonian statistics. \emph{Top Right:} Identical to the top left panel but 
for the galaxies in the intermediate overdensity sample (i.e., infall/intermediate regions of groups/clusters). \emph{Bottom:} Identical to the other two panels 
but for galaxies in the highest densities in ORELSE (i.e., intermediate/core regions of groups/clusters). Galaxy populations in higher density 
environments and at higher stellar masses are generally seen to exhibit a higher incidence of quiescent galaxies from $0.55\le z \le 1.4$. Only the
low-$\mathcal{M}_{\ast}$ galaxies appear to evolve strongly in $f_q$ with redshift, with that of galaxies in the other two stellar mass bins showing 
moderate to no redshift evolution.}
    \label{fig:quiescentfrac}
\end{figure*}

In Figure \ref{fig:quiescentfrac} we show the quiescent fractions, $f_{q}$, of galaxies in our final spectral sample as a function of $z$, $\log(1+\delta_{gal})$,
and $\mathcal{M_{\ast}}$. The quantity $f_{q}$ is simply defined as $N_{q}/(N_{SF}+N_{q})$, where $N_{q}$ and $N_{SF}$ refer to the number of quiescent and 
star-forming galaxies, respectively. Galaxies are separated into star-forming and quiescent types based on the scheme described in \S\ref{SEDfitting}. As noted in 
\S\ref{specrepresent}, this sample is comprised only of 
those ORELSE galaxies with $\log(\mathcal{M}_{\ast})\ge10$ and $M_{NUV}-M_{r}\ge2$ and, as such, represent an upper limit to the true quiescent fractions of
all galaxies with $\log(\mathcal{M}_{\ast})\ge10$ in the redshift range $0.55 \le z \le 1.4$. However, as shown in Figure \ref{fig:representativeness}, galaxies 
with colors bluer than $M_{NUV}-M_{r}=2$ only begin to comprise a non-negligible fraction ($\sim15$\%) of the total galaxy population for stellar masses in the range 
$10.0 \le \log(\mathcal{M}_{\ast})\le10.5$. For higher stellar masses, the contribution of galaxies at such colors is effectively negligible ($\sim1$\%). We 
choose not to make a correction for this effect throughout the paper as we do not know \emph{a priori} the relative number of galaxies with $M_{NUV}-M_{r}<2$ in each redshift
and environmental bin. 
 
The full sample is broken into three different $\mathcal{M_{\ast}}$ bins 
for each redshift and environmental density bin that broadly follow the ranges 10-10.5 (low-$\mathcal{M_{\ast}}$), 10.5-10.9 (med-$\mathcal{M_{\ast}}$), and $>$10.9 
(high-$\mathcal{M_{\ast}}$) in $\log(\mathcal{M}_{\ast}/\mathcal{M}_{\odot})$. Similarly,
galaxies are broken into three different $\log(1+\delta_{gal})$ bins for each that broadly follow the ranges $<0.3$ (low-$\log(1+\delta_{gal})$), 0.3-0.7 
(intermediate-$\log(1+\delta_{gal})$), and $>0.7$ (high-$\log(1+\delta_{gal})$). Physically, these three environmental density bins can be thought of as 
corresponding to field galaxies, the dense field combined with infall or intermediate regions of groups and the infall regions of clusters, and the intermediate to core regions 
of both groups and clusters, respectively. These interpretations are based on the distribution of galaxies in each local overdensity bin in projected
radial and differential velocity space relative to the closest group or cluster (i.e., global density $\eta\equiv R_{proj}/R_{200} \times |\delta v|/\sigma_{v}$).
For precise definitions of infall, intermediate, and core regions as well as our method for calculating global density see \citet{debz18}, \citet{shen19}, and references therein. 
It is important to note also that our VMC $\log(1+\delta_{gal})$ metric also shows a significant ($>>$3$\sigma$) degree of correlation with the global 
metric referred to above. This is true also for the more inveterate environmental metric that simply employs the normalized projected radial distances of galaxies 
relative to the centroid of their nearest group or cluster. In terms of this metric, the high-$\log(1+\delta_{gal})$ bin galaxies
fall within $0 \le R_{proj}/R_{200} \le 1$, with an average $R_{proj}/R_{200}=0.58$. Since all ORELSE groups and clusters are included in this analysis, the average total
mass of structures considered here is $\log(\mathcal{M}_{vir}/\mathcal{M}_{\odot})=14.4$. The exact limits for both the stellar mass and environmental density bins at each 
redshift are slightly modulated ($\pm$0.1 dex) for each bin in order 
to attempt to roughly impose equal numbers of galaxies in each $z$/$\log(1+\delta_{gal})$/$\mathcal{M_{\ast}}$ bin. Uncertainties associated with each quiescent fraction 
are calculated using the formula $N_{q}N_{SF}/(N_{SF}+N_{q})^3$ derived from Poisson statistics. 

The general relationship of $f_{q}$ with respect to each of the three parameters against it is plotted is apparent from a visual inspection of Figure 
\ref{fig:quiescentfrac}. In essentially all epochs observed in ORELSE over the redshift range studied here ($0.55 \le z \le 1.4$), galaxy populations 
with increasing stellar masses are observed with higher quiescent fractions. The behavior of $f_q$ with redshift across all environmental density bins 
appears to be broadly self-similar, with med- to high-$\mathcal{M_{\ast}}$ galaxies showing little to no significant decrease in their $f_q$ over the 
observed redshift range and low-$\mathcal{M_{\ast}}$ galaxies exhibiting a clear monotonic decrease in $f_q$ with redshift. While the behavior appears 
similar for galaxies residing in each type of environment studied here, there is a clear increase in $f_q$ of 0.1-0.2 at essentially all stellar masses
and redshifts when moving from the field to galaxies primarily inhabiting group/cluster environments (i.e., low- to high-$\log(1+\delta_{gal})$). It appears
from these measurements that the ability of high-density environments to quench galaxies of essentially all stellar masses persists from $0.55 \le z \le1.4$, 
a claim that is consistent with other works studying the galaxy populations of groups and clusters at similar redshifts (e.g., \citealt{mcoopz07, mcoopz10, 
olga10, olga17, kovac10, banalbenedetta10, muz12, balogh16, nantais17}). A possible exception to this statement enters for the low-$\mathcal{M}_{\ast}$ 
sample at the highest redshifts observed in ORELSE. At these redshifts and stellar masses the $f_{q}$ of galaxies in the high-$\log(1+\delta_{gal})$ bin 
becomes formally consistent with that of their counterparts inhabiting field environments. 
This lack of a significant difference implies that group/cluster environments at $z\ga1.1$ are either marginally effective or completely ineffectual at 
quenching galaxies at stellar masses of $\log(\mathcal{M}_{\ast}/\mathcal{M}_{\odot})\sim10.2$. 

In Table \ref{tab:fqtab} we list $f_q$ values and their associated uncertainties for all environmental, redshift, and stellar mass bins measured in ORELSE.
Using the values in Table \ref{tab:fqtab} we performed a $\chi^2$ fit of $f_q$ as a function of redshift, local environment, and stellar mass 
following the form:

\begin{equation}
\begin{split}
f_{q}(z, \mathcal{M}_{\ast}, \log(1+\delta_{gal})) = \alpha z^{\zeta} + \beta \log(M_{\ast}/M_{\odot})^{\theta} + \\
\gamma \log(1+\delta_{gal})^{\kappa}
\end{split}
\label{eqn:fqfunc}
\end{equation}

\noindent which yielded best-fit parameters of $\alpha = -1.8544\pm0.0002$, $\beta = 8.3876\pm0.0001$, $\gamma=-11.8321\pm0.0002$, $\zeta=0.1421\pm0.0007$, $\theta=0.2224\pm0.0007$, and 
$\kappa=-0.0076\pm0.0005$, where the uncertainties are taken from the covariance matrix of the fit. This formula is not specific to ORELSE and is generally applicable to the galaxy populations in the range of 
redshifts ($0.55\leq z \leq 1.4$), stellar masses ($10 \leq \log(M_{\ast}/M_{\odot}) \leq 11.6$), and overdensity values ($0.1 \leq  \log(1+\delta_{gal}) \leq 2.0$) 
studied here. We stress that this formula should not be applied outside these ranges. Further, we remind the reader that this formula will tend
to slightly overestimate $f_q$ by perhaps as much as $\sim$0.1 for galaxies in the lower stellar mass range in our sample (i.e., $\log(M_{\ast}/M_{\odot})<10.5$) 
due to the way in which our sample was created (see \S\ref{specrepresent}). 

\begin{table*}
        \centering
        \caption{Quiescent Fractions of the ORELSE Spectral Sample}
        \label{tab:fqtab}
        \begin{tabular}{cccccc}
                \hline
                Sample & $ \widetilde{z}$ & $f_{q}$ & $\log(1+{\widetilde{\delta_{gal}}})$ & $\log({\widetilde{\mathcal{M}_{\ast}}/\mathcal{M}_{\odot}})$ & $N_{\rm{gal}}$ \\ 
                \hline
		 & 0.682 &  0.390$\pm$0.035 &  0.162 &    10.24 &  195 \\
		 & 0.789 &  0.333$\pm$0.040 &  0.165 &    10.22 &  138 \\
		low-$\mathcal{M}_{\ast}$/low-$\log(1+\delta_{gal})$ & 0.890 &  0.217$\pm$0.030 &  0.121 &    10.26 &  189 \\
		 & 1.041 &  0.156$\pm$0.028 &  0.146 &    10.24 &  173 \\
		 & 1.284 & 0.095$\pm$0.045 & 0.088 &    10.27 &   42 \\
		\hline
		 & 0.686 &  0.475$\pm$0.045 &  0.170 &    10.65 &  122 \\
		 & 0.794 &  0.525$\pm$0.046 &  0.172 &    10.61 &  120 \\ 
		med-$\mathcal{M}_{\ast}$/low-$\log(1+\delta_{gal})$ & 0.901 &  0.394$\pm$0.043 &  0.162 &    10.70 &  208 \\
		 & 1.035 &  0.466$\pm$0.036 &  0.131 &    10.65 &  189 \\
		 & 1.262 &  0.361$\pm$0.057 &  0.170 &    10.67 &   72 \\
		\hline
		 & 0.698 &  0.529$\pm$0.061 &  0.192 &    11.10 &   68 \\
		 & 0.791 &  0.402$\pm$0.054 &  0.197 &    11.06 &   82 \\
		high-$\mathcal{M}_{\ast}$/low-$\log(1+\delta_{gal})$ & 0.880 &  0.404$\pm$0.065 &  0.192 &    11.13 &   57 \\
		 & 1.064 &  0.630$\pm$0.054 &  0.140 &    11.05 &   81 \\
		 & 1.254 &  0.429$\pm$0.084 &  0.123 &    11.06 &   35 \\
		\hline
		 & 0.692 &  0.441$\pm$0.043 &  0.471 &    10.26 &  136 \\
		 & 0.797 &  0.360$\pm$0.046 &  0.441 &    10.26 &  111 \\
		low-$\mathcal{M}_{\ast}$/med-$\log(1+\delta_{gal})$ & 0.896 &  0.343$\pm$0.041 &  0.419 &    10.29 &  137 \\
		 & 1.010 &  0.219$\pm$0.042 &  0.418 &    10.24 &   96 \\
		 & 1.260 & 0.045$\pm$0.026 &  0.326 &    10.25 &   66 \\
		\hline
		 & 0.683 &  0.496$\pm$0.042 &  0.463 &    10.69 &  141 \\
		 & 0.798 &  0.532$\pm$0.048 &  0.454 &    10.62 &  109 \\
		med-$\mathcal{M}_{\ast}$/med-$\log(1+\delta_{gal})$ & 0.904 &  0.468$\pm$0.052 &  0.426 &    10.72 &  171 \\
		 & 1.032 &  0.419$\pm$0.040 &  0.395 &    10.65 &  155 \\
		 & 1.211 &  0.348$\pm$0.059 &  0.375 &    10.70 &   66 \\
		\hline
		 & 0.695 &  0.609$\pm$0.059 &  0.442 &    11.02 &   69 \\
		 & 0.797 &  0.450$\pm$0.050 &  0.453 &    11.03 &  100 \\
		high-$\mathcal{M}_{\ast}$/med-$\log(1+\delta_{gal})$ & 0.898 &  0.475$\pm$0.064 &  0.405 &    11.12 &   61 \\
		 & 1.039 &  0.535$\pm$0.059 &  0.403 &    11.12 &   71 \\
		 & 1.245 &  0.619$\pm$0.061 &  0.281 &    11.10 &   63 \\
		\hline
		 & 0.699 &  0.671$\pm$0.051 &  0.941 &    10.30 &   85 \\
		 & 0.807 &  0.570$\pm$0.046 &   1.109 &    10.24 &  114  \\
		low-$\mathcal{M}_{\ast}$/high-$\log(1+\delta_{gal})$ & 0.901 &  0.460$\pm$0.053 &   1.049 &    10.28 &   87 \\
		 & 1.030 &  0.309$\pm$0.062 &  0.794 &    10.22 &   55 \\
		 & 1.258 &  0.143$\pm$0.059 &  0.702 &    10.24 &   35 \\	
		\hline
		 & 0.697 &  0.667$\pm$0.059 &  0.953 &    10.66 &   63 \\
		 & 0.812 &  0.654$\pm$0.046 &   1.173 &    10.60 &  107 \\
		med-$\mathcal{M}_{\ast}$/high-$\log(1+\delta_{gal})$ & 0.902 &  0.507$\pm$0.059 &   1.090 &    10.73 &  146 \\
		 & 1.059 &  0.560$\pm$0.050 &  0.875 &    10.66 & 100 \\
		 & 1.257 &  0.455$\pm$0.075 &  0.812 &    10.62&  44 \\
		\hline
		 & 0.698 &  0.746$\pm$0.057 &  0.958 &    11.09&  59 \\
		 & 0.804 &  0.690$\pm$0.038 &   1.166 &    11.10& 145 \\
		high-$\mathcal{M}_{\ast}$/high-$\log(1+\delta_{gal})$ & 0.874 &  0.644$\pm$0.062 &   1.192 &    11.16&  59 \\
		 & 1.096 &  0.778$\pm$0.049 &  0.829 &    11.08&  72 \\
		 & 1.262 &  0.621$\pm$0.064 &  0.839 &    11.11&  58 \\
		\hline
        \end{tabular}
        \begin{flushleft}
\end{flushleft}
\end{table*}

\begin{figure}
\includegraphics[clip,angle=0,width=0.98\hsize]{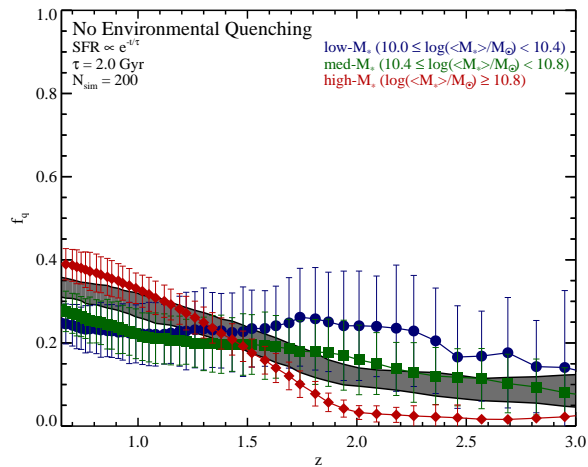}
\caption{Average quiescent fraction ($f_{q}$) vs. redshift for cluster galaxies in three different stellar mass bins simulated using the method described in Appendix \ref{mocks}.
Red diamonds, green squares, and blue circles show the average $f_{q}$ values and their associated NMAD uncertainties for 200 sets of simulated galaxies equivalent to the 
high-, med-, low-$\mathcal{M}$ samples.
In these simulations the effects of environmentally-driven quenching are purposely \emph{not} included. These 
effects are omitted in order to create a reference sample of group/cluster galaxies with which to compare to similar galaxies observed in ORELSE
groups and clusters. As both galaxy populations are, by construction, imprinted with the integrated effect of secular processes, and the latter sample contains
the integrated effect of environmental processes, this comparison can provide an estimate of the latter (see Figure \ref{fig:quencheff}).} 
\label{fig:simulatedfq}
\end{figure}

\subsection{Environmental Quenching Efficiency}
\label{quencheff}

Previous studies of the relationship of $f_q$ with various galaxy properties, including the environment in which galaxies reside, have attempted to 
quantify the efficacy of environmentally-related processes in quenching the star formation of member galaxies by defining a parameter known as the 
``conversion factor" or ``environmental quenching efficiency". This parameter, which we denote $\Psi_{convert}$ for brevity, is defined as the number of galaxies, usually at a 
fixed stellar mass, which have converted from star-forming to quiescent as a result of entering and interacting within a group or cluster environment. 
Practically, $\Psi_{convert}$ is usually defined as the difference in the $f_q$ fractions between the highest and lowest density bins in a given sample divided 
by the fraction of star-forming galaxies in in the lower density sample, i.e., 
$\Psi_{convert}\equiv (f_{q, \, \rm{high}-\log(1+\delta_{gal})}-f_{q, \, \rm{low}-\log(1+\delta_{gal})})/f_{SF, \, \rm{low}-\log(1+\delta_{gal})}$ \citep{vandenbosch08, mcgee14}. 
If the highest density bin in a given sample primarily contains group/cluster galaxies and the galaxies in the lowest density bin can be used as a 
proxy for a progenitor sample for those galaxies in the highest density bin, then $\Psi_{convert}$ yields the fraction of star-forming galaxies which have 
converted to quiescent during coalescence into the group/cluster. These values can then be compared to accretion histories of groups/clusters as measured
from $N$-body simulations (as in, e.g., \citealt{mcgee09, balogh16}) to determine the timescale required to convert the galaxy from star-forming to quiescent, $t_{convert}$, from
the time it leaves the field for the group/cluster environment until the time its rest-frame colors place it in the quenched region of the $NUVrJ$ diagram. 

\begin{figure}
\includegraphics[clip,angle=0,width=0.98\hsize]{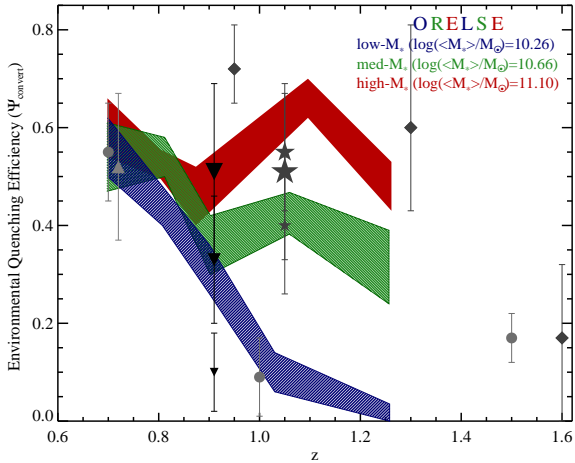}
\caption{Quenching efficiency, defined as $f_{q, \, \rm{high}-\log(1+\delta_{gal}), \, obs}-f_{q, \, \rm{ref.}}/f_{sf, \, \rm{ref.}}$, as measured in a variety of 
different surveys for galaxy populations with different stellar mass ranges. The reference sample is, ideally, a sample of galaxies 
comparable to those galaxies observed in high-density environments but which have been largely untouched by physical effects specific to environment. Different 
schemes are used to define a reference sample. For the ORELSE measurements, shown along with their uncertainties by the red, green, and blue shaded regions, the 
reference sample was generated using the simulations presented in \S\ref{quencheff} and Appendix \ref{mocks}. For other surveys, shown as gray points 
(diamonds: \citealt{nantais17}, circles: \citealt{fossati17}, inverted triangles: \citealt{balogh16}, triangle: \citealt{kovac14}, stars: \citealt{muz12} and 
\citealt{balogh16}), the reference sample is typically chosen as set of field galaxies coeval to and stellar-mass matched to a set of high-density galaxies (see \S\ref{quencheff}). 
The size of the points scale with the stellar mass of the sample, with small, medium, and large points roughly corresponding to the same stellar mass ranges as the low-, med-, 
and high-$\mathcal{M}_{\ast}$ ORELSE samples, respectively.} 
\label{fig:quencheff}
\end{figure}

However, as discussed in Appendix \ref{mocks}, such a concept when applied directly to the data is problematic due to a variety of issues such as the evolution
of the quiescent fraction in the field, differential stellar mass growth, and differential sampling issues. To circumvent these 
issues we have created simulated galaxy populations with which to compare the $f_q$ measurements of the group/cluster galaxies observed in the ORELSE data.
These simulations, meant to simulate the $f_q$ of group and cluster galaxies in the absence of environmental quenching processes, are described in 
detail in Appendix \ref{mocks}. The simulations are performed 200 times, generating 200 unique sets of simulated group/cluster galaxies in the redshift and stellar 
mass ranges $0.65 \le z \le 4$ and $\log(\mathcal{M}_{\ast}/\mathcal{M}_{\odot}) \geq 9$, respectively. The accretion rate of the cluster in each simulation at
each redshift is based on the halo accretion history presented in \citet{mcgee09} for a
$\mathcal{M}_{tot, \, z=0}/\mathcal{M}_{\odot}\ge 14.5$ cluster. This accreted mass is transformed into a galaxy population through the halo-to-stellar mass
relation presented in \citet{moster13} and galaxies are probabilistically assigned a state of quiescence or star-forming based on statistics of the galaxy stellar
mass function of (SMF) \citet{AA14}. For those galaxies assigned a star-forming state, an exponentially declining $\mathcal{SFR}$ is ascribed to each galaxy with
$\tau=2.0$ Gyr and a normalization set by sampling from the relationship between $\mathcal{SFR}$-$\mathcal{M}_{\ast}$ for star-forming galaxies presented in
\citet{AA16}. The average fraction of quiescent galaxies and the normalized median absolute deviation (NMAD; \citealt{hoaglin83}) of the $f_q$ distributions in 
the same three stellar mass bins applied to our final spectral sample (e.g., low-, med-, and high-$\mathcal{M}_{\ast}$, see \S\ref{quiescentfrac}) set 
the values for the low-$\log(1+\delta_{gal})$ populations in the calculation of $\Psi_{convert}$. As a reminder, these $f_{q}$ values are measured on simulated cluster 
galaxy populations in each stellar mass and redshift bin that have been \emph{unaffected by environmental quenching} for their entire evolution beyond what is 
experienced in field samples. The behavior of $f_q$ for simulated galaxies with stellar mass and redshift are shown in Figure \ref{fig:simulatedfq}. 

The simulated $f_{q}$ values are then used in conjunction with the observed $f_{q}$ values measured for the different galaxy populations in the high-$\log(1+\delta_{gal})$ 
bin to estimate the environmental quenching efficiency in the ORELSE survey as a function of redshift and stellar mass. In other words, the simulated $f_{q}$ 
values are used in this calculation instead of our observed low-$\log(1+\delta_{gal})$ $f_q$ values as the former are almost certainly more appropriate when 
calculating quenching efficiencies (see Appendix \ref{mocks}). The errors on the quenching efficiency are calculated from a combination 
of the errors on the $f_q$ values measured for high-$\log(1+\delta_{gal})$ galaxies and the error on the median $f_q$ of all simulations. These environmental quenching efficiencies, 
plotted in Figure \ref{fig:quencheff}, 
exhibit a behavior that largely mirrors that observed in the $f_q$ values shown in Figure \ref{fig:quiescentfrac}. Two specific parallels can be made. First, 
the quenching efficiency decreases precipitously with increasing redshift for low-$\mathcal{M}_{\ast}$ galaxies and becomes formally consistent with group/cluster 
environments having no transformative power at the highest redshifts observed in ORELSE ($z\sim1.4$). Conversely, for galaxies above 
$\log(\mathcal{M}_{\ast}/\mathcal{M}_{\odot})\ge10.4$, $\Psi_{convert}$ appears to decrease gradually or remain constant with increasing redshift. In other words, it 
appears that over the redshift range $0.55\le z \le 1.4$, galaxy clusters and groups 
remain capable of efficiently quenching star formation activity in accreted field galaxies for $\log(\mathcal{M}_{\ast}/\mathcal{M}_{\odot})\ge10.4$, but
that groups and clusters at the higher end of the redshifts studied here ($z\ga1.1$) are incapable of exerting themselves in the same way on galaxies with lower stellar 
masses. 

The values of $\Psi_{convert}$ derived from galaxy populations in other surveys at a variety of different stellar masses and redshifts are overplotted on Figure 
\ref{fig:quencheff}. The size of the points are set to small, medium, and large for galaxy samples with comparable stellar masses to those in the
low-, med-, and high-$\mathcal{M}_{\ast}$ bins in the ORELSE data. Though values associated with other surveys are generally subject to relatively 
large uncertainties, the measurements of $\Psi_{convert}$ made of the ORELSE samples appear generally consistent with other measurements at common redshifts and
stellar masses. Marginally significant tension appears at $z\sim1$ for med-$\mathcal{M}_{\ast}$ galaxies with respect to the study of \citet{fossati17}
based on the 3D-HST survey \citep{brammer12} and the study of \citet{nantais17} based on galaxies selected from the \emph{Spitzer} Adaption of the Red 
Sequence Cluster Survey (SpARCS; \citealt{muz09, wilson09}). Additional moderate tension appears for measurements of low-$\mathcal{M}_{\ast}$ galaxies 
with respect to that measured from the Gemini Cluster Astrophysics Spectroscopic Survey (GCLASS; \citealt{muz12}) as reported in \citet{balogh16}. Though
not plotted, the $\Psi_{convert}$ values derived here are consistently lower at essentially all $\mathcal{M}_{\ast}$ and redshifts than those
measured for groups and cluster populations from initial data taken as part of the Hyper Suprime-Cam Subaru Strategic Program \citep{jian18}. However,
the large number of variants on the method of calculation of this quantity and our imperfect knowledge of any differences that exist the galaxy 
samples presented in these works and those selected here do not allow us to claim anything other than the ORELSE measurements of environmental quenching 
efficiency appear broadly consistent with those from other surveys. 

Though not plotted, the high value of the environmental quenching efficiency observed for low-$\mathcal{M}_{\ast}$ galaxies in our lowest redshift bin is 
broadly consistent with the strong signatures of environmentally-driven quenching seen in the VIPERS Multi-Lambda Survey \citep{thibaud16a} for similar 
galaxies at similar redshifts \citep{thibaud18}. In that study, the authors constrain the quenching timescales 
for such galaxies and settle on ``delayed-then-rapid'' picture of quenching similar to that presented in \S\ref{converttime} in our study. 
Further, it appears that such a quenching channel, while perhaps not the dominant channel for the full population of higher stellar mass galaxies at lower redshifts 
\citep{thibaud18}, appears dominant for galaxies of these stellar masses which inhabit clusters and groups at all redshifts studied here (see \S\ref{converttime}
for more discussion). However, that study lacked the redshift baseline to observe the precipitous drop of environmental quenching efficiency for galaxies 
$\log(\mathcal{M}_{\ast}/\mathcal{M}_{\odot})<10.4$. 
These lines of evidence generally point to a picture in which the conditions in groups and clusters, while still sufficient to be able to efficiently
quench the highest stellar mass galaxies over the redshift range studied here, are changing subtly with redshift. That the environmental quenching
efficiency for lower-$\mathcal{M}_{\ast}$ declines rapidly over this window perhaps suggests a similar decline for higher-$\mathcal{M}_{\ast}$ at even higher redshifts
than are studied here. Such a picture is consistent with several works studying clusters at $z\ga1.4$, in which the average star formation is seen in many cases
to be unperturbed or even enhanced relative to field environments (e.g., \citealt{mbrodz13,alberts14, alberts16,santos14,taowang16,stach17,coogan18}). 
 
That efficient environmental quenching is observed to persist up to $z\la1.4$ for galaxies of stellar masses $\log(\mathcal{M}_{\ast}/\mathcal{M}_{\odot})\ge10.4$
is consistent with results presented in a companion ORELSE study \citep{AA19}. In this study, the specific star formation rate-overdensity for star-forming galaxies and 
$f_q$-overdensity relations from a combined spectroscopic and photometric sample drawn from ORELSE are constructed and used to infer the relative 
number of galaxies undergoing active quenching in different environments in different stellar mass bins across the redshift range $0.8 \la z \la 1.05$. 
For galaxies in a similar stellar mass range, the majority of galaxies ($\sim$70-90\%) inhabiting the highest density environments observed in ORELSE 
($\log(1+\delta_{gal})>1$) are estimated to be undergoing quenching characterized by an $\mathcal{SFR}$ that is exponentially declining with $\tau=0.5$ Gyr. In the next
section, we will combine these results with $N$-body simulations and the $\Psi_{convert}$ valued derived here to create estimates of various timescales required 
to convert a galaxy from star-forming to quiescent after the accretion into a group/cluster environment. Interestingly, the clear persistence of the color-density
relation for higher-stellar mass galaxies across the entire redshift range studied here may help to explain the relatively high quiescent fractions found in distant 
cluster surveys which are predominantly sensitive to galaxies above $\log(\mathcal{M}_{\ast}/\mathcal{M}_{\odot})\ga10.4$ (e.g., \citealt{newman14, cooke16,minirud17,strazz19}). 
This persistence is also broadly consistent with the results of \cite{vanderburg13} who studied the quiescent fraction as a function of stellar mass for galaxies inhabiting 
10 massive $z\sim1$ clusters drawn from the GCLASS survey. In that study, the authors found that quiescent cluster galaxies significantly outnumbered their star-forming 
counterparts for all but the lowest stellar masses probed in their data ($\log(\mathcal{M}_{\ast}/\mathcal{M}_{\odot})\la10.2$). 

Another interesting consequence of the results presented here is that ram pressure stripping and harassment are precluded as the sole mechanisms for driving 
environmental quenching over the redshift range $0.55 \le z \le 1.4$. Because these mechanisms rely on stripping matter, typically atomic or molecular gas, 
\emph{in situ} to group/cluster galaxies, everything else being equal, e.g., galaxy matter density profile, differential velocity, the density of the 
intragroup or intracluster medium (IGM/ICM) in the case of ram pressure stripping, galaxies with a shallower potential well are more susceptible to such processes 
(e.g., \citealt{gunn72, fillingham16}). That $\Psi_{convert}$ is observed to decrease with redshift for the lowest stellar mass galaxies 
(i.e., $\log(\mathcal{M}_{\ast}/\mathcal{M}_{\odot})\sim10.2$) without a corresponding decrease for their higher stellar mass counterparts is a behavior that cannot
be reconciled with ram pressure stripping or harassment acting as the primary quenching mechanism unless lower stellar mass galaxies are accreted with a higher (total) 
amount of gas or are generally more concentrated than galaxies with higher stellar masses. Neither of these differences are observed in studies of field galaxies in the 
redshift range studied here (e.g., \citealt{popping12,vanderwel14, smokynicky17}). Alternatively, if the conditions in groups and clusters change by $z\sim1.4$ to 
reduce the efficacy of ram pressure 
stripping or harassment, i.e., the differential velocity of galaxies relative to one another or the density of the IGM/ICM decreases, 
this change would possibly provide some possibility of explaining the observed behavior. However, such a change in conditions is extremely unlikely for the 
sample of ORELSE clusters as they are observed to contain a well-developed ICM and a galaxy population with a high $\sigma_{v}$ to the highest redshifts
studied here \citep{blanton03, maughan06, meh12, clerc14, rum13, rum18}. In addition, the average virial mass of the detected clusters and groups in the ORELSE
sample does not change appreciably with redshift but rather holds steady in the range $\log(\langle\mathcal{M}_{vir} \rangle) = 14.2-14.5$ (though see discussion
at the end of \S\ref{converttime}). 

\subsection{Conversion Timescales}
\label{converttime} 
In this section we use the values of $\Psi_{convert}$ measured from our final sample in the previous section in conjunction with results from the 
$N$-body simulations to estimate the average amount of time needed for galaxies of various stellar masses to continuously reside in a group or cluster 
environment in order to be quenched by environment-specific processes and for their rest-frame colors to become consistent with quiescence. 
This timescale will be referred to as $t_{convert}$. In order to estimate a 
timescale associated with the conversion of accreted group and cluster galaxies, it is necessary first to create a model of the accretion history 
of galaxies for a given set of groups/clusters. In order to do this, we rely on the results presented \citet{mcgee09} and \citet{balogh16} in which 
the average accretion history of groups and clusters of various masses is estimated from results of the Millennium Simulation \citep{springel05} using 
the methodology of estimating halo merger trees described in \citet{helly03} and \citet{harker06}. For the purposes of this calculation, we adopt the 
average $z=0$ accretion history for a $\log(\mathcal{M}_{tot, \, z=0}/\mathcal{M}_{\odot})>14.5$ cluster and scale it to each redshift of interest
by the redshift dependence of the dynamical timescale (1+$z$)$^{-3/2}$. As shown in \citet{balogh16}, such a scaling provides a close approximation of the true 
cluster accretion history. We choose the accretion history of a massive $z=0$ cluster as the majority of ORELSE structures are predicted to evolve into such structures
by $z=0$ (see Appendix \ref{mocks}). Our results remain largely unchanged if we instead adopt an alternate accretion history. 

\begin{figure*}
\includegraphics[clip,angle=0,width=0.48\hsize]{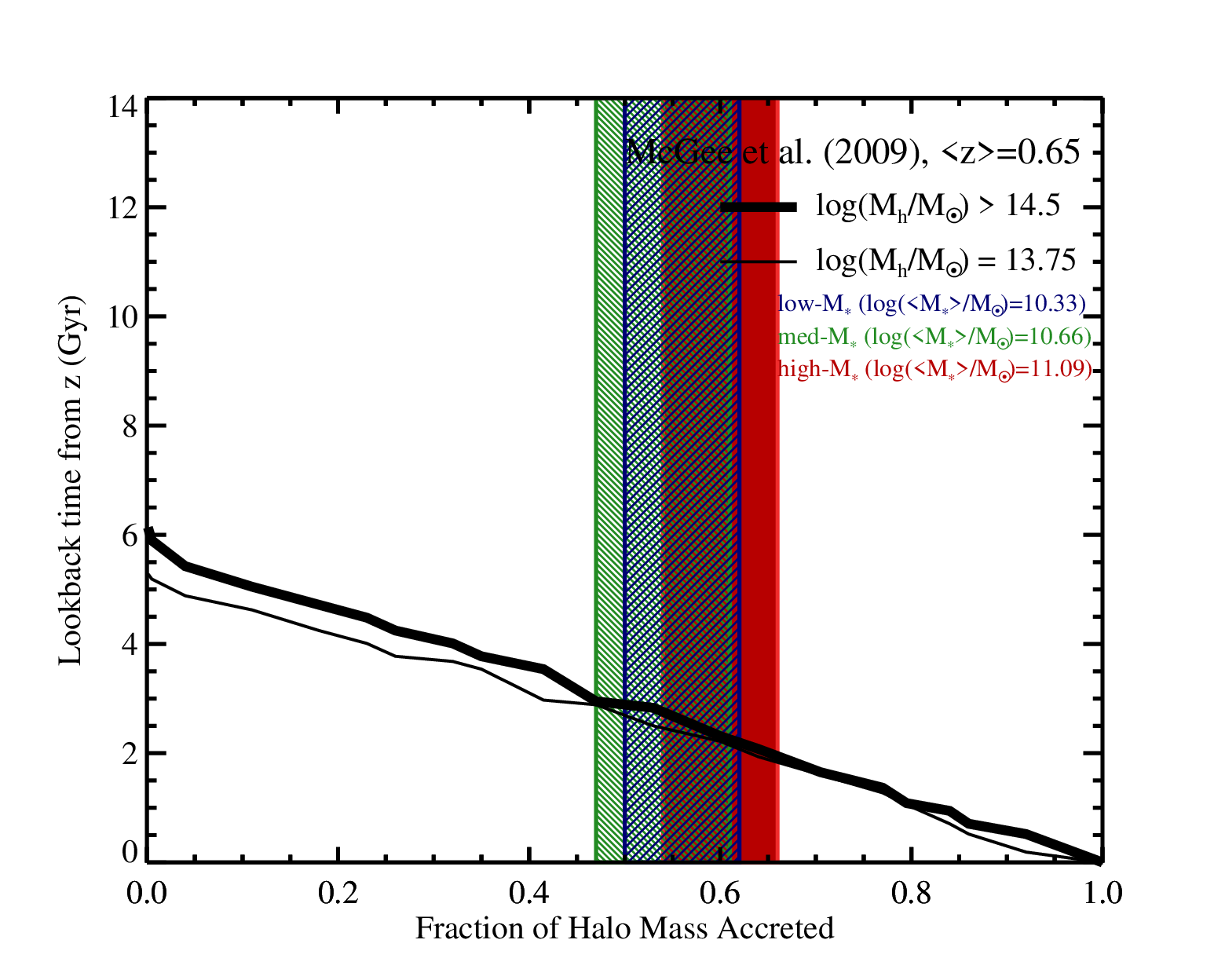}
\includegraphics[clip,angle=0,width=0.48\hsize]{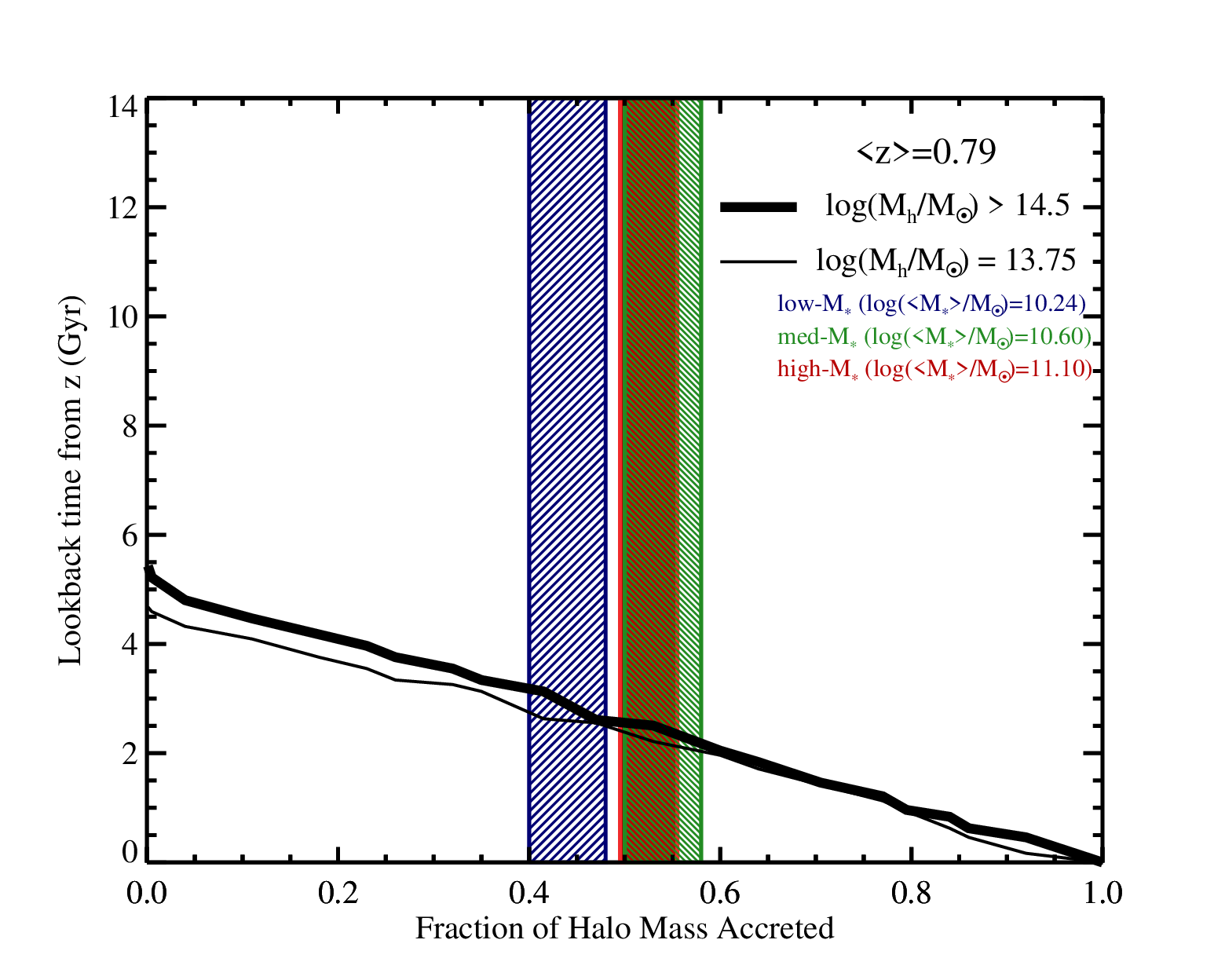}
\includegraphics[clip,angle=0,width=0.48\hsize]{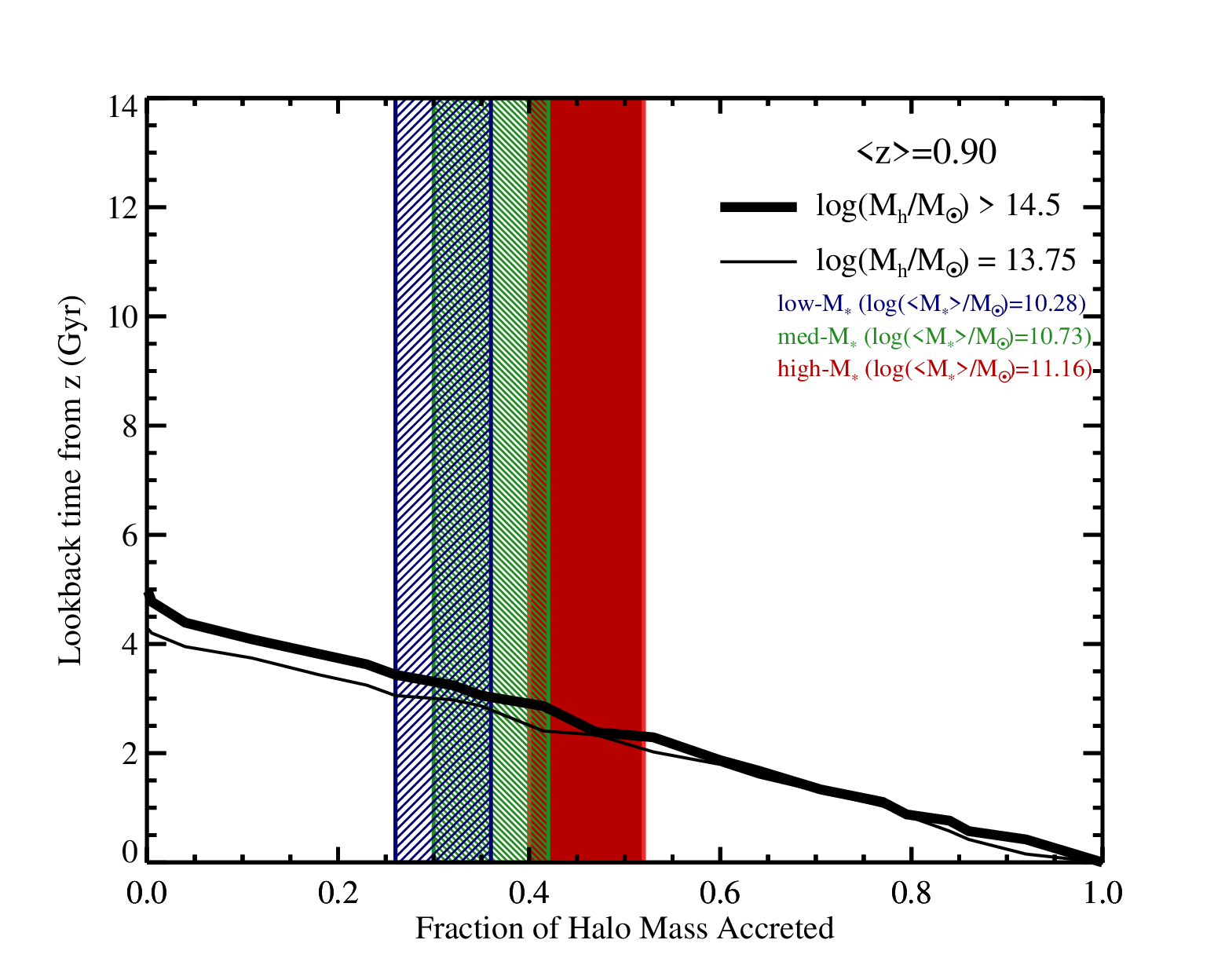}
\includegraphics[clip,angle=0,width=0.48\hsize]{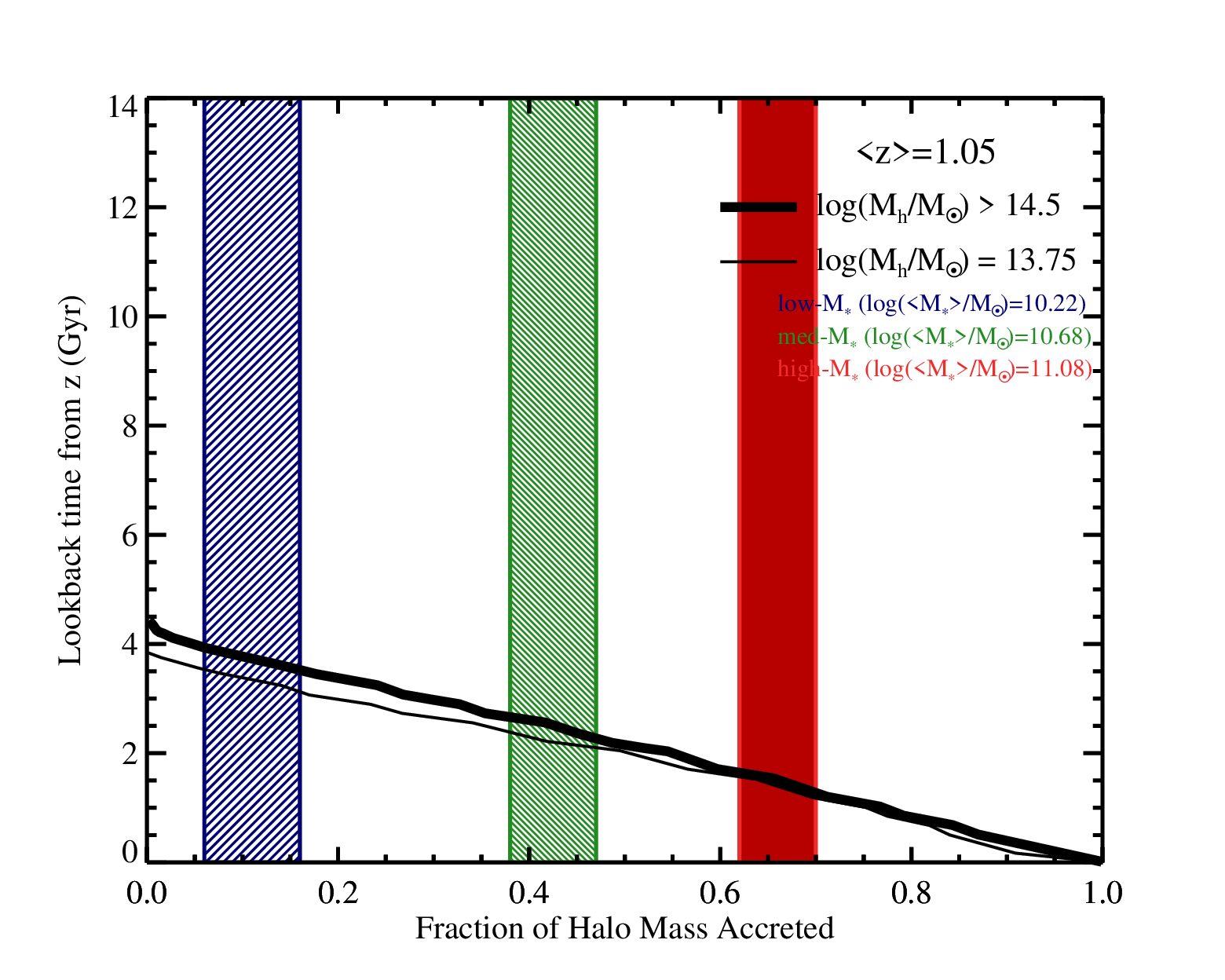}
\includegraphics[clip,angle=0,width=0.48\hsize]{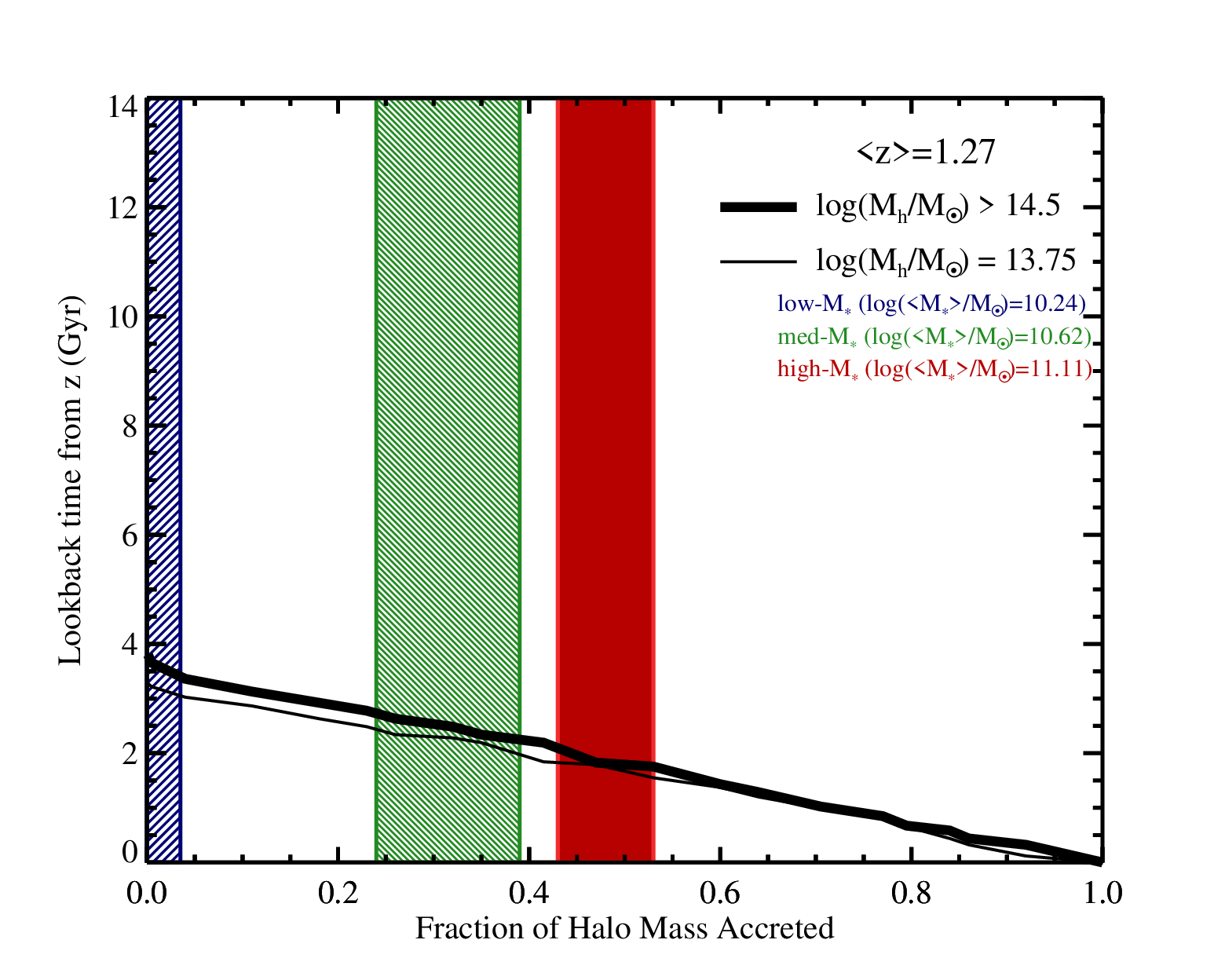}

\caption{Average halo mass accretion histories of two different types of structures differentiated by their $z=0$ mass scaled as estimated from merger trees 
applied to the Millennium $N$-body simulation taken from \citet{mcgee09}. The five panels correspond to measurements on simulated and real galaxies falling in 
the five different redshift ranges applied to the final ORELSE spectral sample, with the average redshift of each sample shown in the top right of each panel. 
Accretion histories are estimated at $z=0$ and scaled to the five average redshift values by the dynamical time (1+$z$)$^{-3/2}$. The shaded color regions in 
each panel denote the $\pm$1$\sigma$ range of environmental quenching efficiencies, $\Psi_{convert}$, of the three different stellar mass samples in each
redshift bin. The average stellar mass of each sample in each redshift bin is given in the top right of each panel. The intersection between the thick solid line 
and each shaded region sets the conversion timescale, $t_{convert}$, of each sample at each redshift. Notice that while $t_{convert}$ remains relatively 
constant for the med- and high-$\mathcal{M}_{\ast}$ samples across all redshifts studied here, $t_{convert}$ is seen to consistently increase with increasing 
redshift for low-$\mathcal{M}_{\ast}$ galaxies, and becomes consistent with being longer than the time since formation of the cluster by $\langle z \rangle =1.27$.} 
 
\label{fig:tconvert}
\end{figure*}

\begin{table*}
        \centering
        \caption{Conversion Timescales, $t_{convert}$, of the ORELSE Spectral Sample}
        \label{tab:tconverttab}
        \begin{tabular}{cccccc}
                \hline
                Sample & $ 0.55 \le z < 0.75 $ & $0.75 \le z < 0.84$ & $0.84\le z < 0.96$ & $0.96 \le z < 1.15$ & $1.15\le z < 1.40$ \\
                & [Gyr] & [Gyr] & [Gyr] & [Gyr] & [Gyr] \\
		\hline
                low-$\mathcal{M}_{\ast}$ & 2.66$^{+0.33}_{-0.41}$ & 2.94$^{+0.28}_{-0.29}$ & 3.25$^{+0.15}_{-0.27}$ & 3.72$^{+0.22}_{-0.21}$ & 3.76$^{a}_{-0.36}$ \\[1.6ex] 
                med-$\mathcal{M}_{\ast}$ & 2.79$^{+0.26}_{-0.49}$ & 2.58$^{+0.11}_{-0.24}$ & 2.98$^{+0.30}_{-0.23}$& 2.60$^{+0.11}_{-0.22}$ & 2.46$^{+0.19}_{-0.29}$ \\[1.6ex] 
                high-$\mathcal{M}_{\ast}$ & 2.37$^{+0.48}_{-0.34}$ & 2.60$^{+0.10}_{-0.19}$ & 2.41$^{+0.39}_{-0.18}$ & 1.64$^{+0.14}_{-0.23}$ & 1.76$^{+0.31}_{-0.08}$ \\
		\hline 
\end{tabular}
        \begin{flushleft}
$a$: Formally consistent with no environmental quenching
\end{flushleft}
\end{table*}

In order to link the measurement of $\Psi_{convert}$ to the simulated cluster accretion history it is necessary to make several assumptions. The first assumption
is that star-forming galaxies that have been in the cluster environment the longest will be the first to convert to a quiescent state. This assumption, by construction,
means that any timescale derived from this exercise will necessarily be an upper limit. For example, suppose we estimated that, at a given redshift for a given galaxy 
sample, 50\% of the star-forming galaxies in the model must convert to quiescent in order to match the observed $f_q$ for that
redshift/galaxy sample. In this case, we make the assumption that the star-forming galaxies which accreted into the structure first are those that have experienced environmental
quenching, and the remaining 50\% have not been in the structure
for a long enough time to experience similar quenching. This assumption makes it, in principle, easy to directly derive a conversion timescale from the measurements
of $\Psi_{convert}$ and the simulated accretion history, as we are able to determine from the latter, the percentage of galaxies accreted by a given time, e.g., 
by $\langle z \rangle = 0.65$, approximately 50\% of star-forming galaxies have been in the cluster environment for longer than $\ga 3$ Gyr (see the 
lookback time corresponding to the point at which the fraction of halo mass accreted is equal to 0.5 in the top left panel of Figure \ref{fig:tconvert}). Notice that 
we make the additional assumption here that the mass accreted in all time steps of the Millennium simulation scales by a constant multiplicative factor with respect to the mass 
accreted in the form of star-forming galaxies. It is likely that this assumption that broadly holds as 
the $f_q$ values of accreted galaxies, i.e., those in the field of the stellar mass ranges considered here, are relatively flat for the majority of the accretion history 
of the cluster (see Figure \ref{fig:simulatedfq} and \citealt{AA14}). Further, we assume on average the amount of stellar mass loss or growth of each star-forming galaxy 
from the time of accretion 
until it becomes quiescent is insufficient to move into a different stellar mass bin. The wideness of the stellar mass bins used here help immensely with the validity 
of this assumption. Finally, as in the simulations presented in Appendix \ref{mocks} the effects of merging processes in modulating the number or stellar mass distribution of 
galaxies since being accreted are again ignored here. 

Because the scaled \citet{mcgee09} group/cluster accretion curves indicate the amount of total mass accreted as a function of time, they provide a direct mapping, 
subject to the assumptions above, between $\Psi_{convert}$ and the earliest time at which the converted galaxies (i.e., the quenched star-forming galaxies) could 
have been accreted. Combining this mapping with our assumption that star-forming galaxies with the earliest accretion times are the first to quench at a given epoch
provides a way to estimate $t_{convert}$ for galaxy populations of different stellar masses at each of the redshifts studied here. The uncertainties in $\Psi_{convert}$
can similarly be mapped to uncertainties in $t_{convert}$. In Figure \ref{fig:tconvert} we plot the scaled simulated accretion histories 
against the backdrop of the $\pm$1$\sigma$ range of allowed values of $\Psi_{convert}$ for each galaxy population in the five different redshift ranges studied here.  
From the intersection of these values we estimate the values of $t_{convert}$ for the 15 different galaxy populations shown in the various panels in Figure 
\ref{fig:tconvert} and list them in Table \ref{tab:tconverttab}. For higher stellar mass galaxies, i.e., those in the med- and high-$\mathcal{M}_{\ast}$ bins,
$t_{convert}$ is seen to largely hold or slightly decrease over the entire redshift range studied here, $0.55 \le z \le 1.4$, varying from 1.6-3.0 Gyr across the various 
samples with an average of 2.4$\pm$0.3 Gyr. Conversely, the conversion timescale for low-$\mathcal{M}_{\ast}$ galaxies is observed to monotonically increase 
with increasing redshift, spanning from 2.7-3.8 Gyr over the same redshift range, with an average of 3.3$\pm$0.3 Gyr, again suggesting that cluster environments struggle 
to process lower stellar mass galaxies at $z\ga 1.1$.  

These results are in broad agreement with the few studies that have attempted to constrain similar or identical timescales within the redshift range probed by the ORELSE
sample. \citet{foltz18} studied the photometric and spectroscopic galaxy population of 14 clusters at $0.85 \le z \le 1.65$ drawn from the GCLASS and SpARCS cluster surveys. 
Using an approach which combines the color distribution of background-subtracted member populations and accretion rates estimated from $N$-body simulations they constrained 
$t_{convert}$ (called $t_{Q}$ in their nomenclature) of member galaxies with $\log(\mathcal{M}_{\ast}/\mathcal{M}_{\odot})>10.5$. For member galaxies 
with mean redshifts of $\langle z \rangle = 1.04$ and 1.55 the conversion timescales were estimated at $t_{convert}=1.50^{+0.19}_{-0.18}$ and $1.24^{+0.23}_{-0.20}$. These timescales
are consistent with the values derived for our high-$\mathcal{M}_{\ast}$ sample at comparable redshifts, though somewhat in tension with those values derived from our
med-$\mathcal{M}_{\ast}$ sample. However, the stellar mass distribution of their final sample appears to be a compromise between that of the ORELSE med- and 
high-$\mathcal{M}_{\ast}$ galaxies. In addition, their sample is comprised exclusively of cluster galaxies while the high-$\log(1+\delta_{gal})$ ORELSE sample contains galaxies 
residing in filaments, groups, and clusters. These differences likely ease this tension considerably. \citet{balogh16} studied the spectroscopic member 
populations in $0.8 \la z \la 1.2$ 
groups and clusters from the Galaxy Environment Evolution Collaboration 2 (GEEC2; \citealt{balogh11,balogh14}) and GCLASS surveys. Combining measurements made from coeval 
field and group/cluster populations along with results from the same simulations that are employed here, they found conversion timescales (called $t_{p}$ in their nomenclature) 
of $t_{convert}=2.4\pm0.6$ and $1.7\pm0.6$ for group and cluster galaxies, respectively, in the stellar mass bin $10.3 < \log(\mathcal{M}_{\ast}/\mathcal{M}_{\odot}) < 10.75$. 
These values are consistent with the $t_{convert}$ values measured in ORELSE at equivalent redshifts and stellar masses. Interestingly, in both the group and cluster
sample, though the former employing $z_{phot}$-selected member galaxies, they found evidence of $t_{convert}$ abruptly increasing with decreasing stellar mass, with
estimated values consistent with those derived from the low-$\mathcal{M}_{\ast}$ sample. Estimates of $t_{convert}$ derived from groups in 3D-HST for galaxies of stellar 
masses comparable to the med-$\mathcal{M}_{\ast}$ are also found to be comparable to the estimates from our sample over the same redshift range ($ \langle t_{convert} 
\rangle \sim$2.5$\pm$0.5 Gyr, \citealt{fossati17}). While considerable uncertainties exist in the conversion timescales of a particular galaxy population, these
estimates of $t_{convert}$ along with our own point to a picture in which galaxies spend a considerable time in the group or cluster environment prior to 
completely ceasing star-formation activity and fading to quiescence.

Such a picture is, however, seemingly difficult to reconcile with the large fractional population of galaxies known as ``post-starburst" or ``K+A" galaxies 
seen in high redshift groups and clusters (e.g., \citealt{dressler99, tran03, muz14, pwu14, lem17a, mildmanneredmiguel18}). Such galaxies contain spectral 
features of colors that necessarily indicate a rapid cessation in their star-formation activity and appear in numbers considerable relative to those of 
star-forming galaxies \citep{pog09, lem17a}, strongly implying that rapid quenching is a common pathway to quiescence for such galaxies. Additionally, 
as mentioned in the previous section, we have observed in ORELSE a strong indication of a fractionally large transition population amongst group/cluster 
galaxies with stellar masses in excess of $\log(\mathcal{M}_{\ast}/\mathcal{M}_{\odot})>10.1$ at $0.85 \le z \le 1.05$. This transition population is 
characterized by an exponentially declining SFH with $\tau=0.5$. Assuming this form of the SFH, a star-forming galaxy at this redshift 
would have colors consistent with quiescence between 0.8-1.4 Gyr after the inception of the initial quenching event, values considerably shorter than
any of the conversion timescales estimated here. Such a timescale is nearly identical to the luminosity-weighted stellar age of the oldest K+A galaxies observed
in the ORELSE survey ($t_{K+A}=1.00^{+0.33}_{-0.21}$ Gyr; \citealt{lem17a}). The only possibility to reconcile the relatively rapid timescale involved in 
moving a galaxy to quiescence once quenching has begun and the relatively large timescale of $t_{convert}$ is to introduce a third timescale during which
a galaxy accreted to a group or cluster environment remains relatively unaffected by quenching processes. We will refer to this time as $t_{delay}$ and 
the time from the inception of the quenching event to the time that a galaxy exhibits colors classifying it as quiescent as $t_{quench}$ such that 
$t_{quench}+t_{delay}\equiv t_{convert}$. 

Adopting $t_{quench}$ as 1.1$\pm$0.3 Gyr from \citet{AA19} and solving for $t_{delay}$ yields an average $t_{delay}$=1.3$\pm$0.4 Gyr and 2.2$\pm$0.4 Gyr for
the combined med- and high-$\mathcal{M}_{\ast}$ and low-$\mathcal{M}_{\ast}$ ORELSE samples, respectively, across all redshifts studied. These delay
times are broadly consistent with similar timescales measured from other surveys in a few comparable redshift and stellar mass ranges
(e.g., \citealt{mok14,muz14, balogh16}). These values are also well in excess of the average time required for galaxies to move to their first pericentric passage
after being accreted by their parent group or cluster at the median redshift of the ORELSE sample \citep{wetzel11,wetzel13}.    
Such a long delay time is supported by recent evidence taken from SDSS and redMaPPer clusters
at low- to intermediate-redshift that show appreciable populations of galaxies far from center of clusters thought to have survived their first passage through the
cluster to make it to their first apogee (i.e., ``splashback" galaxies, \citealt{more16, baxter17, chang18}). The fact that the bulk of group/cluster galaxies remain
largely unperturbed in their star-formation activity at least through their initial pericentric passage and that an additional $\sim$0.5-1.5 Gyr is allowed to elapse prior
to the inception of quenching appears to place most galaxies in the more rarefied regions of their parent groups/clusters when quenching is initiated. This 
inference strongly disfavors ram pressure
stripping and galaxy harassment as the sole mechanisms responsible for quenching. While orbits can take many forms other than purely radial and pericentric passage can occur
far from the group/cluster core, at least some fraction of newly accreted galaxies must pass through $R_{500}$. The lack of a significant quenching signature in this
population makes the possibility that these processes are the sole inducer of quenching activity remote.

Conversely, dynamical processes such as galaxy-galaxy merging 
events or other strong 
interactions are strongly favored by such a scenario, as the $t_{delay}$ timescales estimated here appears to be directly scaled from comparable timescales measured 
at $z=0$ by the dynamical time \citep{wetzel13}. That merging events might be primarily responsible for catalyzing quenching qualitatively conforms with the difference
in the average $t_{delay}$ observed for lower and med- to high-$\mathcal{M}_{\ast}$ galaxies in ORELSE. While the major merging rate at these redshifts is 
not strongly dependent on stellar mass in the ranges of stellar masses studied here at least in the field (e.g., \citealt{carlos13}), field galaxies accreted into 
the group/cluster environment at higher stellar masses are on average more highly clustered than those at lower stellar mass (e.g., \citealt{marulli13, coil17, ania18}). The fact that 
the bulk of the galaxies accreted as star forming survive in that state past their initial pericentric passage provides an intriguing possibility that tidal forces related 
to the overarching halo or interactions with \emph{in situ} galaxies provide the necessary frustration to induce merging between infalling pairs or small groups of 
galaxies. A merging scenario to induce the beginnings of quenching activity is also broadly supported in the analysis of \citet{lem17a} in a study of the K+A 
populations of two ORELSE LSSs. In that study, it was proposed that merging events in the outskirts of groups and clusters induce strong stellar and/or AGN 
feedback that begin the quenching process and this quenching is then finalized by the effects of ram pressure stripping as the remnant galaxy makes it way back 
to pericentric passage. In this scenario, ram pressure stripping is invoked only to prevent the galaxy from rejuvenating its star-formation activity rather than as a primary 
transformative mechanism. The timescales estimated above appear to largely support such a scenario as the additional quenching time, $t_{quench}$, is long enough to 
allow galaxies the possibility of making their second passage through their parent group or cluster. 

This scenario partially mirrors that put forth in \cite{mbrodz13}, in which quenching of galaxies inhabiting higher redshift clusters
($z\ga1$) was suggested to proceed through a combination of major merging and AGN activity. Such a scenario is supported by the increase in both types of activity with 
increasing redshift (see, e.g., \citealt{mancone10, AA17, debz18} and \citealt{galametz09, martini09, martini13}, though see also \citealt{rum17}).  The idea that AGN in such 
environments are triggered by major mergers of infalling field galaxies is supported by the analysis of X-ray-selected AGN hosts of $>100$ massive clusters in the 
redshift range $0.2<z<0.9$ presented in \cite{ehlert15} as well as several ORELSE-specific studies (e.g., \citealt{dirtydale09,rum12,rum17,shen17}). In a future paper,
in which we combine mid-, far-infrared, and radio data along with complex modeling, we will show that ORELSE group and cluster galaxies undergoing strong AGN activity 
are both actively quenching the star formation of their host galaxies and appear to be predominantly infalling galaxies on non-radial orbits (Shen et al.\ \emph{submitted}). 

The above ``merging-then-stripping'' scenario appears to be broadly able to explain the relatively high quenching efficiencies observed at all redshifts 
for the med- and high-$\mathcal{M}_{\ast}$ samples. However, it is not clear such a scenario is appropriate for low-$\mathcal{M}_{\ast}$ galaxies. 
At $z\sim0.7$, $\Psi_{convert}$ for low-$\mathcal{M}_{\ast}$ galaxies is indistinguishable from that of their higher-$\mathcal{M}_{\ast}$ counterparts.
But, unlike higher-$\mathcal{M}_{\ast}$ populations, environmental quenching appears to become less effective with increasing redshift for 
low-$\mathcal{M}_{\ast}$ galaxies and is completely ineffectual at the highest redshifts studied here. Given this dramatically different 
behavior, it naively appears that the merging-then-stripping scenario proposed above cannot explain the behavior 
of both sets of galaxies. 

However, it is possible that various considerations related to the scenario above that allow for efficient quenching of higher-$\mathcal{M}_{\ast}$ galaxies at 
all redshifts operate on low-$\mathcal{M}_{\ast}$ populations in a different way. Further, it is possible that these differences act in such a way 
to explain the decreased quenching efficiency of these galaxies in high-$z$ clusters and groups. The main considerations 
within the scenario are \textbf{a)} the likelihood of undergoing a wet or mixed merging event, \textbf{b)} subsequent to a merging event, the  
likelihood of sufficient stellar and/or AGN feedback to push gas that survives to large galactocentric distance, and \textbf{c)} the likelihood that 
a parent cluster or group has a sufficiently hot and dense medium to strip the excised gas from the merger remnant. If any of these three considerations 
are not the same 
for low-$\mathcal{M}_{\ast}$ as for higher-$\mathcal{M}_{\ast}$ galaxies, it is possible that the merging-then-stripping scenario can still be reconciled
with the behavior observed for galaxies of all stellar masses considered in this study.

Taking each in turn, the merging rate of field galaxies is observed to either stay constant or increase with increasing redshift over the 
redshift range studied here \citep{bundy09,carlos13} and does not depend strongly on stellar mass implying \textbf{a)} is equally or more likely
to happen at high redshift. Thus, it is unlikely that \textbf{a)} is appreciably different for low-$\mathcal{M}_{\ast}$ galaxies at higher $z$ than
for other samples. However, the gas fraction of the galaxies comprising those mergers over that same redshift range increases by a factor of 
$\sim$3 for all stellar masses considered here \citep{smokynicky17, darvish18} and becomes the dominant component of the baryonic content at least
for low-$\mathcal{M}_{\ast}$ galaxies. As such, the feedback required under \textbf{b)} is necessarily stronger than that at lower redshift. 
While stellar feedback is capable of providing some of the required effect, its effect alone is typically insufficient to excise the bulk of the gas to large galactocentric 
distances for all of the galaxies studied here (i.e., $\log(\mathcal{M}_{\ast}/\mathcal{M}_{\odot})\ga10$; \citealt{bahe15}, though see also \citealt{diamond-stanic12}). 
Activity from an AGN can substantially 
increase these large-scale outflows (e.g., \citealt{hopkins16}). While a large fraction of galaxies in the stellar mass range of the low-$\mathcal{M}_{\ast}$ galaxies
studied here are known to host AGN, these AGN are typically of the less powerful variety \citep{juneau13}. As a consequence, it is possible that such AGN 
activity, in tandem with stellar feedback, is sufficient to provide the required feedback in such galaxies at the lower redshift end of our sample, but fails for 
the larger gas reservoirs present at higher redshifts. Incidentally, higher stellar mass galaxies would likely not suffer from the same considerations, as 
AGN activity, when present, is more likely to be powerful (e.g., \citealt{kauff03, pimbblet13, bongiorno16, lanzuisi17}). 

Finally, we consider concerns related to \textbf{c)}. As we have mentioned previously, the median mass of the detected ORELSE structures does not change 
appreciably with redshift, nor do the properties of ICM of those ORELSE clusters detected at X-ray wavelengths \citep{rum18}, implying that \textbf{c)} 
is equally likely at low and high redshift within the sample studied here. However, ORELSE groups are also observed to contain a larger fraction of lower 
stellar mass star-forming galaxies relative to cluster environments (see, e.g., \citealt{AA17}, where local overdensity is used as a proxy for group and 
cluster environments). It is, thus, possible that low-$\mathcal{M}_{\ast}$ galaxies are generally experiencing less compelling ram pressure stripping than their 
higher-$\mathcal{M}_{\ast}$ counterparts. This logic, though, applies to low-$\mathcal{M}_{\ast}$ galaxies across all epochs studied here unless groups at 
lower redshifts, simply by virtue of having a longer available time to form, have been able to produce a denser medium (see, e.g., \citealt{jeltema09} for 
a discussion on this effect over a larger temporal baseline). 

It, thus, appears that the conditions governing \textbf{b)} and \textbf{c)} are sufficiently uncertain to leave open the possibility that  
the merging-then-stripping scenario advocated for quenching group/cluster med- and high-$\mathcal{M}_{\ast}$ galaxies could also generally 
apply to low-$\mathcal{M}_{\ast}$ galaxies in those same environments without inducing similar levels of quenching at the highest redshifts probed in this study. 
The implementation of a variety of AGN selection techniques and obtaining deeper X-ray observations, both to uncover weaker AGN activity and to detect the 
diffuse medium of lower-mass structures, as well as obtaining imaging from the Atacama Large Millimeter/submillimeter Array (ALMA; \citealt{wootten09}) to 
measure the gas content of galaxy samples over a variety of stellar masses, environments, and redshifts would be extremely helpful to validate or 
disprove this scenario. Additionally, while our current detection techniques do not allow us to identify groups in a 
systematic and well-characterized manner up to the highest redshifts studied here, future work \citep{denise19} using more sophisticated group and cluster 
detection algorithms applied to our VMC maps will allow the systematic detection of structures for $\log(\mathcal{M}_{tot}/\mathcal{M}_{\odot})>13.5$ over
the redshift range $0.55 \le z \le 1.4$. Once these environments are identified, we will be able to additionally bin our stellar mass, environmental, and 
redshift sub-samples by total structure mass to further test this scenario.


\section{Summary and Conclusions}

In this study we presented the fraction of different types of galaxies which have ceased to form stars as estimated
from observations taken as part of the Observations of Redshift Evolution in Large Scale Environments (ORELSE) survey. These populations were drawn from a sample of 
$\sim$5000 spectroscopically-confirmed galaxies which span a large range in 
stellar mass, environments, and redshifts and were shown to be broadly representative of the overall galaxy populations at these epochs. Using these fractions, in 
conjunction with a variety of models, we were able to place constraints on the average efficacy of environmental quenching over a 
large baseline in cosmic time for a large variety of galaxy populations, as well as timescales and processes associated with this quenching. Our main conclusions are as follows:

\begin{itemize} 

\renewcommand{\labelitemi}{$\bullet$} 

\item The quiescent fraction, $f_q$, was generally observed to increase with increasing galaxy density, increasing stellar mass, and decreasing redshift for 
the range of environmental densities ($0.1\le \log(1+\delta_{gal} \le 2.0$), stellar masses ($10 \leq \log(M_{\ast}/M_{\odot}) \leq 11.6$), and redshifts
($0.55\leq z \leq 1.4$) studied here. 

\item Except for the very lowest stellar mass bin at the highest redshifts probed, the color-density relation was observed to persist to $z\sim1.4$. For 
intermediate and higher stellar mass galaxies ($\log(M_{\ast}/M_{\odot}) > 10.4$), essentially no significant decrease in $f_q$ was observed 
from $z=0.55$ to $z=1.4$ at fixed stellar mass or environment. For lower stellar mass galaxies, the color-density relation appeared to evolve
rapidly with redshift over the redshift range studied here, essentially disappearing at the highest redshifts. 

\item A general formula to calculate the quiescent fraction of galaxy populations was estimated and took the form: $f_{q}(z, \mathcal{M}_{\ast}, \log(1+\delta_{gal})) 
= \alpha z^{\zeta} + \beta \log(M_{\ast}/M_{\odot})^{\theta} + \gamma \log(1+\delta_{gal})^{\kappa}$, with $\alpha = -1.8544\pm0.0002$, $\beta = 8.3876\pm0.0001$, 
$\gamma=-11.8321\pm0.0002$, $\zeta=0.1421\pm0.0007$, $\theta=0.2224\pm0.0007$, and $\kappa=-0.0076\pm0.0005$. This formula was designed to be applicable to 
general galaxy populations within the range of environments, stellar masses, and redshifts. However, it is possible that this formula slightly overestimates the 
quiescent fraction of lower stellar mass galaxies ($10 \le \log(M_{\ast}/M_{\odot}) \le 10.4$) by up to $\sim$0.1. 

\item A semi-empirical model was used to generate a variety of mock galaxies that inhabit dense environments but which were shielded from the effects of 
environmental quenching. These predicted $f_q$ values, in conjunction with the observed $f_q$ values in dense environments, were used to estimate environmental
quenching efficiencies ($\Psi_{convert}$) for all galaxy types. For intermediate and higher stellar mass galaxies, the efficiency of cluster/group environments to 
inhibit star-formation activity in such galaxies remains roughly constant for all redshifts studied here. For lower stellar mass galaxies, a precipitous drop
in $\Psi_{convert}$ was observed from lower to higher redshift.
 
\item These $\Psi_{convert}$ values along with group/cluster accretion histories as estimated from the Millennium simulation allowed constraints to be placed
on the average time elapsed between the accretion of a galaxy and quiescence ($t_{convert}$). For intermediate and higher stellar mass galaxies, this value
was $\langle t_{convert} \rangle = 2.4\pm0.3$, while for lower stellar mass galaxies this time was estimated to be significantly longer on average
($\langle t_{convert} \rangle = 3.3\pm0.3$). 

\item Using results from a companion ORELSE study, which investigated the relationship between density and specific star-formation rate for a nearly identical
set of galaxies \citep{AA19}, and the $t_{convert}$ timescales estimated here, we found strong support for a scenario in which a galaxy spends a considerable 
amount of time in a group/cluster environment unquenched followed by a rapid truncation of its star-formation, a phenomena known as ``delayed-then-rapid''
quenching. These results were used to place constraints on the average delay time between accretion and the inception of the quenching process, finding 
$t_{delay}=1.3\pm0.4$ and $t_{delay}=2.2\pm0.4$ Gyr for the combined intermediate- to high-stellar mass and lower-stellar mass samples, respectively.

\item The behavior of $f_q$ with redshift as well as the long delay times estimated for all galaxy populations were both used to argue against ram pressure stripping
and galaxy harassment as primary quenching mechanisms. 
Galaxies appear to make their passage through the group/cluster essentially unperturbed and
the inception of the quenching event appears to occur in the more rarefied regions of the group/cluster environment. These results are, rather, consistent 
with a scenario of galaxies merging in the outskirts of clusters and groups prior to their second pericentric passage at least 
for galaxies $\log(M_{\ast}/M_{\odot}) > 10.45$. It is possible that ram pressure stripping and/or cluster tidal effects aid the quenching process as the 
merger remnant returns to pericenter, stripping any remaining gas ejected to large galacto-centric distances. It appears, however, that these mechanisms 
cannot solely be responsible for the observed behavior.   

\end{itemize}

While the redshift baseline of this study is large and extends to a lookback time of $\sim$9 Gyr, it appears that the processes that serve to quench galaxies
in higher density environments are still broadly efficient for the bulk of the massive galaxy population. However, the mere existence of a large population of 
relatively massive quiescent galaxies in the ORELSE cluster/groups across all redshifts studied here, their relative scarcity at higher redshift,
and the presence of extremely massive star-forming galaxies in high-density environments at higher redshift (e.g., \citealt{lem14a, lem18, darvish16, taowang16, 
miettinen17, vernesa17}), strongly argues that the efficiency of environmental quenching must subside quickly, and perhaps even reverse its effects, in 
forming groups and clusters at earlier epochs ($z\ga1.4$). Studies of such forming structures will allow for investigations on the progenitor 
galaxies of those massive galaxies observed to be pervasively quenched at $z<1.4$. Making statistical connections between these forming structures and more established 
structures at lower redshift, as well as their constituent galaxy populations, will allow for a comprehensive view of the average life of galaxies in dense environments 
from inception until quiescence.  

\section*{Acknowledgements}

{\footnotesize 
 This material is based upon work supported by the National Science Foundation under Grant No. 1411943. Part of the work 
presented herein is supported by NASA Grant Number NNX15AK92G. This work was additionally supported by the France-Berkeley Fund, a 
joint venture between UC Berkeley, UC Davis, and le Centre National de la Recherche Scientifique de France promoting lasting institutional
and intellectual cooperation between France and the United States. PFW acknowledges funding through the H2020 ERC Consolidator Grant 683184 and 
the support of an EACOA Fellowship from the East Asian Core Observatories Association. 
BCL gratefully acknowledges Julie Nantais, Susmita Adhikari, and Olga Cucciati for discussions 
that helped improve the paper. This paper is dedicated to Shari and Gary Goodenow, whose generosity and humor had a deep effect on the formative years of the 
first author, and whose influence and love is still profoundly felt. This study is based, in part, on data collected 
at the Subaru Telescope and obtained from the SMOKA, which is operated by the Astronomy Data Center, National Astronomical Observatory of 
Japan. This work is based, in part, on observations made with the Spitzer Space Telescope, which is operated by the Jet Propulsion Laboratory, California 
Institute of Technology under a contract with NASA. UKIRT is supported by NASA and operated under an agreement among the University of Hawaii, the 
University of Arizona, and Lockheed Martin Advanced Technology Center; operations are enabled through the cooperation of the East Asian Observatory.  
When the data reported here were acquired, UKIRT was operated by the Joint Astronomy Centre on behalf of the Science and Technology Facilities Council 
of the U.K. This study is also based, in part, on observations obtained with WIRCam, a joint project of CFHT, Taiwan, Korea, Canada, France, and the 
Canada-France-Hawaii Telescope which is operated by the National Research Council (NRC) of Canada, the Institut National des Sciences de l'Univers 
of the Centre National de la Recherche Scientifique of France, and the University of Hawai'i. Some portion of the spectrographic data presented herein was based on observations
obtained with the European Southern Observatory Very Large Telescope, Paranal, Chile, under Large Programs 070.A-9007 and 177.A-0837. The remainder of the spectrographic 
data presented herein were obtained at the W.M. Keck Observatory, which is operated as a scientific partnership among the California Institute of Technology, the University of 
California, and the National Aeronautics and Space Administration. The Observatory was made possible by the generous financial support of the W.M. Keck 
Foundation. We thank the indigenous Hawaiian community for allowing us to be guests on their sacred mountain, a privilege, without which, this 
work would not have been possible. We are most fortunate to be able to conduct observations from this site.}





\bibliographystyle{mnras}
\bibliography{quiescent} 

\begin{thebibliography}{}
\makeatletter
\relax
\def\mn@urlcharsother{\let\do\@makeother \do\$\do\&\do\#\do\^\do\_\do\%\do\~}
\def\mn@doi{\begingroup\mn@urlcharsother \@ifnextchar [ {\mn@doi@}
  {\mn@doi@[]}}
\def\mn@doi@[#1]#2{\def\@tempa{#1}\ifx\@tempa\@empty \href
  {http://dx.doi.org/#2} {doi:#2}\else \href {http://dx.doi.org/#2} {#1}\fi
  \endgroup}
\def\mn@eprint#1#2{\mn@eprint@#1:#2::\@nil}
\def\mn@eprint@arXiv#1{\href {http://arxiv.org/abs/#1} {{\tt arXiv:#1}}}
\def\mn@eprint@dblp#1{\href {http://dblp.uni-trier.de/rec/bibtex/#1.xml}
  {dblp:#1}}
\def\mn@eprint@#1:#2:#3:#4\@nil{\def\@tempa {#1}\def\@tempb {#2}\def\@tempc
  {#3}\ifx \@tempc \@empty \let \@tempc \@tempb \let \@tempb \@tempa \fi \ifx
  \@tempb \@empty \def\@tempb {arXiv}\fi \@ifundefined
  {mn@eprint@\@tempb}{\@tempb:\@tempc}{\expandafter \expandafter \csname
  mn@eprint@\@tempb\endcsname \expandafter{\@tempc}}}

\bibitem[\protect\citeauthoryear{{Alberts} et~al.,}{{Alberts}
  et~al.}{2014}]{alberts14}
{Alberts} S.,  et~al., 2014, \mn@doi [\mnras] {10.1093/mnras/stt1897}, \href
  {https://ui.adsabs.harvard.edu/abs/2014MNRAS.437..437A} {437, 437}

\bibitem[\protect\citeauthoryear{{Alberts} et~al.,}{{Alberts}
  et~al.}{2016}]{alberts16}
{Alberts} S.,  et~al., 2016, \mn@doi [\apj] {10.3847/0004-637X/825/1/72}, \href
  {http://adsabs.harvard.edu/abs/2016ApJ...825...72A} {825, 72}

\bibitem[\protect\citeauthoryear{{Arnouts} et~al.,}{{Arnouts}
  et~al.}{2007}]{animatedarnouts07}
{Arnouts} S.,  et~al., 2007, \mn@doi [\aap] {10.1051/0004-6361:20077632}, \href
  {http://adsabs.harvard.edu/abs/2007A%26A...476..137A} {476, 137}

\bibitem[\protect\citeauthoryear{{Arnouts} et~al.,}{{Arnouts}
  et~al.}{2013}]{animatedarnouts13}
{Arnouts} S.,  et~al., 2013, \mn@doi [\aap] {10.1051/0004-6361/201321768},
  \href {http://adsabs.harvard.edu/abs/2013A%26A...558A..67A} {558, A67}

\bibitem[\protect\citeauthoryear{{Ascaso}, {Lemaux}, {Lubin}, {Gal},
  {Kocevski}, {Rumbaugh}  \& {Squires}}{{Ascaso} et~al.}{2014}]{begona14}
{Ascaso} B.,  {Lemaux} B.~C.,  {Lubin} L.~M.,  {Gal} R.~R.,  {Kocevski} D.~D.,
  {Rumbaugh} N.,   {Squires} G.,  2014, \mn@doi [\mnras]
  {10.1093/mnras/stu877}, \href
  {http://adsabs.harvard.edu/abs/2014MNRAS.442..589A} {442, 589}

\bibitem[\protect\citeauthoryear{{Bah{\'e}} \& {McCarthy}}{{Bah{\'e}} \&
  {McCarthy}}{2015}]{bahe15}
{Bah{\'e}} Y.~M.,  {McCarthy} I.~G.,  2015, \mn@doi [\mnras]
  {10.1093/mnras/stu2293}, \href
  {http://adsabs.harvard.edu/abs/2015MNRAS.447..969B} {447, 969}

\bibitem[\protect\citeauthoryear{{Balogh} et~al.,}{{Balogh}
  et~al.}{2011}]{balogh11}
{Balogh} M.~L.,  et~al., 2011, \mn@doi [\mnras]
  {10.1111/j.1365-2966.2010.18052.x}, \href
  {http://adsabs.harvard.edu/abs/2011MNRAS.412.2303B} {412, 2303}

\bibitem[\protect\citeauthoryear{{Balogh} et~al.,}{{Balogh}
  et~al.}{2014}]{balogh14}
{Balogh} M.~L.,  et~al., 2014, \mn@doi [\mnras] {10.1093/mnras/stu1332}, \href
  {http://adsabs.harvard.edu/abs/2014MNRAS.443.2679B} {443, 2679}

\bibitem[\protect\citeauthoryear{{Balogh} et~al.,}{{Balogh}
  et~al.}{2016}]{balogh16}
{Balogh} M.~L.,  et~al., 2016, \mn@doi [\mnras] {10.1093/mnras/stv2949}, \href
  {http://adsabs.harvard.edu/abs/2016MNRAS.456.4364B} {456, 4364}

\bibitem[\protect\citeauthoryear{{Baxter} et~al.,}{{Baxter}
  et~al.}{2017}]{baxter17}
{Baxter} E.,  et~al., 2017, \mn@doi [\apj] {10.3847/1538-4357/aa6ff0}, \href
  {http://adsabs.harvard.edu/abs/2017ApJ...841...18B} {841, 18}

\bibitem[\protect\citeauthoryear{{Becker}, {White}  \& {Helfand}}{{Becker}
  et~al.}{1995}]{thebob95}
{Becker} R.~H.,  {White} R.~L.,   {Helfand} D.~J.,  1995, \mn@doi [\apj]
  {10.1086/176166}, \href {http://adsabs.harvard.edu/abs/1995ApJ...450..559B}
  {450, 559}

\bibitem[\protect\citeauthoryear{{Belfiore} et~al.,}{{Belfiore}
  et~al.}{2016}]{belfiore16}
{Belfiore} F.,  et~al., 2016, \mn@doi [\mnras] {10.1093/mnras/stw1234}, \href
  {http://adsabs.harvard.edu/abs/2016MNRAS.461.3111B} {461, 3111}

\bibitem[\protect\citeauthoryear{{Belfiore} et~al.,}{{Belfiore}
  et~al.}{2017}]{belfiore17}
{Belfiore} F.,  et~al., 2017, \mn@doi [\mnras] {10.1093/mnras/stw3211}, \href
  {http://adsabs.harvard.edu/abs/2017MNRAS.466.2570B} {466, 2570}

\bibitem[\protect\citeauthoryear{{Bell} et~al.,}{{Bell} et~al.}{2012}]{bell12}
{Bell} E.~F.,  et~al., 2012, \mn@doi [\apj] {10.1088/0004-637X/753/2/167},
  \href {http://adsabs.harvard.edu/abs/2012ApJ...753..167B} {753, 167}

\bibitem[\protect\citeauthoryear{{Bertin} \& {Arnouts}}{{Bertin} \&
  {Arnouts}}{1996}]{BertinArn96}
{Bertin} E.,  {Arnouts} S.,  1996, \mn@doi [\aaps] {10.1051/aas:1996164}, \href
  {http://adsabs.harvard.edu/abs/1996A%26AS..117..393B} {117, 393}

\bibitem[\protect\citeauthoryear{{Birnboim} \& {Dekel}}{{Birnboim} \&
  {Dekel}}{2003}]{birnboim03}
{Birnboim} Y.,  {Dekel} A.,  2003, \mn@doi [\mnras]
  {10.1046/j.1365-8711.2003.06955.x}, \href
  {http://adsabs.harvard.edu/abs/2003MNRAS.345..349B} {345, 349}

\bibitem[\protect\citeauthoryear{{Blanton}, {Gregg}, {Helfand}, {Becker}  \&
  {White}}{{Blanton} et~al.}{2003}]{blanton03}
{Blanton} E.~L.,  {Gregg} M.~D.,  {Helfand} D.~J.,  {Becker} R.~H.,   {White}
  R.~L.,  2003, \mn@doi [\aj] {10.1086/368140}, \href
  {http://adsabs.harvard.edu/abs/2003AJ....125.1635B} {125, 1635}

\bibitem[\protect\citeauthoryear{{Bongiorno} et~al.,}{{Bongiorno}
  et~al.}{2016}]{bongiorno16}
{Bongiorno} A.,  et~al., 2016, \mn@doi [\aap] {10.1051/0004-6361/201527436},
  \href {http://adsabs.harvard.edu/abs/2016A%26A...588A..78B} {588, A78}

\bibitem[\protect\citeauthoryear{{Boulade} et~al.,}{{Boulade}
  et~al.}{2003}]{boulade03}
{Boulade} O.,  et~al., 2003, in {Iye} M.,  {Moorwood} A.~F.~M.,  eds,
  \procspie Vol. 4841, Instrument Design and Performance for Optical/Infrared
  Ground-based Telescopes. pp 72--81, \mn@doi{10.1117/12.459890}

\bibitem[\protect\citeauthoryear{{Brammer}, {van Dokkum}  \& {Coppi}}{{Brammer}
  et~al.}{2008}]{brammer08}
{Brammer} G.~B.,  {van Dokkum} P.~G.,   {Coppi} P.,  2008, \mn@doi [\apj]
  {10.1086/591786}, \href {http://adsabs.harvard.edu/abs/2008ApJ...686.1503B}
  {686, 1503}

\bibitem[\protect\citeauthoryear{{Brammer} et~al.,}{{Brammer}
  et~al.}{2011}]{brammer11}
{Brammer} G.~B.,  et~al., 2011, \mn@doi [\apj] {10.1088/0004-637X/739/1/24},
  \href {http://adsabs.harvard.edu/abs/2011ApJ...739...24B} {739, 24}

\bibitem[\protect\citeauthoryear{{Brammer} et~al.,}{{Brammer}
  et~al.}{2012}]{brammer12}
{Brammer} G.~B.,  et~al., 2012, \mn@doi [\apjs] {10.1088/0067-0049/200/2/13},
  \href {http://adsabs.harvard.edu/abs/2012ApJS..200...13B} {200, 13}

\bibitem[\protect\citeauthoryear{{Brodwin} et~al.,}{{Brodwin}
  et~al.}{2013}]{mbrodz13}
{Brodwin} M.,  et~al., 2013, \mn@doi [\apj] {10.1088/0004-637X/779/2/138},
  \href {https://ui.adsabs.harvard.edu/abs/2013ApJ...779..138B} {779, 138}

\bibitem[\protect\citeauthoryear{{Bruzual} \& {Charlot}}{{Bruzual} \&
  {Charlot}}{2003}]{bc03}
{Bruzual} G.,  {Charlot} S.,  2003, \mn@doi [\mnras]
  {10.1046/j.1365-8711.2003.06897.x}, \href
  {http://adsabs.harvard.edu/abs/2003MNRAS.344.1000B} {344, 1000}

\bibitem[\protect\citeauthoryear{{Bundy}, {Fukugita}, {Ellis}, {Targett},
  {Belli}  \& {Kodama}}{{Bundy} et~al.}{2009}]{bundy09}
{Bundy} K.,  {Fukugita} M.,  {Ellis} R.~S.,  {Targett} T.~A.,  {Belli} S.,
  {Kodama} T.,  2009, \mn@doi [\apj] {10.1088/0004-637X/697/2/1369}, \href
  {http://adsabs.harvard.edu/abs/2009ApJ...697.1369B} {697, 1369}

\bibitem[\protect\citeauthoryear{{Calzetti}, {Armus}, {Bohlin}, {Kinney},
  {Koornneef}  \& {Storchi-Bergmann}}{{Calzetti} et~al.}{2000}]{calz00}
{Calzetti} D.,  {Armus} L.,  {Bohlin} R.~C.,  {Kinney} A.~L.,  {Koornneef} J.,
   {Storchi-Bergmann} T.,  2000, \mn@doi [\apj] {10.1086/308692}, \href
  {http://adsabs.harvard.edu/abs/2000ApJ...533..682C} {533, 682}

\bibitem[\protect\citeauthoryear{{Casali} et~al.,}{{Casali}
  et~al.}{2007}]{casali07}
{Casali} M.,  et~al., 2007, \mn@doi [\aap] {10.1051/0004-6361:20066514}, \href
  {http://adsabs.harvard.edu/abs/2007A%26A...467..777C} {467, 777}

\bibitem[\protect\citeauthoryear{{Chabrier}}{{Chabrier}}{2003}]{chab03}
{Chabrier} G.,  2003, \mn@doi [\pasp] {10.1086/376392}, \href
  {http://adsabs.harvard.edu/abs/2003PASP..115..763C} {115, 763}

\bibitem[\protect\citeauthoryear{{Chang} et~al.,}{{Chang}
  et~al.}{2018}]{chang18}
{Chang} C.,  et~al., 2018, \mn@doi [\apj] {10.3847/1538-4357/aad5e7}, \href
  {https://ui.adsabs.harvard.edu/abs/2018ApJ...864...83C} {864, 83}

\bibitem[\protect\citeauthoryear{{Clerc} et~al.,}{{Clerc}
  et~al.}{2014}]{clerc14}
{Clerc} N.,  et~al., 2014, \mn@doi [\mnras] {10.1093/mnras/stu1625}, \href
  {http://adsabs.harvard.edu/abs/2014MNRAS.444.2723C} {444, 2723}

\bibitem[\protect\citeauthoryear{{Coil}, {Mendez}, {Eisenstein}  \&
  {Moustakas}}{{Coil} et~al.}{2017}]{coil17}
{Coil} A.~L.,  {Mendez} A.~J.,  {Eisenstein} D.~J.,   {Moustakas} J.,  2017,
  \mn@doi [\apj] {10.3847/1538-4357/aa63ec}, \href
  {http://adsabs.harvard.edu/abs/2017ApJ...838...87C} {838, 87}

\bibitem[\protect\citeauthoryear{{Coogan} et~al.,}{{Coogan}
  et~al.}{2018}]{coogan18}
{Coogan} R.~T.,  et~al., 2018, \mn@doi [\mnras] {10.1093/mnras/sty1446}, \href
  {https://ui.adsabs.harvard.edu/abs/2018MNRAS.479..703C} {479, 703}

\bibitem[\protect\citeauthoryear{{Cooke} et~al.,}{{Cooke}
  et~al.}{2016}]{cooke16}
{Cooke} E.~A.,  et~al., 2016, \mn@doi [\apj] {10.3847/0004-637X/816/2/83},
  \href {http://adsabs.harvard.edu/abs/2016ApJ...816...83C} {816, 83}

\bibitem[\protect\citeauthoryear{{Cooper} et~al.,}{{Cooper}
  et~al.}{2007}]{mcoopz07}
{Cooper} M.~C.,  et~al., 2007, \mn@doi [\mnras]
  {10.1111/j.1365-2966.2007.11534.x}, \href
  {http://adsabs.harvard.edu/abs/2007MNRAS.376.1445C} {376, 1445}

\bibitem[\protect\citeauthoryear{{Cooper} et~al.,}{{Cooper}
  et~al.}{2010}]{mcoopz10}
{Cooper} M.~C.,  et~al., 2010, \mn@doi [\mnras]
  {10.1111/j.1365-2966.2010.17312.x}, \href
  {http://adsabs.harvard.edu/abs/2010MNRAS.409..337C} {409, 337}

\bibitem[\protect\citeauthoryear{{Cucciati} et~al.,}{{Cucciati}
  et~al.}{2006}]{olga06}
{Cucciati} O.,  et~al., 2006, \mn@doi [\aap] {10.1051/0004-6361:20065161},
  \href {http://adsabs.harvard.edu/abs/2006A%26A...458...39C} {458, 39}

\bibitem[\protect\citeauthoryear{{Cucciati} et~al.,}{{Cucciati}
  et~al.}{2010}]{olga10}
{Cucciati} O.,  et~al., 2010, \mn@doi [\aap] {10.1051/0004-6361/200912585},
  \href {http://adsabs.harvard.edu/abs/2010A%26A...524A...2C} {524, A2}

\bibitem[\protect\citeauthoryear{{Cucciati} et~al.,}{{Cucciati}
  et~al.}{2017}]{olga17}
{Cucciati} O.,  et~al., 2017, \mn@doi [\aap] {10.1051/0004-6361/201630113},
  \href {https://ui.adsabs.harvard.edu/abs/2017A&A...602A..15C} {602, A15}

\bibitem[\protect\citeauthoryear{{Cucciati} et~al.,}{{Cucciati}
  et~al.}{2018}]{olga18}
{Cucciati} O.,  et~al., 2018, \mn@doi [\aap] {10.1051/0004-6361/201833655},
  \href {http://adsabs.harvard.edu/abs/2018A%26A...619A..49C} {619, A49}

\bibitem[\protect\citeauthoryear{{Darvish}, {Mobasher}, {Sobral}, {Rettura},
  {Scoville}, {Faisst}  \& {Capak}}{{Darvish} et~al.}{2016}]{darvish16}
{Darvish} B.,  {Mobasher} B.,  {Sobral} D.,  {Rettura} A.,  {Scoville} N.,
  {Faisst} A.,   {Capak} P.,  2016, \mn@doi [\apj]
  {10.3847/0004-637X/825/2/113}, \href
  {http://adsabs.harvard.edu/abs/2016ApJ...825..113D} {825, 113}

\bibitem[\protect\citeauthoryear{{Darvish}, {Scoville}, {Martin}, {Mobasher},
  {Diaz-Santos}  \& {Shen}}{{Darvish} et~al.}{2018}]{darvish18}
{Darvish} B.,  {Scoville} N.~Z.,  {Martin} C.,  {Mobasher} B.,  {Diaz-Santos}
  T.,   {Shen} L.,  2018, \mn@doi [\apj] {10.3847/1538-4357/aac836}, \href
  {http://adsabs.harvard.edu/abs/2018ApJ...860..111D} {860, 111}

\bibitem[\protect\citeauthoryear{{Davis} et~al.,}{{Davis}
  et~al.}{2003}]{davis03}
{Davis} M.,  et~al., 2003, in {Guhathakurta} P.,  ed.,  \procspie Vol. 4834,
  Discoveries and Research Prospects from 6- to 10-Meter-Class Telescopes II.
  pp 161--172 (\mn@eprint {} {astro-ph/0209419}), \mn@doi{10.1117/12.457897}

\bibitem[\protect\citeauthoryear{{Diamond-Stanic}, {Moustakas}, {Tremonti},
  {Coil}, {Hickox}, {Robaina}, {Rudnick}  \& {Sell}}{{Diamond-Stanic}
  et~al.}{2012}]{diamond-stanic12}
{Diamond-Stanic} A.~M.,  {Moustakas} J.,  {Tremonti} C.~A.,  {Coil} A.~L.,
  {Hickox} R.~C.,  {Robaina} A.~R.,  {Rudnick} G.~H.,   {Sell} P.~H.,  2012,
  \mn@doi [\apjl] {10.1088/2041-8205/755/2/L26}, \href
  {https://ui.adsabs.harvard.edu/abs/2012ApJ...755L..26D} {755, L26}

\bibitem[\protect\citeauthoryear{{Dressler}, {Smail}, {Poggianti}, {Butcher},
  {Couch}, {Ellis}  \& {Oemler}}{{Dressler} et~al.}{1999}]{dressler99}
{Dressler} A.,  {Smail} I.,  {Poggianti} B.~M.,  {Butcher} H.,  {Couch} W.~J.,
  {Ellis} R.~S.,   {Oemler} Jr. A.,  1999, \mn@doi [\apjs] {10.1086/313213},
  \href {http://adsabs.harvard.edu/abs/1999ApJS..122...51D} {122, 51}

\bibitem[\protect\citeauthoryear{{Dubois}, {Gavazzi}, {Peirani}  \&
  {Silk}}{{Dubois} et~al.}{2013}]{ofthewood13}
{Dubois} Y.,  {Gavazzi} R.,  {Peirani} S.,   {Silk} J.,  2013, \mn@doi [\mnras]
  {10.1093/mnras/stt997}, \href
  {http://adsabs.harvard.edu/abs/2013MNRAS.433.3297D} {433, 3297}

\bibitem[\protect\citeauthoryear{{Durkalec} et~al.,}{{Durkalec}
  et~al.}{2018}]{ania18}
{Durkalec} A.,  et~al., 2018, \mn@doi [\aap] {10.1051/0004-6361/201730734},
  \href {http://adsabs.harvard.edu/abs/2018A%26A...612A..42D} {612, A42}

\bibitem[\protect\citeauthoryear{{Ehlert} et~al.,}{{Ehlert}
  et~al.}{2015}]{ehlert15}
{Ehlert} S.,  et~al., 2015, \mn@doi [\mnras] {10.1093/mnras/stu2091}, \href
  {https://ui.adsabs.harvard.edu/abs/2015MNRAS.446.2709E} {446, 2709}

\bibitem[\protect\citeauthoryear{{Emami}, {Siana}, {Weisz}, {Johnson}, {Ma}  \&
  {El-Badry}}{{Emami} et~al.}{2019}]{emami19}
{Emami} N.,  {Siana} B.,  {Weisz} D.~R.,  {Johnson} B.~D.,  {Ma} X.,
  {El-Badry} K.,  2019, \mn@doi [\apj] {10.3847/1538-4357/ab211a}, \href
  {https://ui.adsabs.harvard.edu/abs/2019ApJ...881...71E} {881, 71}

\bibitem[\protect\citeauthoryear{{Faber} et~al.,}{{Faber} et~al.}{2003}]{fab03}
{Faber} S.~M.,  et~al., 2003, in {Iye} M.,  {Moorwood} A.~F.~M.,  eds,  Society
  of Photo-Optical Instrumentation Engineers (SPIE) Conference Series Vol.
  4841, Instrument Design and Performance for Optical/Infrared Ground-based
  Telescopes. pp 1657--1669, \mn@doi{10.1117/12.460346}

\bibitem[\protect\citeauthoryear{{Fanaroff} \& {Riley}}{{Fanaroff} \&
  {Riley}}{1974}]{FR74}
{Fanaroff} B.~L.,  {Riley} J.~M.,  1974, \mn@doi [\mnras]
  {10.1093/mnras/167.1.31P}, \href
  {http://adsabs.harvard.edu/abs/1974MNRAS.167P..31F} {167, 31P}

\bibitem[\protect\citeauthoryear{{Fazio} et~al.,}{{Fazio}
  et~al.}{2004}]{fazio04}
{Fazio} G.~G.,  et~al., 2004, \mn@doi [\apjs] {10.1086/422843}, \href
  {http://adsabs.harvard.edu/abs/2004ApJS..154...10F} {154, 10}

\bibitem[\protect\citeauthoryear{{Fillingham}, {Cooper}, {Pace},
  {Boylan-Kolchin}, {Bullock}, {Garrison-Kimmel}  \& {Wheeler}}{{Fillingham}
  et~al.}{2016}]{fillingham16}
{Fillingham} S.~P.,  {Cooper} M.~C.,  {Pace} A.~B.,  {Boylan-Kolchin} M.,
  {Bullock} J.~S.,  {Garrison-Kimmel} S.,   {Wheeler} C.,  2016, \mn@doi
  [\mnras] {10.1093/mnras/stw2131}, \href
  {http://adsabs.harvard.edu/abs/2016MNRAS.463.1916F} {463, 1916}

\bibitem[\protect\citeauthoryear{{Foltz} et~al.,}{{Foltz}
  et~al.}{2018}]{foltz18}
{Foltz} R.,  et~al., 2018, \mn@doi [\apj] {10.3847/1538-4357/aad80d}, \href
  {https://ui.adsabs.harvard.edu/abs/2018ApJ...866..136F} {866, 136}

\bibitem[\protect\citeauthoryear{{Fossati} et~al.,}{{Fossati}
  et~al.}{2017}]{fossati17}
{Fossati} M.,  et~al., 2017, \mn@doi [\apj] {10.3847/1538-4357/835/2/153},
  \href {http://adsabs.harvard.edu/abs/2017ApJ...835..153F} {835, 153}

\bibitem[\protect\citeauthoryear{{Fukugita}, {Ichikawa}, {Gunn}, {Doi},
  {Shimasaku}  \& {Schneider}}{{Fukugita} et~al.}{1996}]{fukugita96}
{Fukugita} M.,  {Ichikawa} T.,  {Gunn} J.~E.,  {Doi} M.,  {Shimasaku} K.,
  {Schneider} D.~P.,  1996, \mn@doi [\aj] {10.1086/117915}, \href
  {http://adsabs.harvard.edu/abs/1996AJ....111.1748F} {111, 1748}

\bibitem[\protect\citeauthoryear{{Gal} \& {Lubin}}{{Gal} \&
  {Lubin}}{2004}]{galnlub04}
{Gal} R.~R.,  {Lubin} L.~M.,  2004, \mn@doi [\apjl] {10.1086/421463}, \href
  {http://adsabs.harvard.edu/abs/2004ApJ...607L...1G} {607, L1}

\bibitem[\protect\citeauthoryear{{Gal}, {Lemaux}, {Lubin}, {Kocevski}  \&
  {Squires}}{{Gal} et~al.}{2008}]{gal08}
{Gal} R.~R.,  {Lemaux} B.~C.,  {Lubin} L.~M.,  {Kocevski} D.,   {Squires}
  G.~K.,  2008, \mn@doi [\apj] {10.1086/590416}, \href
  {http://adsabs.harvard.edu/abs/2008ApJ...684..933G} {684, 933}

\bibitem[\protect\citeauthoryear{{Galametz} et~al.,}{{Galametz}
  et~al.}{2009}]{galametz09}
{Galametz} A.,  et~al., 2009, \mn@doi [\apj] {10.1088/0004-637X/694/2/1309},
  \href {https://ui.adsabs.harvard.edu/abs/2009ApJ...694.1309G} {694, 1309}

\bibitem[\protect\citeauthoryear{{Garilli} et~al.,}{{Garilli}
  et~al.}{2014}]{bianca14}
{Garilli} B.,  et~al., 2014, \mn@doi [\aap] {10.1051/0004-6361/201322790},
  \href {http://adsabs.harvard.edu/abs/2014A%26A...562A..23G} {562, A23}

\bibitem[\protect\citeauthoryear{{Genzel} et~al.,}{{Genzel}
  et~al.}{2008}]{genzel08}
{Genzel} R.,  et~al., 2008, \mn@doi [\apj] {10.1086/591840}, \href
  {http://adsabs.harvard.edu/abs/2008ApJ...687...59G} {687, 59}

\bibitem[\protect\citeauthoryear{{Genzel} et~al.,}{{Genzel}
  et~al.}{2014}]{genzel14}
{Genzel} R.,  et~al., 2014, \mn@doi [\apj] {10.1088/0004-637X/785/1/75}, \href
  {http://adsabs.harvard.edu/abs/2014ApJ...785...75G} {785, 75}

\bibitem[\protect\citeauthoryear{{Gioia}, {Wolter}, {Mullis}, {Henry},
  {B{\"o}hringer}  \& {Briel}}{{Gioia} et~al.}{2004}]{gioia04}
{Gioia} I.~M.,  {Wolter} A.,  {Mullis} C.~R.,  {Henry} J.~P.,  {B{\"o}hringer}
  H.,   {Briel} U.~G.,  2004, \mn@doi [\aap] {10.1051/0004-6361:20041426},
  \href {http://adsabs.harvard.edu/abs/2004A%26A...428..867G} {428, 867}

\bibitem[\protect\citeauthoryear{{Gunn} \& {Gott}}{{Gunn} \&
  {Gott}}{1972}]{gunn72}
{Gunn} J.~E.,  {Gott} III J.~R.,  1972, \mn@doi [\apj] {10.1086/151605}, \href
  {http://adsabs.harvard.edu/abs/1972ApJ...176....1G} {176, 1}

\bibitem[\protect\citeauthoryear{{Guzzo} et~al.,}{{Guzzo}
  et~al.}{2014}]{guzzo14}
{Guzzo} L.,  et~al., 2014, \mn@doi [\aap] {10.1051/0004-6361/201321489}, \href
  {http://adsabs.harvard.edu/abs/2014A%26A...566A.108G} {566, A108}

\bibitem[\protect\citeauthoryear{{Harker}, {Cole}, {Helly}, {Frenk}  \&
  {Jenkins}}{{Harker} et~al.}{2006}]{harker06}
{Harker} G.,  {Cole} S.,  {Helly} J.,  {Frenk} C.,   {Jenkins} A.,  2006,
  \mn@doi [\mnras] {10.1111/j.1365-2966.2006.10022.x}, \href
  {http://adsabs.harvard.edu/abs/2006MNRAS.367.1039H} {367, 1039}

\bibitem[\protect\citeauthoryear{{Helly}, {Cole}, {Frenk}, {Baugh}, {Benson}
  \& {Lacey}}{{Helly} et~al.}{2003}]{helly03}
{Helly} J.~C.,  {Cole} S.,  {Frenk} C.~S.,  {Baugh} C.~M.,  {Benson} A.,
  {Lacey} C.,  2003, \mn@doi [\mnras] {10.1046/j.1365-8711.2003.06151.x}, \href
  {http://adsabs.harvard.edu/abs/2003MNRAS.338..903H} {338, 903}

\bibitem[\protect\citeauthoryear{{Hoaglin}, {Mosteller}  \& {Tukey}}{{Hoaglin}
  et~al.}{1983}]{hoaglin83}
{Hoaglin} D.~C.,  {Mosteller} F.,   {Tukey} J.~W.,  1983, {Understanding robust
  and exploratory data anlysis}

\bibitem[\protect\citeauthoryear{{Hopkins}, {Cox}, {Hernquist}, {Narayanan},
  {Hayward}  \& {Murray}}{{Hopkins} et~al.}{2013}]{hopkins13}
{Hopkins} P.~F.,  {Cox} T.~J.,  {Hernquist} L.,  {Narayanan} D.,  {Hayward}
  C.~C.,   {Murray} N.,  2013, \mn@doi [\mnras] {10.1093/mnras/stt017}, \href
  {http://adsabs.harvard.edu/abs/2013MNRAS.430.1901H} {430, 1901}

\bibitem[\protect\citeauthoryear{{Hopkins}, {Torrey}, {Faucher-Gigu{\`e}re},
  {Quataert}  \& {Murray}}{{Hopkins} et~al.}{2016}]{hopkins16}
{Hopkins} P.~F.,  {Torrey} P.,  {Faucher-Gigu{\`e}re} C.-A.,  {Quataert} E.,
  {Murray} N.,  2016, \mn@doi [\mnras] {10.1093/mnras/stw289}, \href
  {http://adsabs.harvard.edu/abs/2016MNRAS.458..816H} {458, 816}

\bibitem[\protect\citeauthoryear{{Horne}}{{Horne}}{1986}]{horne86}
{Horne} K.,  1986, \mn@doi [\pasp] {10.1086/131801}, \href
  {http://adsabs.harvard.edu/abs/1986PASP...98..609H} {98, 609}

\bibitem[\protect\citeauthoryear{{Hung} et~al.,}{{Hung} et~al.}{2018}]{hung18}
{Hung} C.-L.,  et~al., 2018, \mn@doi [\mnras] {10.1093/mnras/sty2970}, \href
  {http://adsabs.harvard.edu/abs/2018MNRAS.tmp.2832H} {}

\bibitem[\protect\citeauthoryear{{Hung} et~al.,}{{Hung}
  et~al.}{2019}]{denise19}
{Hung} D.,  et~al., 2019, arXiv e-prints, \href
  {https://ui.adsabs.harvard.edu/abs/2019arXiv190509298H} {p. arXiv:1905.09298}

\bibitem[\protect\citeauthoryear{{Ilbert} et~al.,}{{Ilbert}
  et~al.}{2006}]{ilbert06}
{Ilbert} O.,  et~al., 2006, \mn@doi [\aap] {10.1051/0004-6361:20065138}, \href
  {http://adsabs.harvard.edu/abs/2006A%26A...457..841I} {457, 841}

\bibitem[\protect\citeauthoryear{{Ilbert} et~al.,}{{Ilbert}
  et~al.}{2010}]{ilbert10}
{Ilbert} O.,  et~al., 2010, \mn@doi [\apj] {10.1088/0004-637X/709/2/644}, \href
  {http://adsabs.harvard.edu/abs/2010ApJ...709..644I} {709, 644}

\bibitem[\protect\citeauthoryear{{Ilbert} et~al.,}{{Ilbert}
  et~al.}{2013}]{ilbert13}
{Ilbert} O.,  et~al., 2013, \mn@doi [\aap] {10.1051/0004-6361/201321100}, \href
  {http://adsabs.harvard.edu/abs/2013A%26A...556A..55I} {556, A55}

\bibitem[\protect\citeauthoryear{{Jeltema} et~al.,}{{Jeltema}
  et~al.}{2009}]{jeltema09}
{Jeltema} T.~E.,  et~al., 2009, \mn@doi [\mnras]
  {10.1111/j.1365-2966.2009.15377.x}, \href
  {http://adsabs.harvard.edu/abs/2009MNRAS.399..715J} {399, 715}

\bibitem[\protect\citeauthoryear{{Jian} et~al.,}{{Jian} et~al.}{2018}]{jian18}
{Jian} H.-Y.,  et~al., 2018, \mn@doi [\pasj] {10.1093/pasj/psx096}, \href
  {http://adsabs.harvard.edu/abs/2018PASJ...70S..23J} {70, S23}

\bibitem[\protect\citeauthoryear{{Juneau} et~al.,}{{Juneau}
  et~al.}{2013}]{juneau13}
{Juneau} S.,  et~al., 2013, \mn@doi [\apj] {10.1088/0004-637X/764/2/176}, \href
  {http://adsabs.harvard.edu/abs/2013ApJ...764..176J} {764, 176}

\bibitem[\protect\citeauthoryear{{Kannappan}, {Guie}  \& {Baker}}{{Kannappan}
  et~al.}{2009}]{krazykannappan09}
{Kannappan} S.~J.,  {Guie} J.~M.,   {Baker} A.~J.,  2009, \mn@doi [\aj]
  {10.1088/0004-6256/138/2/579}, \href
  {http://adsabs.harvard.edu/abs/2009AJ....138..579K} {138, 579}

\bibitem[\protect\citeauthoryear{{Kannappan} et~al.,}{{Kannappan}
  et~al.}{2013}]{krazykannappan13}
{Kannappan} S.~J.,  et~al., 2013, \mn@doi [\apj] {10.1088/0004-637X/777/1/42},
  \href {http://adsabs.harvard.edu/abs/2013ApJ...777...42K} {777, 42}

\bibitem[\protect\citeauthoryear{{Kauffmann}}{{Kauffmann}}{2014}]{kauffmann14}
{Kauffmann} G.,  2014, \mn@doi [\mnras] {10.1093/mnras/stu752}, \href
  {http://adsabs.harvard.edu/abs/2014MNRAS.441.2717K} {441, 2717}

\bibitem[\protect\citeauthoryear{{Kauffmann} et~al.,}{{Kauffmann}
  et~al.}{2003}]{kauff03}
{Kauffmann} G.,  et~al., 2003, \mn@doi [\mnras]
  {10.1111/j.1365-2966.2003.07154.x}, \href
  {http://adsabs.harvard.edu/abs/2003MNRAS.346.1055K} {346, 1055}

\bibitem[\protect\citeauthoryear{{Kaviraj}}{{Kaviraj}}{2014}]{kaviraj14}
{Kaviraj} S.,  2014, \mn@doi [\mnras] {10.1093/mnras/stu338}, \href
  {http://adsabs.harvard.edu/abs/2014MNRAS.440.2944K} {440, 2944}

\bibitem[\protect\citeauthoryear{{Kawinwanichakij} et~al.,}{{Kawinwanichakij}
  et~al.}{2016}]{kawinwanichakij16}
{Kawinwanichakij} L.,  et~al., 2016, \mn@doi [\apj]
  {10.3847/0004-637X/817/1/9}, \href
  {http://adsabs.harvard.edu/abs/2016ApJ...817....9K} {817, 9}

\bibitem[\protect\citeauthoryear{{Kawinwanichakij} et~al.,}{{Kawinwanichakij}
  et~al.}{2017}]{kawinwanichakij17}
{Kawinwanichakij} L.,  et~al., 2017, \mn@doi [\apj] {10.3847/1538-4357/aa8b75},
  \href {http://adsabs.harvard.edu/abs/2017ApJ...847..134K} {847, 134}

\bibitem[\protect\citeauthoryear{{Kere{\v s}}, {Katz}, {Weinberg}  \&
  {Dav{\'e}}}{{Kere{\v s}} et~al.}{2005}]{keres05}
{Kere{\v s}} D.,  {Katz} N.,  {Weinberg} D.~H.,   {Dav{\'e}} R.,  2005, \mn@doi
  [\mnras] {10.1111/j.1365-2966.2005.09451.x}, \href
  {http://adsabs.harvard.edu/abs/2005MNRAS.363....2K} {363, 2}

\bibitem[\protect\citeauthoryear{{Kocevski}, {Lubin}, {Lemaux}, {Gal},
  {Fassnacht}, {Lin}  \& {Squires}}{{Kocevski} et~al.}{2009}]{dirtydale09}
{Kocevski} D.~D.,  {Lubin} L.~M.,  {Lemaux} B.~C.,  {Gal} R.~R.,  {Fassnacht}
  C.~D.,  {Lin} R.,   {Squires} G.~K.,  2009, \mn@doi [\apj]
  {10.1088/0004-637X/700/2/901}, \href
  {http://adsabs.harvard.edu/abs/2009ApJ...700..901K} {700, 901}

\bibitem[\protect\citeauthoryear{{Kocevski} et~al.,}{{Kocevski}
  et~al.}{2011}]{dirtydale11b}
{Kocevski} D.~D.,  et~al., 2011, \mn@doi [\apj] {10.1088/0004-637X/736/1/38},
  \href {http://adsabs.harvard.edu/abs/2011ApJ...736...38K} {736, 38}

\bibitem[\protect\citeauthoryear{{Kova{\v c}} et~al.,}{{Kova{\v c}}
  et~al.}{2010}]{kovac10}
{Kova{\v c}} K.,  et~al., 2010, \mn@doi [\apj] {10.1088/0004-637X/718/1/86},
  \href {http://adsabs.harvard.edu/abs/2010ApJ...718...86K} {718, 86}

\bibitem[\protect\citeauthoryear{{Kova{\v c}} et~al.,}{{Kova{\v c}}
  et~al.}{2014}]{kovac14}
{Kova{\v c}} K.,  et~al., 2014, \mn@doi [\mnras] {10.1093/mnras/stt2241}, \href
  {http://adsabs.harvard.edu/abs/2014MNRAS.438..717K} {438, 717}

\bibitem[\protect\citeauthoryear{{Koyama} et~al.,}{{Koyama}
  et~al.}{2013}]{koyama13}
{Koyama} Y.,  et~al., 2013, \mn@doi [\mnras] {10.1093/mnras/stt1035}, \href
  {http://adsabs.harvard.edu/abs/2013MNRAS.434..423K} {434, 423}

\bibitem[\protect\citeauthoryear{{Kraljic} et~al.,}{{Kraljic}
  et~al.}{2018}]{katarina18}
{Kraljic} K.,  et~al., 2018, \mn@doi [\mnras] {10.1093/mnras/stx2638}, \href
  {http://adsabs.harvard.edu/abs/2018MNRAS.474..547K} {474, 547}

\bibitem[\protect\citeauthoryear{{Kriek}, {van Dokkum}, {Labb{\'e}}, {Franx},
  {Illingworth}, {Marchesini}  \& {Quadri}}{{Kriek} et~al.}{2009}]{kriek09}
{Kriek} M.,  {van Dokkum} P.~G.,  {Labb{\'e}} I.,  {Franx} M.,  {Illingworth}
  G.~D.,  {Marchesini} D.,   {Quadri} R.~F.,  2009, \mn@doi [\apj]
  {10.1088/0004-637X/700/1/221}, \href
  {http://adsabs.harvard.edu/abs/2009ApJ...700..221K} {700, 221}

\bibitem[\protect\citeauthoryear{{Kronberger}, {Kapferer}, {Ferrari},
  {Unterguggenberger}  \& {Schindler}}{{Kronberger}
  et~al.}{2008}]{kronberger08}
{Kronberger} T.,  {Kapferer} W.,  {Ferrari} C.,  {Unterguggenberger} S.,
  {Schindler} S.,  2008, \mn@doi [\aap] {10.1051/0004-6361:20078904}, \href
  {http://adsabs.harvard.edu/abs/2008A%26A...481..337K} {481, 337}

\bibitem[\protect\citeauthoryear{{Kuutma}, {Tamm}  \& {Tempel}}{{Kuutma}
  et~al.}{2017}]{kuutma17}
{Kuutma} T.,  {Tamm} A.,   {Tempel} E.,  2017, \mn@doi [\aap]
  {10.1051/0004-6361/201730526}, \href
  {http://adsabs.harvard.edu/abs/2017A%26A...600L...6K} {600, L6}

\bibitem[\protect\citeauthoryear{{Laigle} et~al.,}{{Laigle}
  et~al.}{2016}]{laigle16}
{Laigle} C.,  et~al., 2016, \mn@doi [\apjs] {10.3847/0067-0049/224/2/24}, \href
  {http://adsabs.harvard.edu/abs/2016ApJS..224...24L} {224, 24}

\bibitem[\protect\citeauthoryear{{Laigle} et~al.,}{{Laigle}
  et~al.}{2018}]{laigle18}
{Laigle} C.,  et~al., 2018, \mn@doi [\mnras] {10.1093/mnras/stx3055}, \href
  {http://adsabs.harvard.edu/abs/2018MNRAS.474.5437L} {474, 5437}

\bibitem[\protect\citeauthoryear{{Landolt}}{{Landolt}}{1992}]{landolt92}
{Landolt} A.~U.,  1992, \mn@doi [\aj] {10.1086/116242}, \href
  {http://adsabs.harvard.edu/abs/1992AJ....104..340L} {104, 340}

\bibitem[\protect\citeauthoryear{{Lanzuisi} et~al.,}{{Lanzuisi}
  et~al.}{2017}]{lanzuisi17}
{Lanzuisi} G.,  et~al., 2017, \mn@doi [\aap] {10.1051/0004-6361/201629955},
  \href {http://adsabs.harvard.edu/abs/2017A%26A...602A.123L} {602, A123}

\bibitem[\protect\citeauthoryear{{Le F{\`e}vre} et~al.,}{{Le F{\`e}vre}
  et~al.}{2005}]{dong05}
{Le F{\`e}vre} O.,  et~al., 2005, \mn@doi [\aap] {10.1051/0004-6361:20041960},
  \href {http://adsabs.harvard.edu/abs/2005A%26A...439..845L} {439, 845}

\bibitem[\protect\citeauthoryear{{Le F{\`e}vre} et~al.,}{{Le F{\`e}vre}
  et~al.}{2013}]{dong13}
{Le F{\`e}vre} O.,  et~al., 2013, \mn@doi [\aap] {10.1051/0004-6361/201322179},
  \href {http://adsabs.harvard.edu/abs/2013A%26A...559A..14L} {559, A14}

\bibitem[\protect\citeauthoryear{{Le F{\`e}vre} et~al.,}{{Le F{\`e}vre}
  et~al.}{2015}]{dong15}
{Le F{\`e}vre} O.,  et~al., 2015, \mn@doi [\aap] {10.1051/0004-6361/201423829},
  \href {http://adsabs.harvard.edu/abs/2015A%26A...576A..79L} {576, A79}

\bibitem[\protect\citeauthoryear{{Lee-Brown} et~al.,}{{Lee-Brown}
  et~al.}{2017}]{minirud17}
{Lee-Brown} D.~B.,  et~al., 2017, \mn@doi [\apj] {10.3847/1538-4357/aa7948},
  \href {https://ui.adsabs.harvard.edu/abs/2017ApJ...844...43L} {844, 43}

\bibitem[\protect\citeauthoryear{{Lemaux} et~al.,}{{Lemaux}
  et~al.}{2009}]{lem09}
{Lemaux} B.~C.,  et~al., 2009, \mn@doi [\apj] {10.1088/0004-637X/700/1/20},
  \href {http://adsabs.harvard.edu/abs/2009ApJ...700...20L} {700, 20}

\bibitem[\protect\citeauthoryear{{Lemaux}, {Lubin}, {Shapley}, {Kocevski},
  {Gal}  \& {Squires}}{{Lemaux} et~al.}{2010}]{lem10}
{Lemaux} B.~C.,  {Lubin} L.~M.,  {Shapley} A.,  {Kocevski} D.,  {Gal} R.~R.,
  {Squires} G.~K.,  2010, \mn@doi [\apj] {10.1088/0004-637X/716/2/970}, \href
  {http://adsabs.harvard.edu/abs/2010ApJ...716..970L} {716, 970}

\bibitem[\protect\citeauthoryear{{Lemaux} et~al.,}{{Lemaux}
  et~al.}{2012}]{lem12}
{Lemaux} B.~C.,  et~al., 2012, \mn@doi [\apj] {10.1088/0004-637X/745/2/106},
  \href {http://adsabs.harvard.edu/abs/2012ApJ...745..106L} {745, 106}

\bibitem[\protect\citeauthoryear{{Lemaux} et~al.,}{{Lemaux}
  et~al.}{2014a}]{lem14a}
{Lemaux} B.~C.,  et~al., 2014a, \mn@doi [\aap] {10.1051/0004-6361/201423828},
  \href {http://adsabs.harvard.edu/abs/2014A%26A...572A..41L} {572, A41}

\bibitem[\protect\citeauthoryear{{Lemaux} et~al.,}{{Lemaux}
  et~al.}{2014b}]{lem14}
{Lemaux} B.~C.,  et~al., 2014b, \mn@doi [\aap] {10.1051/0004-6361/201323089},
  \href {http://adsabs.harvard.edu/abs/2014A%26A...572A..90L} {572, A90}

\bibitem[\protect\citeauthoryear{{Lemaux}, {Tomczak}, {Lubin}, {Wu}, {Gal},
  {Rumbaugh}, {Kocevski}  \& {Squires}}{{Lemaux} et~al.}{2017}]{lem17a}
{Lemaux} B.~C.,  {Tomczak} A.~R.,  {Lubin} L.~M.,  {Wu} P.-F.,  {Gal} R.~R.,
  {Rumbaugh} N.,  {Kocevski} D.~D.,   {Squires} G.~K.,  2017, \mn@doi [\mnras]
  {10.1093/mnras/stx1579}, \href
  {http://adsabs.harvard.edu/abs/2017MNRAS.472..419L} {472, 419}

\bibitem[\protect\citeauthoryear{{Lemaux} et~al.,}{{Lemaux}
  et~al.}{2018}]{lem18}
{Lemaux} B.~C.,  et~al., 2018, \mn@doi [\aap] {10.1051/0004-6361/201730870},
  \href {http://adsabs.harvard.edu/abs/2018A%26A...615A..77L} {615, A77}

\bibitem[\protect\citeauthoryear{{Lilly} et~al.,}{{Lilly}
  et~al.}{2007}]{lilly07}
{Lilly} S.~J.,  et~al., 2007, \mn@doi [\apjs] {10.1086/516589}, \href
  {http://adsabs.harvard.edu/abs/2007ApJS..172...70L} {172, 70}

\bibitem[\protect\citeauthoryear{{Lilly} et~al.,}{{Lilly}
  et~al.}{2009}]{lilly09}
{Lilly} S.~J.,  et~al., 2009, \mn@doi [\apjs] {10.1088/0067-0049/184/2/218},
  \href {http://adsabs.harvard.edu/abs/2009ApJS..184..218L} {184, 218}

\bibitem[\protect\citeauthoryear{{L{\'o}pez-Sanjuan}
  et~al.,}{{L{\'o}pez-Sanjuan} et~al.}{2013}]{carlos13}
{L{\'o}pez-Sanjuan} C.,  et~al., 2013, \mn@doi [\aap]
  {10.1051/0004-6361/201220286}, \href
  {http://adsabs.harvard.edu/abs/2013A%26A...553A..78L} {553, A78}

\bibitem[\protect\citeauthoryear{{Lubin}, {Gal}, {Lemaux}, {Kocevski}  \&
  {Squires}}{{Lubin} et~al.}{2009}]{lub09}
{Lubin} L.~M.,  {Gal} R.~R.,  {Lemaux} B.~C.,  {Kocevski} D.~D.,   {Squires}
  G.~K.,  2009, \mn@doi [\aj] {10.1088/0004-6256/137/6/4867}, \href
  {http://adsabs.harvard.edu/abs/2009AJ....137.4867L} {137, 4867}

\bibitem[\protect\citeauthoryear{{Makovoz} \& {Marleau}}{{Makovoz} \&
  {Marleau}}{2005}]{makovoz06}
{Makovoz} D.,  {Marleau} F.~R.,  2005, \mn@doi [\pasp] {10.1086/432977}, \href
  {http://adsabs.harvard.edu/abs/2005PASP..117.1113M} {117, 1113}

\bibitem[\protect\citeauthoryear{{Mancone}, {Gonzalez}, {Brodwin}, {Stanford},
  {Eisenhardt}, {Stern}  \& {Jones}}{{Mancone} et~al.}{2010}]{mancone10}
{Mancone} C.~L.,  {Gonzalez} A.~H.,  {Brodwin} M.,  {Stanford} S.~A.,
  {Eisenhardt} P. R.~M.,  {Stern} D.,   {Jones} C.,  2010, \mn@doi [\apj]
  {10.1088/0004-637X/720/1/284}, \href
  {https://ui.adsabs.harvard.edu/abs/2010ApJ...720..284M} {720, 284}

\bibitem[\protect\citeauthoryear{{Martig}, {Bournaud}, {Teyssier}  \&
  {Dekel}}{{Martig} et~al.}{2009}]{martig09}
{Martig} M.,  {Bournaud} F.,  {Teyssier} R.,   {Dekel} A.,  2009, \mn@doi
  [\apj] {10.1088/0004-637X/707/1/250}, \href
  {http://adsabs.harvard.edu/abs/2009ApJ...707..250M} {707, 250}

\bibitem[\protect\citeauthoryear{{Martin} et~al.,}{{Martin}
  et~al.}{2007}]{cmart07}
{Martin} D.~C.,  et~al., 2007, \mn@doi [\apjs] {10.1086/522088}, \href
  {http://adsabs.harvard.edu/abs/2007ApJS..173..415M} {173, 415}

\bibitem[\protect\citeauthoryear{{Martini}, {Sivakoff}  \&
  {Mulchaey}}{{Martini} et~al.}{2009}]{martini09}
{Martini} P.,  {Sivakoff} G.~R.,   {Mulchaey} J.~S.,  2009, \mn@doi [\apj]
  {10.1088/0004-637X/701/1/66}, \href
  {https://ui.adsabs.harvard.edu/abs/2009ApJ...701...66M} {701, 66}

\bibitem[\protect\citeauthoryear{{Martini} et~al.,}{{Martini}
  et~al.}{2013}]{martini13}
{Martini} P.,  et~al., 2013, \mn@doi [\apj] {10.1088/0004-637X/768/1/1}, \href
  {http://adsabs.harvard.edu/abs/2013ApJ...768....1M} {768, 1}

\bibitem[\protect\citeauthoryear{{Marulli} et~al.,}{{Marulli}
  et~al.}{2013}]{marulli13}
{Marulli} F.,  et~al., 2013, \mn@doi [\aap] {10.1051/0004-6361/201321476},
  \href {http://adsabs.harvard.edu/abs/2013A%26A...557A..17M} {557, A17}

\bibitem[\protect\citeauthoryear{{Maughan}, {Ellis}, {Jones}, {Mason},
  {C{\'o}rdova}  \& {Priedhorsky}}{{Maughan} et~al.}{2006}]{maughan06}
{Maughan} B.~J.,  {Ellis} S.~C.,  {Jones} L.~R.,  {Mason} K.~O.,  {C{\'o}rdova}
  F.~A.,   {Priedhorsky} W.,  2006, \mn@doi [\apj] {10.1086/499939}, \href
  {http://adsabs.harvard.edu/abs/2006ApJ...640..219M} {640, 219}

\bibitem[\protect\citeauthoryear{{McGee}, {Balogh}, {Bower}, {Font}  \&
  {McCarthy}}{{McGee} et~al.}{2009}]{mcgee09}
{McGee} S.~L.,  {Balogh} M.~L.,  {Bower} R.~G.,  {Font} A.~S.,   {McCarthy}
  I.~G.,  2009, \mn@doi [\mnras] {10.1111/j.1365-2966.2009.15507.x}, \href
  {http://adsabs.harvard.edu/abs/2009MNRAS.400..937M} {400, 937}

\bibitem[\protect\citeauthoryear{{McGee}, {Balogh}, {Wilman}, {Bower},
  {Mulchaey}, {Parker}  \& {Oemler}}{{McGee} et~al.}{2011}]{mcgee11}
{McGee} S.~L.,  {Balogh} M.~L.,  {Wilman} D.~J.,  {Bower} R.~G.,  {Mulchaey}
  J.~S.,  {Parker} L.~C.,   {Oemler} A.,  2011, \mn@doi [\mnras]
  {10.1111/j.1365-2966.2010.18189.x}, \href
  {http://adsabs.harvard.edu/abs/2011MNRAS.413..996M} {413, 996}

\bibitem[\protect\citeauthoryear{{McGee}, {Bower}  \& {Balogh}}{{McGee}
  et~al.}{2014}]{mcgee14}
{McGee} S.~L.,  {Bower} R.~G.,   {Balogh} M.~L.,  2014, \mn@doi [\mnras]
  {10.1093/mnrasl/slu066}, \href
  {http://adsabs.harvard.edu/abs/2014MNRAS.442L.105M} {442, L105}

\bibitem[\protect\citeauthoryear{{Mehrtens} et~al.,}{{Mehrtens}
  et~al.}{2012}]{meh12}
{Mehrtens} N.,  et~al., 2012, \mn@doi [\mnras]
  {10.1111/j.1365-2966.2012.20931.x}, \href
  {http://adsabs.harvard.edu/abs/2012MNRAS.423.1024M} {423, 1024}

\bibitem[\protect\citeauthoryear{{Mei} et~al.,}{{Mei} et~al.}{2012}]{mei12}
{Mei} S.,  et~al., 2012, \mn@doi [\apj] {10.1088/0004-637X/754/2/141}, \href
  {http://adsabs.harvard.edu/abs/2012ApJ...754..141M} {754, 141}

\bibitem[\protect\citeauthoryear{{Merlin} et~al.,}{{Merlin}
  et~al.}{2015}]{merlin15}
{Merlin} E.,  et~al., 2015, \mn@doi [\aap] {10.1051/0004-6361/201526471}, \href
  {http://adsabs.harvard.edu/abs/2015A%26A...582A..15M} {582, A15}

\bibitem[\protect\citeauthoryear{{Miettinen} et~al.,}{{Miettinen}
  et~al.}{2017}]{miettinen17}
{Miettinen} O.,  et~al., 2017, \mn@doi [\aap] {10.1051/0004-6361/201730762},
  \href {http://adsabs.harvard.edu/abs/2017A%26A...606A..17M} {606, A17}

\bibitem[\protect\citeauthoryear{{Miyazaki} et~al.,}{{Miyazaki}
  et~al.}{2002}]{miyazaki02}
{Miyazaki} S.,  et~al., 2002, \mn@doi [\pasj] {10.1093/pasj/54.6.833}, \href
  {http://adsabs.harvard.edu/abs/2002PASJ...54..833M} {54, 833}

\bibitem[\protect\citeauthoryear{{Mok} et~al.,}{{Mok} et~al.}{2014}]{mok14}
{Mok} A.,  et~al., 2014, \mn@doi [\mnras] {10.1093/mnras/stt2419}, \href
  {http://adsabs.harvard.edu/abs/2014MNRAS.438.3070M} {438, 3070}

\bibitem[\protect\citeauthoryear{{Moorman}, {Moreno}, {White}, {Vogeley},
  {Hoyle}, {Giovanelli}  \& {Haynes}}{{Moorman} et~al.}{2016}]{moorman16}
{Moorman} C.~M.,  {Moreno} J.,  {White} A.,  {Vogeley} M.~S.,  {Hoyle} F.,
  {Giovanelli} R.,   {Haynes} M.~P.,  2016, \mn@doi [\apj]
  {10.3847/0004-637X/831/2/118}, \href
  {http://adsabs.harvard.edu/abs/2016ApJ...831..118M} {831, 118}

\bibitem[\protect\citeauthoryear{{More} et~al.,}{{More} et~al.}{2016}]{more16}
{More} S.,  et~al., 2016, \mn@doi [\apj] {10.3847/0004-637X/825/1/39}, \href
  {http://adsabs.harvard.edu/abs/2016ApJ...825...39M} {825, 39}

\bibitem[\protect\citeauthoryear{{Moster}, {Somerville}, {Maulbetsch}, {van den
  Bosch}, {Macci{\`o}}, {Naab}  \& {Oser}}{{Moster} et~al.}{2010}]{moster10}
{Moster} B.~P.,  {Somerville} R.~S.,  {Maulbetsch} C.,  {van den Bosch} F.~C.,
  {Macci{\`o}} A.~V.,  {Naab} T.,   {Oser} L.,  2010, \mn@doi [\apj]
  {10.1088/0004-637X/710/2/903}, \href
  {http://adsabs.harvard.edu/abs/2010ApJ...710..903M} {710, 903}

\bibitem[\protect\citeauthoryear{{Moster}, {Naab}  \& {White}}{{Moster}
  et~al.}{2013}]{moster13}
{Moster} B.~P.,  {Naab} T.,   {White} S.~D.~M.,  2013, \mn@doi [\mnras]
  {10.1093/mnras/sts261}, \href
  {http://adsabs.harvard.edu/abs/2013MNRAS.428.3121M} {428, 3121}

\bibitem[\protect\citeauthoryear{{Moutard} et~al.,}{{Moutard}
  et~al.}{2016a}]{thibaud16a}
{Moutard} T.,  et~al., 2016a, \mn@doi [\aap] {10.1051/0004-6361/201527945},
  \href {https://ui.adsabs.harvard.edu/abs/2016A&A...590A.102M} {590, A102}

\bibitem[\protect\citeauthoryear{{Moutard} et~al.,}{{Moutard}
  et~al.}{2016b}]{thibaud16}
{Moutard} T.,  et~al., 2016b, \mn@doi [\aap] {10.1051/0004-6361/201527294},
  \href {http://adsabs.harvard.edu/abs/2016A%26A...590A.103M} {590, A103}

\bibitem[\protect\citeauthoryear{{Moutard}, {Sawicki}, {Arnouts}, {Golob},
  {Malavasi}, {Adami}, {Coupon}  \& {Ilbert}}{{Moutard}
  et~al.}{2018}]{thibaud18}
{Moutard} T.,  {Sawicki} M.,  {Arnouts} S.,  {Golob} A.,  {Malavasi} N.,
  {Adami} C.,  {Coupon} J.,   {Ilbert} O.,  2018, \mn@doi [\mnras]
  {10.1093/mnras/sty1543}, \href
  {https://ui.adsabs.harvard.edu/abs/2018MNRAS.479.2147M} {479, 2147}

\bibitem[\protect\citeauthoryear{{Muzzin} et~al.,}{{Muzzin}
  et~al.}{2009}]{muz09}
{Muzzin} A.,  et~al., 2009, \mn@doi [\apj] {10.1088/0004-637X/698/2/1934},
  \href {http://adsabs.harvard.edu/abs/2009ApJ...698.1934M} {698, 1934}

\bibitem[\protect\citeauthoryear{{Muzzin} et~al.,}{{Muzzin}
  et~al.}{2012}]{muz12}
{Muzzin} A.,  et~al., 2012, \mn@doi [\apj] {10.1088/0004-637X/746/2/188}, \href
  {http://adsabs.harvard.edu/abs/2012ApJ...746..188M} {746, 188}

\bibitem[\protect\citeauthoryear{{Muzzin} et~al.,}{{Muzzin}
  et~al.}{2013}]{lunchpailmcgee13}
{Muzzin} A.,  et~al., 2013, \mn@doi [\apj] {10.1088/0004-637X/777/1/18}, \href
  {http://adsabs.harvard.edu/abs/2013ApJ...777...18M} {777, 18}

\bibitem[\protect\citeauthoryear{{Muzzin} et~al.,}{{Muzzin}
  et~al.}{2014}]{muz14}
{Muzzin} A.,  et~al., 2014, \mn@doi [\apj] {10.1088/0004-637X/796/1/65}, \href
  {http://adsabs.harvard.edu/abs/2014ApJ...796...65M} {796, 65}

\bibitem[\protect\citeauthoryear{{Nantais} et~al.,}{{Nantais}
  et~al.}{2016}]{nantais16}
{Nantais} J.~B.,  et~al., 2016, \mn@doi [\aap] {10.1051/0004-6361/201628663},
  \href {http://adsabs.harvard.edu/abs/2016A%26A...592A.161N} {592, A161}

\bibitem[\protect\citeauthoryear{{Nantais} et~al.,}{{Nantais}
  et~al.}{2017}]{nantais17}
{Nantais} J.~B.,  et~al., 2017, \mn@doi [\mnras] {10.1093/mnrasl/slw224}, \href
  {http://adsabs.harvard.edu/abs/2017MNRAS.465L.104N} {465, L104}

\bibitem[\protect\citeauthoryear{{Newman} et~al.,}{{Newman}
  et~al.}{2013}]{new13}
{Newman} J.~A.,  et~al., 2013, \mn@doi [\apjs] {10.1088/0067-0049/208/1/5},
  \href {http://adsabs.harvard.edu/abs/2013ApJS..208....5N} {208, 5}

\bibitem[\protect\citeauthoryear{{Newman}, {Ellis}, {Andreon}, {Treu},
  {Raichoor}  \& {Trinchieri}}{{Newman} et~al.}{2014}]{newman14}
{Newman} A.~B.,  {Ellis} R.~S.,  {Andreon} S.,  {Treu} T.,  {Raichoor} A.,
  {Trinchieri} G.,  2014, \mn@doi [\apj] {10.1088/0004-637X/788/1/51}, \href
  {https://ui.adsabs.harvard.edu/abs/2014ApJ...788...51N} {788, 51}

\bibitem[\protect\citeauthoryear{{O'Donoghue}, {Eilek}  \& {Owen}}{{O'Donoghue}
  et~al.}{1993}]{odon93}
{O'Donoghue} A.~A.,  {Eilek} J.~A.,   {Owen} F.~N.,  1993, \mn@doi [\apj]
  {10.1086/172600}, \href {http://adsabs.harvard.edu/abs/1993ApJ...408..428O}
  {408, 428}

\bibitem[\protect\citeauthoryear{{Oke} \& {Gunn}}{{Oke} \&
  {Gunn}}{1983}]{okengunn83}
{Oke} J.~B.,  {Gunn} J.~E.,  1983, \mn@doi [\apj] {10.1086/160817}, \href
  {http://adsabs.harvard.edu/abs/1983ApJ...266..713O} {266, 713}

\bibitem[\protect\citeauthoryear{{Oke} et~al.,}{{Oke} et~al.}{1995}]{oke95}
{Oke} J.~B.,  et~al., 1995, \mn@doi [\pasp] {10.1086/133562}, \href
  {http://adsabs.harvard.edu/abs/1995PASP..107..375O} {107, 375}

\bibitem[\protect\citeauthoryear{{Oke}, {Postman}  \& {Lubin}}{{Oke}
  et~al.}{1998}]{oke98}
{Oke} J.~B.,  {Postman} M.,   {Lubin} L.~M.,  1998, \mn@doi [\aj]
  {10.1086/300462}, \href {http://adsabs.harvard.edu/abs/1998AJ....116..549O}
  {116, 549}

\bibitem[\protect\citeauthoryear{{Ouchi} et~al.,}{{Ouchi}
  et~al.}{2004}]{ouchi04}
{Ouchi} M.,  et~al., 2004, \mn@doi [\apj] {10.1086/422207}, \href
  {http://adsabs.harvard.edu/abs/2004ApJ...611..660O} {611, 660}

\bibitem[\protect\citeauthoryear{{Owen} \& {Rudnick}}{{Owen} \&
  {Rudnick}}{1976}]{owen76}
{Owen} F.~N.,  {Rudnick} L.,  1976, \mn@doi [\apjl] {10.1086/182077}, \href
  {http://adsabs.harvard.edu/abs/1976ApJ...205L...1O} {205, L1}

\bibitem[\protect\citeauthoryear{{Pacifici}, {Kassin}, {Weiner}, {Charlot}  \&
  {Gardner}}{{Pacifici} et~al.}{2013}]{pacifici13}
{Pacifici} C.,  {Kassin} S.~A.,  {Weiner} B.,  {Charlot} S.,   {Gardner} J.~P.,
   2013, \mn@doi [\apjl] {10.1088/2041-8205/762/1/L15}, \href
  {http://adsabs.harvard.edu/abs/2013ApJ...762L..15P} {762, L15}

\bibitem[\protect\citeauthoryear{{Patel}, {Kelson}, {Holden}, {Franx}  \&
  {Illingworth}}{{Patel} et~al.}{2011}]{patel11}
{Patel} S.~G.,  {Kelson} D.~D.,  {Holden} B.~P.,  {Franx} M.,   {Illingworth}
  G.~D.,  2011, \mn@doi [\apj] {10.1088/0004-637X/735/1/53}, \href
  {http://adsabs.harvard.edu/abs/2011ApJ...735...53P} {735, 53}

\bibitem[\protect\citeauthoryear{{Pelliccia} et~al.,}{{Pelliccia}
  et~al.}{2018}]{debz18}
{Pelliccia} D.,  et~al., 2018, \mn@doi [\mnras] {10.1093/mnras/sty2876}, \href
  {http://adsabs.harvard.edu/abs/2018MNRAS.tmp.2738P} {}

\bibitem[\protect\citeauthoryear{{Peng} et~al.,}{{Peng} et~al.}{2010}]{peng10}
{Peng} Y.-j.,  et~al., 2010, \mn@doi [\apj] {10.1088/0004-637X/721/1/193},
  \href {http://adsabs.harvard.edu/abs/2010ApJ...721..193P} {721, 193}

\bibitem[\protect\citeauthoryear{{Pickles}}{{Pickles}}{1998}]{pickles98}
{Pickles} A.~J.,  1998, \mn@doi [\pasp] {10.1086/316197}, \href
  {http://adsabs.harvard.edu/abs/1998PASP..110..863P} {110, 863}

\bibitem[\protect\citeauthoryear{{Pimbblet}, {Shabala}, {Haines},
  {Fraser-McKelvie}  \& {Floyd}}{{Pimbblet} et~al.}{2013}]{pimbblet13}
{Pimbblet} K.~A.,  {Shabala} S.~S.,  {Haines} C.~P.,  {Fraser-McKelvie} A.,
  {Floyd} D.~J.~E.,  2013, \mn@doi [\mnras] {10.1093/mnras/sts470}, \href
  {http://adsabs.harvard.edu/abs/2013MNRAS.429.1827P} {429, 1827}

\bibitem[\protect\citeauthoryear{{Poggianti} et~al.,}{{Poggianti}
  et~al.}{2009}]{pog09}
{Poggianti} B.~M.,  et~al., 2009, \mn@doi [\apj] {10.1088/0004-637X/693/1/112},
  \href {http://adsabs.harvard.edu/abs/2009ApJ...693..112P} {693, 112}

\bibitem[\protect\citeauthoryear{{Popping}, {Caputi}, {Somerville}  \&
  {Trager}}{{Popping} et~al.}{2012}]{popping12}
{Popping} G.,  {Caputi} K.~I.,  {Somerville} R.~S.,   {Trager} S.~C.,  2012,
  \mn@doi [\mnras] {10.1111/j.1365-2966.2012.21702.x}, \href
  {http://adsabs.harvard.edu/abs/2012MNRAS.425.2386P} {425, 2386}

\bibitem[\protect\citeauthoryear{{Puget} et~al.,}{{Puget}
  et~al.}{2004}]{puget04}
{Puget} P.,  et~al., 2004, in {Moorwood} A.~F.~M.,  {Iye} M.,  eds,  \procspie
  Vol. 5492, Ground-based Instrumentation for Astronomy. pp 978--987,
  \mn@doi{10.1117/12.551097}

\bibitem[\protect\citeauthoryear{{Quadri}, {Williams}, {Franx}  \&
  {Hildebrandt}}{{Quadri} et~al.}{2012}]{quadri12}
{Quadri} R.~F.,  {Williams} R.~J.,  {Franx} M.,   {Hildebrandt} H.,  2012,
  \mn@doi [\apj] {10.1088/0004-637X/744/2/88}, \href
  {http://adsabs.harvard.edu/abs/2012ApJ...744...88Q} {744, 88}

\bibitem[\protect\citeauthoryear{{Rayner}, {Cushing}  \& {Vacca}}{{Rayner}
  et~al.}{2009}]{rayner09}
{Rayner} J.~T.,  {Cushing} M.~C.,   {Vacca} W.~D.,  2009, \mn@doi [\apjs]
  {10.1088/0067-0049/185/2/289}, \href
  {http://adsabs.harvard.edu/abs/2009ApJS..185..289R} {185, 289}

\bibitem[\protect\citeauthoryear{{Rumbaugh}, {Kocevski}, {Gal}, {Lemaux},
  {Lubin}, {Fassnacht}, {McGrath}  \& {Squires}}{{Rumbaugh}
  et~al.}{2012}]{rum12}
{Rumbaugh} N.,  {Kocevski} D.~D.,  {Gal} R.~R.,  {Lemaux} B.~C.,  {Lubin}
  L.~M.,  {Fassnacht} C.~D.,  {McGrath} E.~J.,   {Squires} G.~K.,  2012, \apj,
  \href {http://adsabs.harvard.edu/abs/2012ApJ...746..155R} {746, 155}

\bibitem[\protect\citeauthoryear{{Rumbaugh}, {Kocevski}, {Gal}, {Lemaux},
  {Lubin}, {Fassnacht}  \& {Squires}}{{Rumbaugh} et~al.}{2013}]{rum13}
{Rumbaugh} N.,  {Kocevski} D.~D.,  {Gal} R.~R.,  {Lemaux} B.~C.,  {Lubin}
  L.~M.,  {Fassnacht} C.~D.,   {Squires} G.~K.,  2013, \mn@doi [\apj]
  {10.1088/0004-637X/763/2/124}, \href
  {http://adsabs.harvard.edu/abs/2013ApJ...763..124R} {763, 124}

\bibitem[\protect\citeauthoryear{{Rumbaugh} et~al.,}{{Rumbaugh}
  et~al.}{2017}]{rum17}
{Rumbaugh} N.,  et~al., 2017, \mn@doi [\mnras] {10.1093/mnras/stw3091}, \href
  {http://adsabs.harvard.edu/abs/2017MNRAS.466..496R} {466, 496}

\bibitem[\protect\citeauthoryear{{Rumbaugh} et~al.,}{{Rumbaugh}
  et~al.}{2018}]{rum18}
{Rumbaugh} N.,  et~al., 2018, \mn@doi [\mnras] {10.1093/mnras/sty1181}, \href
  {http://adsabs.harvard.edu/abs/2018MNRAS.478.1403R} {478, 1403}

\bibitem[\protect\citeauthoryear{{Ryan} Jr. et~al.,}{{Ryan}
  et~al.}{2014}]{Rusl13}
{Ryan} Jr. R.~E.,  et~al., 2014, \mn@doi [\apjl] {10.1088/2041-8205/786/1/L4},
  \href {http://adsabs.harvard.edu/abs/2014ApJ...786L...4R} {786, L4}

\bibitem[\protect\citeauthoryear{{Sanders} et~al.,}{{Sanders}
  et~al.}{2018}]{sanders18}
{Sanders} R.~L.,  et~al., 2018, \mn@doi [\apj] {10.3847/1538-4357/aabcbd},
  \href {http://adsabs.harvard.edu/abs/2018ApJ...858...99S} {858, 99}

\bibitem[\protect\citeauthoryear{{Santos} et~al.,}{{Santos}
  et~al.}{2014}]{santos14}
{Santos} J.~S.,  et~al., 2014, \mn@doi [\mnras] {10.1093/mnras/stt2376}, \href
  {http://adsabs.harvard.edu/abs/2014MNRAS.438.2565S} {438, 2565}

\bibitem[\protect\citeauthoryear{{Scoville} et~al.,}{{Scoville}
  et~al.}{2016}]{smokynicky16}
{Scoville} N.,  et~al., 2016, \mn@doi [\apj] {10.3847/0004-637X/820/2/83},
  \href {http://adsabs.harvard.edu/abs/2016ApJ...820...83S} {820, 83}

\bibitem[\protect\citeauthoryear{{Scoville} et~al.,}{{Scoville}
  et~al.}{2017}]{smokynicky17}
{Scoville} N.,  et~al., 2017, \mn@doi [\apj] {10.3847/1538-4357/aa61a0}, \href
  {http://adsabs.harvard.edu/abs/2017ApJ...837..150S} {837, 150}

\bibitem[\protect\citeauthoryear{{Shen} et~al.,}{{Shen} et~al.}{2017}]{shen17}
{Shen} L.,  et~al., 2017, \mn@doi [\mnras] {10.1093/mnras/stx1984}, \href
  {http://adsabs.harvard.edu/abs/2017MNRAS.472..998S} {472, 998}

\bibitem[\protect\citeauthoryear{{Shen} et~al.,}{{Shen} et~al.}{2019}]{shen19}
{Shen} L.,  et~al., 2019, \mn@doi [\mnras] {10.1093/mnras/stz152}, \href
  {https://ui.adsabs.harvard.edu/abs/2019MNRAS.484.2433S} {484, 2433}

\bibitem[\protect\citeauthoryear{{Simcoe}, {Metzger}, {Small}  \&
  {Araya}}{{Simcoe} et~al.}{2000}]{simcoe00}
{Simcoe} R.~A.,  {Metzger} M.~R.,  {Small} T.~A.,   {Araya} G.,  2000, in
  American Astronomical Society Meeting Abstracts \#196. p.~758

\bibitem[\protect\citeauthoryear{{Skrutskie} et~al.,}{{Skrutskie}
  et~al.}{2006}]{skrutskie06}
{Skrutskie} M.~F.,  et~al., 2006, \mn@doi [\aj] {10.1086/498708}, \href
  {http://adsabs.harvard.edu/abs/2006AJ....131.1163S} {131, 1163}

\bibitem[\protect\citeauthoryear{{Smol{\v c}i{\'c}} et~al.,}{{Smol{\v c}i{\'c}}
  et~al.}{2017}]{vernesa17}
{Smol{\v c}i{\'c}} V.,  et~al., 2017, \mn@doi [\aap]
  {10.1051/0004-6361/201526989}, \href
  {http://adsabs.harvard.edu/abs/2017A%26A...597A...4S} {597, A4}

\bibitem[\protect\citeauthoryear{{Snyder}, {Cox}, {Hayward}, {Hernquist}  \&
  {Jonsson}}{{Snyder} et~al.}{2011}]{snyder11}
{Snyder} G.~F.,  {Cox} T.~J.,  {Hayward} C.~C.,  {Hernquist} L.,   {Jonsson}
  P.,  2011, \mn@doi [\apj] {10.1088/0004-637X/741/2/77}, \href
  {http://adsabs.harvard.edu/abs/2011ApJ...741...77S} {741, 77}

\bibitem[\protect\citeauthoryear{{Socolovsky}, {Almaini}, {Hatch}, {Wild},
  {Maltby}, {Hartley}  \& {Simpson}}{{Socolovsky}
  et~al.}{2018}]{mildmanneredmiguel18}
{Socolovsky} M.,  {Almaini} O.,  {Hatch} N.~A.,  {Wild} V.,  {Maltby} D.~T.,
  {Hartley} W.~G.,   {Simpson} C.,  2018, \mn@doi [\mnras]
  {10.1093/mnras/sty312}, \href
  {http://adsabs.harvard.edu/abs/2018MNRAS.476.1242S} {476, 1242}

\bibitem[\protect\citeauthoryear{{Sparre}, {Hayward}, {Feldmann},
  {Faucher-Gigu{\`e}re}, {Muratov}, {Kere{\v s}}  \& {Hopkins}}{{Sparre}
  et~al.}{2017}]{sparre17}
{Sparre} M.,  {Hayward} C.~C.,  {Feldmann} R.,  {Faucher-Gigu{\`e}re} C.-A.,
  {Muratov} A.~L.,  {Kere{\v s}} D.,   {Hopkins} P.~F.,  2017, \mn@doi [\mnras]
  {10.1093/mnras/stw3011}, \href
  {http://adsabs.harvard.edu/abs/2017MNRAS.466...88S} {466, 88}

\bibitem[\protect\citeauthoryear{{Springel} et~al.,}{{Springel}
  et~al.}{2005}]{springel05}
{Springel} V.,  et~al., 2005, \mn@doi [\nat] {10.1038/nature03597}, \href
  {http://adsabs.harvard.edu/abs/2005Natur.435..629S} {435, 629}

\bibitem[\protect\citeauthoryear{{Stach}, {Swinbank}, {Smail}, {Hilton},
  {Simpson}  \& {Cooke}}{{Stach} et~al.}{2017}]{stach17}
{Stach} S.~M.,  {Swinbank} A.~M.,  {Smail} I.,  {Hilton} M.,  {Simpson} J.~M.,
   {Cooke} E.~A.,  2017, \mn@doi [\apj] {10.3847/1538-4357/aa93f6}, \href
  {https://ui.adsabs.harvard.edu/abs/2017ApJ...849..154S} {849, 154}

\bibitem[\protect\citeauthoryear{{Strazzullo} et~al.,}{{Strazzullo}
  et~al.}{2010}]{strazzullo10}
{Strazzullo} V.,  et~al., 2010, \mn@doi [\aap] {10.1051/0004-6361/201015251},
  \href {http://adsabs.harvard.edu/abs/2010A%26A...524A..17S} {524, A17}

\bibitem[\protect\citeauthoryear{{Strazzullo} et~al.,}{{Strazzullo}
  et~al.}{2019}]{strazz19}
{Strazzullo} V.,  et~al., 2019, \mn@doi [\aap] {10.1051/0004-6361/201833944},
  \href {https://ui.adsabs.harvard.edu/abs/2019A&A...622A.117S} {622, A117}

\bibitem[\protect\citeauthoryear{{Tal} et~al.,}{{Tal} et~al.}{2014}]{tal14}
{Tal} T.,  et~al., 2014, \mn@doi [\apj] {10.1088/0004-637X/789/2/164}, \href
  {http://adsabs.harvard.edu/abs/2014ApJ...789..164T} {789, 164}

\bibitem[\protect\citeauthoryear{{Tanaka} et~al.,}{{Tanaka}
  et~al.}{2008}]{tanaka08}
{Tanaka} M.,  et~al., 2008, \mn@doi [\aap] {10.1051/0004-6361:200810440}, \href
  {http://adsabs.harvard.edu/abs/2008A%26A...489..571T} {489, 571}

\bibitem[\protect\citeauthoryear{{Teyssier}, {Pontzen}, {Dubois}  \&
  {Read}}{{Teyssier} et~al.}{2013}]{teyssier13}
{Teyssier} R.,  {Pontzen} A.,  {Dubois} Y.,   {Read} J.~I.,  2013, \mn@doi
  [\mnras] {10.1093/mnras/sts563}, \href
  {http://adsabs.harvard.edu/abs/2013MNRAS.429.3068T} {429, 3068}

\bibitem[\protect\citeauthoryear{{Tody}}{{Tody}}{1993}]{tody93}
{Tody} D.,  1993, in {Hanisch} R.~J.,  {Brissenden} R.~J.~V.,   {Barnes} J.,
  eds,  Astronomical Society of the Pacific Conference Series Vol. 52,
  Astronomical Data Analysis Software and Systems II. p.~173

\bibitem[\protect\citeauthoryear{{Tomczak} et~al.,}{{Tomczak}
  et~al.}{2014}]{AA14}
{Tomczak} A.~R.,  et~al., 2014, \mn@doi [\apj] {10.1088/0004-637X/783/2/85},
  \href {http://adsabs.harvard.edu/abs/2014ApJ...783...85T} {783, 85}

\bibitem[\protect\citeauthoryear{{Tomczak} et~al.,}{{Tomczak}
  et~al.}{2016}]{AA16}
{Tomczak} A.~R.,  et~al., 2016, \mn@doi [\apj] {10.3847/0004-637X/817/2/118},
  \href {http://adsabs.harvard.edu/abs/2016ApJ...817..118T} {817, 118}

\bibitem[\protect\citeauthoryear{{Tomczak} et~al.,}{{Tomczak}
  et~al.}{2017}]{AA17}
{Tomczak} A.~R.,  et~al., 2017, \mn@doi [\mnras] {10.1093/mnras/stx2245}, \href
  {http://adsabs.harvard.edu/abs/2017MNRAS.472.3512T} {472, 3512}

\bibitem[\protect\citeauthoryear{{Tomczak} et~al.,}{{Tomczak}
  et~al.}{2019}]{AA19}
{Tomczak} A.~R.,  et~al., 2019, \mn@doi [\mnras] {10.1093/mnras/stz342}, \href
  {https://ui.adsabs.harvard.edu/abs/2019MNRAS.484.4695T} {484, 4695}

\bibitem[\protect\citeauthoryear{{Tran}, {Franx}, {Illingworth}, {Kelson}  \&
  {van Dokkum}}{{Tran} et~al.}{2003}]{tran03}
{Tran} K.-V.~H.,  {Franx} M.,  {Illingworth} G.,  {Kelson} D.~D.,   {van
  Dokkum} P.,  2003, \mn@doi [\apj] {10.1086/379804}, \href
  {http://adsabs.harvard.edu/abs/2003ApJ...599..865T} {599, 865}

\bibitem[\protect\citeauthoryear{{Tran} et~al.,}{{Tran} et~al.}{2015}]{tran15}
{Tran} K.-V.~H.,  et~al., 2015, \mn@doi [\apj] {10.1088/0004-637X/811/1/28},
  \href {http://adsabs.harvard.edu/abs/2015ApJ...811...28T} {811, 28}

\bibitem[\protect\citeauthoryear{{Vulcani}, {Poggianti}, {Finn}, {Rudnick},
  {Desai}  \& {Bamford}}{{Vulcani} et~al.}{2010}]{banalbenedetta10}
{Vulcani} B.,  {Poggianti} B.~M.,  {Finn} R.~A.,  {Rudnick} G.,  {Desai} V.,
  {Bamford} S.,  2010, \mn@doi [\apjl] {10.1088/2041-8205/710/1/L1}, \href
  {http://adsabs.harvard.edu/abs/2010ApJ...710L...1V} {710, L1}

\bibitem[\protect\citeauthoryear{{Vulcani} et~al.,}{{Vulcani}
  et~al.}{2017}]{vulcani17}
{Vulcani} B.,  et~al., 2017, \mn@doi [\apj] {10.3847/1538-4357/aa618b}, \href
  {http://adsabs.harvard.edu/abs/2017ApJ...837..126V} {837, 126}

\bibitem[\protect\citeauthoryear{{Wagner}, {Courteau}, {Brodwin}, {Stanford},
  {Snyder}  \& {Stern}}{{Wagner} et~al.}{2017}]{wagner17}
{Wagner} C.~R.,  {Courteau} S.,  {Brodwin} M.,  {Stanford} S.~A.,  {Snyder}
  G.~F.,   {Stern} D.,  2017, \mn@doi [\apj] {10.3847/1538-4357/834/1/53},
  \href {http://adsabs.harvard.edu/abs/2017ApJ...834...53W} {834, 53}

\bibitem[\protect\citeauthoryear{{Wang} et~al.,}{{Wang}
  et~al.}{2016}]{taowang16}
{Wang} T.,  et~al., 2016, \mn@doi [\apj] {10.3847/0004-637X/828/1/56}, \href
  {https://ui.adsabs.harvard.edu/abs/2016ApJ...828...56W} {828, 56}

\bibitem[\protect\citeauthoryear{{Wang} et~al.,}{{Wang} et~al.}{2018}]{wang18}
{Wang} T.,  et~al., 2018, \mn@doi [\apjl] {10.3847/2041-8213/aaeb2c}, \href
  {http://adsabs.harvard.edu/abs/2018ApJ...867L..29W} {867, L29}

\bibitem[\protect\citeauthoryear{{Werner} et~al.,}{{Werner}
  et~al.}{2004}]{wer04}
{Werner} M.~W.,  et~al., 2004, \mn@doi [\apjs] {10.1086/422992}, \href
  {http://adsabs.harvard.edu/abs/2004ApJS..154....1W} {154, 1}

\bibitem[\protect\citeauthoryear{{Wetzel}}{{Wetzel}}{2011}]{wetzel11}
{Wetzel} A.~R.,  2011, \mn@doi [\mnras] {10.1111/j.1365-2966.2010.17877.x},
  \href {http://adsabs.harvard.edu/abs/2011MNRAS.412...49W} {412, 49}

\bibitem[\protect\citeauthoryear{{Wetzel}, {Tinker}, {Conroy}  \& {van den
  Bosch}}{{Wetzel} et~al.}{2013}]{wetzel13}
{Wetzel} A.~R.,  {Tinker} J.~L.,  {Conroy} C.,   {van den Bosch} F.~C.,  2013,
  \mn@doi [\mnras] {10.1093/mnras/stt469}, \href
  {http://adsabs.harvard.edu/abs/2013MNRAS.432..336W} {432, 336}

\bibitem[\protect\citeauthoryear{{Whitaker} et~al.,}{{Whitaker}
  et~al.}{2017}]{whitaker17}
{Whitaker} K.~E.,  et~al., 2017, \mn@doi [\apj] {10.3847/1538-4357/aa6258},
  \href {http://adsabs.harvard.edu/abs/2017ApJ...838...19W} {838, 19}

\bibitem[\protect\citeauthoryear{{Wijesinghe} et~al.,}{{Wijesinghe}
  et~al.}{2012}]{wijesinghe12}
{Wijesinghe} D.~B.,  et~al., 2012, \mn@doi [\mnras]
  {10.1111/j.1365-2966.2012.21164.x}, \href
  {http://adsabs.harvard.edu/abs/2012MNRAS.423.3679W} {423, 3679}

\bibitem[\protect\citeauthoryear{{Williams}, {Quadri}, {Franx}, {van Dokkum}
  \& {Labb{\'e}}}{{Williams} et~al.}{2009}]{williams09}
{Williams} R.~J.,  {Quadri} R.~F.,  {Franx} M.,  {van Dokkum} P.,   {Labb{\'e}}
  I.,  2009, \mn@doi [\apj] {10.1088/0004-637X/691/2/1879}, \href
  {http://adsabs.harvard.edu/abs/2009ApJ...691.1879W} {691, 1879}

\bibitem[\protect\citeauthoryear{{Wilson} et~al.,}{{Wilson}
  et~al.}{2009}]{wilson09}
{Wilson} G.,  et~al., 2009, \mn@doi [\apj] {10.1088/0004-637X/698/2/1943},
  \href {http://adsabs.harvard.edu/abs/2009ApJ...698.1943W} {698, 1943}

\bibitem[\protect\citeauthoryear{{Wootten} \& {Thompson}}{{Wootten} \&
  {Thompson}}{2009}]{wootten09}
{Wootten} A.,  {Thompson} A.~R.,  2009, \mn@doi [IEEE Proceedings]
  {10.1109/JPROC.2009.2020572}, \href
  {http://adsabs.harvard.edu/abs/2009IEEEP..97.1463W} {97, 1463}

\bibitem[\protect\citeauthoryear{{Wu}, {Gal}, {Lemaux}, {Kocevski}, {Lubin},
  {Rumbaugh}  \& {Squires}}{{Wu} et~al.}{2014}]{pwu14}
{Wu} P.-F.,  {Gal} R.~R.,  {Lemaux} B.~C.,  {Kocevski} D.~D.,  {Lubin} L.~M.,
  {Rumbaugh} N.,   {Squires} G.~K.,  2014, \mn@doi [\apj]
  {10.1088/0004-637X/792/1/16}, \href
  {http://adsabs.harvard.edu/abs/2014ApJ...792...16W} {792, 16}

\bibitem[\protect\citeauthoryear{{Wuyts} et~al.,}{{Wuyts}
  et~al.}{2013}]{wuyts13}
{Wuyts} S.,  et~al., 2013, \mn@doi [\apj] {10.1088/0004-637X/779/2/135}, \href
  {https://ui.adsabs.harvard.edu/abs/2013ApJ...779..135W} {779, 135}

\bibitem[\protect\citeauthoryear{{Wyder} et~al.,}{{Wyder}
  et~al.}{2007}]{wyder07}
{Wyder} T.~K.,  et~al., 2007, \mn@doi [\apjs] {10.1086/521402}, \href
  {http://adsabs.harvard.edu/abs/2007ApJS..173..293W} {173, 293}

\bibitem[\protect\citeauthoryear{{Yan}, {Newman}, {Faber}, {Konidaris}, {Koo}
  \& {Davis}}{{Yan} et~al.}{2006}]{yan06}
{Yan} R.,  {Newman} J.~A.,  {Faber} S.~M.,  {Konidaris} N.,  {Koo} D.,
  {Davis} M.,  2006, \mn@doi [\apj] {10.1086/505629}, \href
  {http://adsabs.harvard.edu/abs/2006ApJ...648..281Y} {648, 281}

\bibitem[\protect\citeauthoryear{{Zeimann} et~al.,}{{Zeimann}
  et~al.}{2013}]{zeimann13}
{Zeimann} G.~R.,  et~al., 2013, \mn@doi [\apj] {10.1088/0004-637X/779/2/137},
  \href {http://adsabs.harvard.edu/abs/2013ApJ...779..137Z} {779, 137}

\bibitem[\protect\citeauthoryear{{Ziparo} et~al.,}{{Ziparo}
  et~al.}{2014}]{ziparo14}
{Ziparo} F.,  et~al., 2014, \mn@doi [\mnras] {10.1093/mnras/stt1901}, \href
  {http://adsabs.harvard.edu/abs/2014MNRAS.437..458Z} {437, 458}

\bibitem[\protect\citeauthoryear{{van den Bosch}, {Aquino}, {Yang}, {Mo},
  {Pasquali}, {McIntosh}, {Weinmann}  \& {Kang}}{{van den Bosch}
  et~al.}{2008}]{vandenbosch08}
{van den Bosch} F.~C.,  {Aquino} D.,  {Yang} X.,  {Mo} H.~J.,  {Pasquali} A.,
  {McIntosh} D.~H.,  {Weinmann} S.~M.,   {Kang} X.,  2008, \mn@doi [\mnras]
  {10.1111/j.1365-2966.2008.13230.x}, \href
  {http://adsabs.harvard.edu/abs/2008MNRAS.387...79V} {387, 79}

\bibitem[\protect\citeauthoryear{{van der Burg} et~al.,}{{van der Burg}
  et~al.}{2013}]{vanderburg13}
{van der Burg} R.~F.~J.,  et~al., 2013, \mn@doi [\aap]
  {10.1051/0004-6361/201321237}, \href
  {http://adsabs.harvard.edu/abs/2013A%26A...557A..15V} {557, A15}

\bibitem[\protect\citeauthoryear{{van der Wel} et~al.,}{{van der Wel}
  et~al.}{2014}]{vanderwel14}
{van der Wel} A.,  et~al., 2014, \mn@doi [\apj] {10.1088/0004-637X/788/1/28},
  \href {http://adsabs.harvard.edu/abs/2014ApJ...788...28V} {788, 28}

\makeatother
\end{thebibliography}




\appendix

\section{\normalsize{Creation of Mock Group/Cluster Galaxies Without Environmental Quenching}}
\label{mocks} 
When calculating environmental quenching efficiencies it is necessary to define a sample of galaxies that has had the chance to experience environmental 
quenching, if such quenching exists, and a sample of galaxies that has evolved on a nearly identical track, but have been completely sheltered from processes 
related to environmental quenching. Defining the former set of galaxies becomes a trivial task with the data available to the ORELSE survey as large number
of galaxies have been spectroscopically confirmed in group and cluster environments. The sample that was adopted for this part of the calculation was 
referred to as the high-$\log(1+\delta_{gal})$ sample in the main body of the paper. The definition of the latter population is more tricky. In many studies,
the population that has evolved on a similar track as group/cluster galaxies, though without the possibility of experiencing pervasive environmental quenching, 
is selected as coeval field populations. While this set of galaxies can serve as an approximation of what group/cluster galaxies would have evolved into 
in the absence of environmental quenching processes, there are several issues that make this approximation a crude one. The most pernicious of these issues 
is that field galaxies at a given redshift cannot be the progenitors of and group/cluster galaxies at that same redshift, and it is, rather, field galaxies at
earlier epochs that seed the group/cluster population at a given redshift. Further, it is not possible to fully expiate this issue by substituting a 
field galaxy population observed at a fixed epoch earlier than that of the group/cluster population as group/cluster progenitor galaxies originate from a 
spectrum of earlier epochs. Other issues include the differential growth of stellar mass between field and group/cluster populations and differing 
selections of galaxies in field vs. group/cluster environments. 

In an attempt to circumvent these issues, we generated a suite of mock galaxies that were intended to mimic the properties of group and cluster 
galaxies in absence of all environmental quenching processes. As such, in this simulation we completely ignore the effects of merging, ram-pressure stripping, 
tidal stripping, harassment, strangulation, and all other galaxy-galaxy and galaxy-group/cluster processes that might effect the star formation and stellar 
mass properties in excess of what is experienced in the field. Because we are primarily focused on the quiescent fraction across relatively large
($\sim$0.5 dex) bins in stellar mass, we assume for the purposes of this calculation that none of these processes strongly modulate the quiescent fraction 
in a given bin or cause appreciable cross-talk between the various bins (e.g., a major merger, of which only $\sim$25\% of group/cluster galaxies 
experience during their accretion history \citealt{AA17, debz18}, increases the stellar mass of the more massive component of the merger by, at most, 
$\sim$0.3 dex). Still, since some of these processes are known, in certain cases, to enhance the occurrence and level of star formation in galaxies (e.g., 
\citealt{kronberger08, hopkins13, kaviraj14}) for at least some period of time, the environmental quenching efficiencies derived from these simulations 
in conjunction with our data will still remain lower limits, though much stricter ones than those derived directly from the data. 

The simulation begins by adopting the group/cluster merging histories presented in \citet{mcgee09} and \citet{balogh16}, which are based on the
merger trees estimated from the $N$-body Millennium Simulation as discussed in section \ref{converttime}. 
A linear fit was made to the average accretion history of a cluster with a $z=0$ total mass $\log(\mathcal{M}_{tot}/\mathcal{M}_{\odot})>14.5$, as
the majority of ORELSE groups and cluster are predicted to fall within this mass range by $z=0$ if we assume the accretion history characterized
by those in \citet{mcgee09} and \citet{balogh16} and $M_{vir}$ as the total mass of ORELSE groups/clusters at the redshift they are observed 
(see \citealt{AA17,AA19, rum18} for $M_{vir}$ values of groups/clusters in specific ORELSE fields). Regardless, the results of the simulation 
do not change meaningfully if we instead adopt average merging histories for lower mass structures. The simulation begins at $z=4$ and takes 50
steps equally spaced in time to arrive to $z=0.65$, the average redshift of the lowest redshift sample studied in this paper. In the first time
step, the fraction of halo mass that is predicted to accrete over this time step is computed and multiplied by the total mass at $z=0$ 
($\log(\mathcal{M}_{tot}/\mathcal{M}_{\odot})=14.5$). This value sets the amount of mass accreted into the (proto-)cluster environment in this
time step. In order to translate this amount of halo mass into a galaxy population, we begin sampling from the galaxy total SMF 
presented in \citet{AA14} appropriate for the median redshift of the step (in this case, $z\sim3.85$). For each galaxy sampled 
in this way, a galaxy was assigned:

\begin{itemize}

\item A stellar mass, $\mathcal{M}_{\ast}$

\item A halo mass, $\mathcal{M}_{\rm{h}}$, estimated from inverting the redshift-dependent halo-to-stellar mass relation of \citet{moster10,moster13}.

\item An initial state of quiescence or ongoing star formation, a state that is chosen from sampling a random uniform variable and comparing that
sample to the fraction of quiescent to star-forming galaxies in the field at that redshift and $\mathcal{M}_{\ast}$ as estimated by \citet{AA14} 
(e.g., if the random sample for a simulated galaxy was 0.61 and the fraction of star-forming to total galaxies at that redshift and 
$\mathcal{M}_{\ast}$ was 0.6, the galaxy is considered quiescent) 

\item For those galaxies assigned a state of ongoing star formation, a star formation rate ($\mathcal{SFR}$) was assigned based on the redshift-dependent
$\mathcal{SFR}$-$M_{\ast}$ relation for star-forming galaxies of \citet{AA16} with an imposed scatter of 0.15 dex.

\item An exponentially declining star formation future with a normalization set by the assigned SFR at the time of accretion and an e-folding time 
equal to 2 Gyr, a value appropriate for the general field population at the stellar masses of interest here (e.g., \citealt{AA19}) 

\end{itemize}

Only those galaxies with stellar masses in excess of $\log(\mathcal{M}_{\ast}/\mathcal{M}_{\odot})\ge9$ are considered in the simulation. This limit
is roughly the stellar mass limit of the SMFs presented in \citet{AA14} at all of the redshifts considered here and is sufficient to generate 
essentially all galaxies in the stellar mass and redshift ranges of interest. For this first step, the SMF continues to be sampled until the amount
of halo mass assembled reaches or exceeds 95\% of the halo mass predicted to be accreted in that time step, a threshold that typically requires $\sim$10
galaxies to be generated. At the conclusion of this first step, all mock galaxies are binned into the same three stellar mass bins considered throughout 
this paper, and $f_{q}$ is calculated for each bin. 

In subsequent steps, two galaxy populations are considered: newly accreted galaxies and those galaxies that are already 
coalesced into the cluster environment in previous steps. For the former population, the same steps as listed above are repeated and these
galaxies are incorporated into the full simulated cluster galaxy population. For the latter population, the amount of stellar mass generated 
through \emph{in situ} star formation in that step is added to the previous stellar mass of the galaxy by integrating star formation rate 
over that step and subtracting the amount of stellar mass lost over that same time period resulting from stellar evolution (BC03, \citealt{chab03, moster13}). 
The value of $f_{q}$ is then calculated for the combined sample of galaxies in the three $\mathcal{M}_{\ast}$ bins considered throughout 
this paper.

At the conclusion of this process, the $\mathcal{SFR}$ of each galaxy is evolved by the time elapsed in the step subject to its star formation future set at the moment of 
coalescence. Finally, we compute the specific star formation rate ($\mathcal{SSFR}$) of all galaxies at the end of this time step and set the 
star formation state of any galaxies with a $\log(\mathcal{SSFR}/\rm{yr}^{-1})\le-11$ to quiescent. This $\mathcal{SSFR}$ threshold is chosen as it has been shown that galaxies
below this limit are almost exclusively found in the $NUVrJ$ quiescent region \citep{ilbert13, olga17}. A similar exercise is not performed for accreted 
field galaxies as those galaxies which have their initial state set as star forming, their $\mathcal{SSFR}$s are set by the star-forming 
$\mathcal{SFR}$-$M_{\ast}$ relation, and, thus, by construction, cannot be considered quiescent. Once quiescent, the simulation ignores the potential for
rejuvenation (e.g., \citealt{krazykannappan09, krazykannappan13}) of the star formation for each galaxy for the remainder of the time of the simulation. 
The results of these simulations are shown in the main body of the paper in Figure \ref{fig:simulatedfq}. 


\bsp	
\label{lastpage}
\end{document}